\documentclass[a4paper,11pt]{article}
\pdfoutput=1 
\usepackage{jcappub}
\usepackage{amsmath}
\usepackage{graphicx}
\usepackage{latexsym}
\usepackage{xspace}
\usepackage{color}
\usepackage{hyperref} 
\usepackage{bm}
\usepackage{relsize}
\usepackage{tabularx}
\usepackage{multirow}
\usepackage{amssymb}
\usepackage[table]{xcolor}
\usepackage{slashbox}
\usepackage{braket}
\usepackage{comment}
\usepackage[utf8]{inputenc}
\usepackage{slashed}
\usepackage{empheq}

\allowdisplaybreaks

\makeatletter
\gdef\@fpheader{}
\g@addto@macro\bfseries{\boldmath}
\makeatother

\makeatletter
\renewcommand*\env@matrix[1][\arraystretch]{%
  \edef\arraystretch{#1}%
  \hskip -\arraycolsep
  \let\@ifnextchar\new@ifnextchar
  \array{*\c@MaxMatrixCols c}}
\makeatother

\newcommand{\ds}{\displaystyle}

\newcommand{\spn}[1]{$\textsl{span}$\left\{#1\right\}}
\newcommand{\dz}{ \delta\vec{z}_{\vec{k}} }
\newcommand{\dzdot}{  \dot{\delta\vec{z}}_{\vec{k}} }
\newcommand{\dztilde}{ \widetilde{\delta\vec{z}}_{\vec{k}} }
\newcommand{\dq}{ \delta\vec{q}_{\vec{k}} }
\newcommand{\dqi}[1]{\delta q_{\vec{k}\,#1}}
\newcommand{\B}[1]{B_{#1}}
\newcommand{\Pbasis}{physical basis\xspace}
\newcommand{\Dbasis}{dynamical basis\xspace}

\newcommand{\ie}{\textsl{i.e.~}}

\newcommand{\eg}{\textsl{e.g.~}}

\newcommand{\dd}{\mathrm{d}}

\newcommand{\sss}[1]{{\scriptscriptstyle{#1}}}
\newcommand{\boldmathsymbol}[1]{{\ensuremath{\boldsymbol{#1}}}}

\newcommand{\uPl}{\mathrm{Pl}}
\newcommand{\uin}{\mathrm{in}}

\newcommand{\usssPl}{\sss{\uPl}}

\newcommand{\Mp}{M_\usssPl}

\newcommand{\beq}{\begin{equation}}
\newcommand{\eeq}{\end{equation}}
\newcommand{\bea}{\begin{eqnarray}}
\newcommand{\eea}{\end{eqnarray}}

\newlength{\wsingfig}
\newlength{\wdblefig}
\newlength{\wquadfig}
\newlength{\wtriplefig}
\newcommand{\Eq}[1]{Eq.~(\ref{#1})}
\newcommand{\Eqs}[1]{Eqs.~(\ref{#1})}
\newcommand{\Fig}[1]{Fig.~{\ref{#1}}}

\newcommand{\Refc}[1]{Ref.~{\cite{#1}}}
\newcommand{\Refs}[1]{Refs.~{\cite{#1}}}

\newcommand{\Sec}[1]{Sec.~\ref{#1}}
\newcommand{\App}[1]{App.~\ref{#1}}

\subheader{}

\title{Hamiltonian formalism for cosmological perturbations: fixing the gauge}

\author[a,b,c]{Danilo Artigas,}
\emailAdd{artigas@tap.scphys.kyoto-u.ac.jp}

\author[a]{Julien Grain,}
\emailAdd{julien.grain@universite-paris-saclay.fr}

\author[d]{Vincent Vennin}
\emailAdd{vincent.vennin@ens.fr}

\affiliation[a]{Universit\'e Paris-Saclay, CNRS, Institut d'Astrophysique Spatiale, 91405, Orsay, France}
\affiliation[b]{Institute of Theoretical Physics, Jagiellonian University, \L ojasiewicza 11, 
30-348 Cracow, Poland, EU}
\affiliation[c]{Department of Physics, Kyoto University, Kyoto 606-8502, Japan}
\affiliation[d]{Laboratoire de Physique de l'Ecole Normale Sup\'erieure, ENS, CNRS, Universit\'e PSL, Sorbonne Universit\'e, Universit\'e Paris Cit\'e, F-75005, Paris, France}

\date{today}

\begin{document}
\sloppy
\begin{flushright}
{\small{\textsc{Matches published version in JCAP.}}}
\end{flushright}

\abstract{Cosmological perturbation theory is an example of a gauge theory, where gauge transformations correspond to changes in the space-time coordinate system. To determine physical quantities, one is free to introduce gauge conditions (\ie to work with specific space-time coordinates), and such conditions are often used to simplify technical aspects of the calculation or to facilitate the interpretation of the physical degrees of freedom. Some of the prescriptions introduced in the literature are known to fix the gauge only partially, but it is commonly assumed that the remaining gauge degrees of freedom can be fixed somehow. In this work, we show that this is not necessarily the case, and that some of these gauges are indeed pathological. We derive a systematic procedure to determine whether a gauge is pathological or not, and to complete partially-fixed gauges into healthy gauges when this is possible. In this approach, the Lagrange multipliers (\ie the perturbed lapse and shift in the ADM formalism) cannot appear in the off-shell definition of the gauges, they necessarily arise as on-shell consequences of the gauge conditions. As illustrative applications, we propose an alternative, non-pathological formulation of the synchronous gauge, and we show that the uniform-expansion gauge (as well as any gauge ensuring vanishing lapse perturbations) can hardly be made healthy. Our methodology also allows us to construct all gauge-invariant variables. We further show that our non-pathological criterion for gauges is also the one that ensures Dirac brackets to be properly defined. This allows cosmological perturbations to be quantised in a gauge-fixed way. We finally discuss possible generalisations of our formalism.}


\arxivnumber{2309.17184}

\maketitle

\flushbottom

\section{Introduction}
Cosmological-perturbation theory (CPT) is a key ingredient of modern cosmology~\cite{Mukhanov:1990me,Malik:2008im}. It lays the ground for our understanding of cosmic structures, from their origin during inflation to their evolution at large scales later on in the cosmic history. In this approach, inhomogeneities are described by small deviations from the homogeneous and isotropic background space-time by means of perturbative techniques, and CPT inherits the fundamental invariance of General Relativity under the change of the coordinate system. 

In the phase-space formulation of the theory, this gauge symmetry yields a constrained Hamiltonian with Lagrange multipliers being the lapse function labelling time and the shift vector labelling space. As a consequence, not all degrees of freedom can be considered physical. Indeed, take for instance a system made of $2n$ degrees of freedom in the phase space, whose dynamics is generated by a Hamiltonian containing $m<n$ constraints. The space of solution has to meet the constraints, which removes $m$ degrees of freedom. Among the $2n-m$ remaining ones, $m$ of them will be fixed by the Lagrange multipliers, which are arbitrarily chosen. These are called gauge degrees of freedom for they are the ones changing under gauge transformations generated by the constraints. Since physical solutions should hold for any arbitrary choice of the Lagrange multipliers, the gauge degrees of freedom cannot carry physical information. Hence, the system is truly composed of $2(n-m)$ physical degrees of freedom~\cite{doi:10.1063/1.529065}. 

CPT applied in the context of inflation with $n$ scalar fields is exactly of the type described above. The phase space of scalar perturbations is composed of $2n+4$ degrees of freedom (where the four additional degrees of freedom come from the gravitational sector) and the Hamiltonian contains 2 algebraic constraints -- the two Lagrange multipliers being the perturbed lapse function and the scalar part of the perturbed shift vector. This leaves $2n$ physical degrees of freedom out of the $2n+4$. Hence in CPT, removing gauge degrees of freedom is necessary to make physical predictions free of any arbitrariness and to define gauge-invariant observables to be compared with cosmological observations. 

To this end, one approach consists in fixing the gauge. This exploits our freedom to tune the Lagrange multipliers such that the gauge degrees of freedom are unequivocally fixed by the physical degrees of freedom. Gauges are usually selected in order to simplify the equations of motion or because they lead to a simple understanding of the coordinate system to work in. However, they are rarely chosen on the {\textsl{a priori}} basis of removing the gauge degrees of freedom. Instead, this property is  checked to hold or not {\textsl{a posteriori}} and on a case-by-case basis. Several gauges are known to be pathological for they do not remove all the gauge degrees of freedom, and several examples of non-pathological gauges are known as well. Despite these examples, a systematic prescription to build non-pathological gauges has not been proposed yet, which is the main goal of this paper. 

Another motivation of this analysis is to formulate the process of gauge-fixing in the Hamiltonian formalism, which will prove particularly handy when it comes to selecting non-pathological gauges. It is also worth stressing that gauge-fixing in the Hamiltonian framework is necessary for all situations relying on the phase-space formulation of CPT, of which we now give a few examples.

First, the phase-space formulation is the most natural framework to adopt when it comes to the quantum-mechanical aspects of cosmological perturbations. For instance, the choice of the initial state is intimately related to the choice of the phase-space parameterisation~\cite{Grain:2019vnq}. Though our analysis is classical, formalising gauge-fixing in the Hamiltonian approach lays the ground for in-depth studies of gauge-fixing and gauge transformations at the quantum level. 

A second situation in which the phase-space description of cosmological perturbations is essential is the separate-universe approach~\cite{Salopek:1990jq,Sasaki:1995aw,Wands:2000dp,PhysRevD.68.103515,Rigopoulos:2003ak,Lyth:2005fi, Tanaka:2021dww} (or ``quasi-isotropic picture''~\cite{Lifshitz:1960,Starobinsky:1982ee,PhysRevD.49.2759,Khalatnikov_2002}). This applies to scales much larger than the Hubble radius if the inhomogeneous universe can be described by a set of independent, locally isotropic and homogeneous patches. When valid, this picture greatly simplifies technical calculations of the evolution of cosmic inhomogeneities, and captures relevant non-linear effects at large scales. The conditions for the separate-universe approximation to hold have been discussed in \Refs{Rigopoulos:2003ak, Nambu:2005ne, Abolhasani:2019cqw, Pattison:2019hef, Tanaka:2021dww, Artigas:2021zdk, Cruces:2022dom}. They allow one to tackle situations that go far beyond slow-roll inflation, including for instance inflationary phases in the ultra-slow-roll regime~\cite{Pattison:2018bct}, contracting universes~\cite{Miranda:2019ara,Grain:2020wro}, or bouncing cosmologies~\cite{Wands:1998yp,Finelli:2001sr,Khoury:2001wf,Barrau:2013ula,Brandenberger:2016vhg,Agullo:2016tjh}. All these situations demand to keep track of the entire phase space. However, for this approximation to effectively work in a gauge-fixed manner, it is necessary that the space-time labelling of the separate universes matches the one fixed by the gauge chosen at the level of CPT -- at least in some large-scale limit (see \Refs{Pattison:2019hef, Artigas:2021zdk}). For example, such a matching holds using the synchronous gauge which is nonetheless notoriously known to be pathological in CPT, but fails in other gauges unless specific prescriptions are implemented~\cite{Artigas:2021zdk}. The Hamiltonian take on gauge-fixing presented in this work is a mandatory step for building a generic prescription that properly matches the separate-universe picture to CPT in a gauge-fixed way.

A third situation is the so-called stochastic-inflation formalism~\cite{Starobinsky:1982ee,Starobinsky:1986fx}, which rests on the separate-universe approximation and models the backreaction of quantum fluctuations onto the local background dynamics. In this approach, cosmological perturbations stretched well beyond the Hubble radius act as a stochastic noise on the large-scale dynamics of the universe. This formalism can be combined with the $\delta N$-formalism to extract the statistics of curvature perturbations~\cite{Fujita:2013cna, Vennin:2015hra}. This allows one to describe non-perturbative effects (for instance, non-perturbative deviation from Gaussian statistics) that are particularly relevant to the formation of primordial black holes~\cite{Pattison:2017mbe,Biagetti:2018pjj,Ezquiaga:2019ftu} and other extreme objects such as heavy dark-matter haloes~\cite{Ezquiaga:2022qpw}. Large stochastic effects mainly occur in models where deviations from slow roll are also encountered and the stochastic-$\delta N$ formalism has been extended to these situations where the full phase-space needs to be accounted for~\cite{Nakao:1988yi, PhysRevD.46.2408, Rigopoulos:2005ae, Tolley:2008na, Weenink:2011dd, Grain:2017dqa, Firouzjahi:2018vet, Ezquiaga:2018gbw, Pattison:2019hef, Pattison:2021oen}. Moreover, the stochastic noise is most naturally expressed in the phase space since it is derived from the quantum-mechanical description of cosmological perturbations. So far, the stochastic-inflation formalism has been built in a gauge-fixed manner (for instance in the uniform-expansion gauge when combined with the $\delta N$-formalism~\cite{Pattison:2019hef}), which again stresses the importance of having a clear understanding of the gauge-fixing procedure in the Hamiltonian framework.

For all these reasons, our analysis will be carried out in the phase space of CPT, contrary to most studies of CPT gauge fixing, which are done in the Lagrangian formalism (note however formal studies of gauge-fixing in constrained Hamiltonian system in \Refs{Pons:1995su,PhysRevD.55.658,Pons:1998ht}). Part of our approach follows similar lines of thoughts -- with a less formal tone -- as the one presented in \Refc{Boldrin:2022vcp}, which treats cosmological perturbations in a Kasner universe using the Kucha\v{r} decomposition~\cite{Boldrin:2023jqc,doi:10.1063/1.1666050}.

Let us finally mention that removing gauge degrees of freedom can also be done by identifying gauge-invariant variables to work with~\cite{Bardeen:1983qw}. By construction, they are free from any gauge degrees of freedom, otherwise they would change under gauge transformations. Several gauge-invariant variables have been proposed in the literature. However, a systematic procedure to build them is still lacking. One of our motivations is thus to lay the ground for a generic prescription to build gauge-invariant variables, alternative to the one based on generating functions \cite{Langlois:1994ec, Domenech:2017ems}, as well as for defining the set of gauge-invariant quantum states.

Our analysis is performed in the context of Friedmann-Lema\^itre-Robertson-Walker (FLRW) space-times filled with a single scalar field, and we will work at leading order in perturbations. The rest of the paper is organised as follows. The theory of cosmological perturbations in the Hamiltonian framework is presented in \Sec{sec:cosmo} where special attention is paid to the role of constraints and to gauge transformations. In order to prepare for the rest of our analysis, convenient vectorial notations are also introduced there. In \Sec{sec:PSgen} we build a set of basis vectors on the phase space of cosmological perturbations which clearly separates the physical degrees of freedom from the constraints and from the gauge degrees of freedom. This new parameterisation of the phase space is studied at the kinematical level and at the dynamical level. In \Sec{sec:GF} this parameterisation is then used to identify the properties that gauge conditions must fulfil for the gauge choice to be non-pathological. This sheds new light on the role played by Lagrange multipliers in the process of gauge-fixing and we show that gauges that are uniquely fixed are non-pathological. In \Sec{sec:WorkEx}, our formalism is applied to examples of well-known gauges -- showing for instance how to build the synchronous gauge in a non-pathological way. We also introduce in this section new classes of gauge. \Sec{sec:GIvar} is devoted to the construction of gauge-invariant variables using this new parameterisation of the phase space that we further extend to the quantum-mechanical framework. Finally, we conclude by summarising our results and by opening on a few additional applications of our formalism in \Sec{sec:Conclusion}. The paper ends with technical appendices to which we defer the derivation of some of the results used in the main text.

\section{Cosmology in the phase space}
\label{sec:cosmo}

This section is devoted to a presentation of cosmology, including perturbations, in the Hamiltonian formalism. This approach has been widely used  in the literature and we refer the interested reader to \eg \Refs{thieman_book,Langlois:1994ec,Artigas:2021zdk} for more details (see also \eg \Refs{Mukhanov:1990me,Malik:2008im} for reviews in the Lagrangian framework). Our notations and definitions follow the ones of \Refc{Artigas:2021zdk}.

In the Arnowitt-Deser-Misner (ADM) formalism~\cite{Arnowitt:1962hi}, the line element is expressed as
\bea
\label{eq:metric:ADM}
	\dd s^2=-N^2(t,\vec{x})\dd\tau^2+\gamma_{ij}(\tau,\vec{x})\left[\dd x^i+N^i(\tau,\vec{x})\dd\tau\right]\left[\dd x^j+N^j(\tau,\vec{x})\dd\tau\right],
\eea
where $N$ is the lapse function, $N^i$ the (three-dimensional) shift vector, and $\gamma_{ij}$ is the induced metric on the spatial hypersurfaces (indices are lowered by $\gamma_{ij}$ and raised by its inverse $\gamma^{ij}$). Fixing $N$ is equivalent to setting a time parameterisation,\footnote{In cosmology, commonly-used choices are $N=1$ corresponding to cosmic time, denoted $t$ hereafter, and $N=a$ with $a$ the scale factor corresponding to conformal time, denoted $\eta$. Another convenient time-slicing is to set $N$ equal to the Hubble rate $H$, hence the time variable is the number of e-folds $\mathcal{N}$.} while choosing $N^i$ fixes a space parameterisation. 
 
When the matter content of the universe is made of a single scalar field $\phi$, General Relativity is described through a total Hamiltonian of the form
\bea
C[N,N^i]= \int \dd^3\vec{x} \left[ N\left(\mathcal{S}^{(G)}+\mathcal{S}^{(\phi)}\right) + N^i\left(\mathcal{D}_i^{(G)} + \mathcal{D}_i^{(\phi)}\right)\right] .
\label{HamGR}
\eea
Variation of the action with respect to the lapse and the shift lead respectively to the scalar constraint $\mathcal{S}^{(G)}+\mathcal{S}^{(\phi)}=0$ and to the diffeomorphism constraints $\mathcal{D}_i^{(G)} + \mathcal{D}_i^{(\phi)}=0$, where the superscript ``$(G)$'' denotes the gravitational contribution. These constraints are all functionals of the phase space, defined by a symmetric space metric $\gamma_{ij}$, its momentum $\pi^{ij}$, a scalar field $\phi$ and its momentum $\pi_\phi$. 

Note that here, we follow the convention where the lapse $N$ and shift $N^i$ are treated as Lagrange multipliers~\cite{thieman_book,Pons:1998ht}. Alternatively, they may also be included in the list of phase-space variables, see \App{app:GTHam} for further comparison between the two viewpoints.

\subsection{Background dynamics}
 We assume that space-time is homogeneous and isotropic at the background level, so it can be described by a FLRW metric where the shift vector $N^i(\tau,\vec{x})$ vanishes. Moreover, we assume that spatial hypersurfaces are flat (since inflation makes any initial amount of spatial curvature decay away very rapidly), so $\gamma_{ij}= v^{2/3}\widetilde{\gamma}_{ij}$ with $\widetilde{\gamma}_{ij}=\mathrm{diag}[1,1,1]$ denotes the euclidean metric and $\widetilde{\gamma}^{im}\widetilde{\gamma}_{mj}=\delta^i_j$. The scalar constraint thus reduces to
\bea
\mathcal{S}^{(0)} = - \frac{3}{4 \Mp^2} v\theta^2 + \frac{1}{2v} \pi_\phi^2 + vV(\phi) =0 \,.
\eea
This constraint is also known as the Friedmann equation, more commonly rearranged under the following form:
\bea
\theta^2=\frac{4\Mp^2}{3}\left[\frac{\pi_\phi^2}{2v^2}+V(\phi)\right] . \label{eq:Friedman}
\eea
It involves the background degrees of freedom: the gravitational contribution is described by the covolume $v(\tau)$ and its canonical momentum, the expansion rate $\theta(\tau)$.\footnote{In terms of the usual scale factor $a$ and Hubble factor $H$, one can define these variables as $v:=a^3$ and $\theta:=-2H\Mp^2/N$, with $\Mp$ the reduced Planck mass.} The variables $\phi(\tau)$ and $\pi_\phi(\tau)$ denote the scalar field and its canonical momentum respectively, and $V(\phi)$ is the potential energy of the scalar field.

The dynamics of these background degrees of freedom is obtained from the Poisson bracket $\dot{z}=\{z,C^{(0)}\}$, where we have defined
\bea
	\left\{F,G\right\}&=&\ds\sum_{A=1}^2\left(\frac{\partial F}{\partial f_A}\frac{\partial G}{\partial\pi_A}-\frac{\partial G}{\partial f_A}\frac{\partial F}{\partial\pi_A}\right),
\eea
for two arbitrary functionals of the phase space $F$ and $G$, with the configuration variables $f_A=(v,\phi)$ and their momenta $\pi_A=(\theta, \pi_\phi)$. One can thus derive the equations of motion for the background
\bea
	\left\{\begin{array}{l}
		\ds\dot\phi = N \frac{\pi_\phi}{v}  \\
		\ds\dot\pi_\phi  =  -NvV_{,\phi}
	\end{array}\right. \label{eq:BckgFull} 
\quad \quad \text{and}\quad \quad 
	\left\{\begin{array}{l}
		\ds\dot{v}=- \frac{3N}{2\Mp^2} v\,\theta \\
		\ds\dot\theta=N\left(\frac{\pi_\phi}{v}\right)^2
	\end{array}\right. \quad \, ,
\eea
where the constraint, \Eq{eq:Friedman}, has been used to simplify the right-hand side of the last equation.

\subsection{Perturbation dynamics}
\label{section:Perturbations}
\subsubsection{Definition of the perturbations}
In order to achieve a more realistic description of the universe, we now take the whole system defined by General Relativity and subtract from it the FLRW background. This allows one to define perturbations as deviations from a homogeneous and isotropic spacetime
\bea
	&&\left\{\begin{array}{l}
		\delta \phi(\tau, \vec{x}) = \phi(\tau, \vec{x}) - \phi(\tau)\,, \\
		\delta \pi_\phi(\tau, \vec{x})  =  \pi_\phi(\tau, \vec{x}) - \pi_\phi(\tau)\,,
	\end{array}\right. 
	\quad\quad	\quad\quad
	\left\{\begin{array}{l}
		\delta \gamma_{ij} (\tau,\vec{x}) =\gamma_{ij}(\tau,\vec{x}) -\gamma_{ij}(\tau)  \,, \\
		\delta \pi^{ij}(\tau,\vec{x}) =\pi^{ij}(\tau,\vec{x}) - \pi^{ij}(\tau) \,, \nonumber
	\end{array}\right. 
\eea
and one can act similarly with the lapse and shift
\bea
	&&\left\{\begin{array}{l}
		\delta N(\tau, \vec{x}) = N(\tau, \vec{x}) - N(\tau), \\
		\delta N^i(\tau, \vec{x})  =  N^i(\tau, \vec{x}) \,.
	\end{array}\right.
\eea

Since background hypersurfaces are homogeneous and isotropic, all the perturbed quantities will be preferentially defined in Fourier space. The Fourier transform and its inverse are
\bea
	\delta\phi(\vec{k},\tau)&=&\ds\int\frac{\dd^3x}{(2\pi)^{3/2}}\delta\phi(\vec{x},\tau)e^{-i\vec{k}\cdot\vec{x}}\, , \\
	\delta\phi(\vec{x},\tau)&=&\ds\int\frac{\dd^3k}{(2\pi)^{3/2}}\delta\phi(\vec{k},\tau)e^{i\vec{k}\cdot\vec{x}}\, ,
\eea
with similar expressions for other perturbation fields. This leads to $\int\dd^3k\,e^{i\vec{k}\cdot\vec{x}}=(2\pi)^3\delta^3(\vec{x})$ and $\int\dd^3x\,e^{i\vec{k}\cdot\vec{x}}=(2\pi)^3\delta^3(\vec{k})$ with $\delta^3$ the Dirac distribution. The wavevector is defined with respect to the flat euclidean metric, hence it is the comoving wavenumber. We thus have $k^i=\widetilde{\gamma}^{ij}k_j$ and its norm is $k^2=\widetilde{\gamma}^{ij}k_ik_j$. We finally stress that Fourier components are subject to the reality constraint, $\phi^\star(\vec{k},\tau)=\phi(-\vec{k},\tau)$, since all quantities considered here are real-valued fields. 

For the metric fields, the Fourier modes of $\delta\gamma_{ij}$ are further decomposed into
\bea
\delta\gamma_{ij}(\tau,\vec{k}) = \sum_{A=1}^2\delta\gamma_A(\tau,\vec{k}) M^{A}_{ij}(\vec{k})\, 
\, ,\\
\delta\pi^{ij}(\tau,\vec{k}) = \sum_{A=1}^2\delta\pi_A(\tau,\vec{k}) M_{A}^{ij}(\vec{k})\, 
\, ,
\eea
where we have kept scalar degrees of freedom only, and we have introduced the orthonormal basis
\bea
	M^1_{ij}(\vec{k})=\frac{1}{\sqrt{3}}\widetilde{\gamma}_{ij} &~~~\mathrm{and}~~~& M^2_{ij}(\vec{k})=\sqrt{\frac{3}{2}}\left(\frac{k_ik_j}{k^2}-\frac{\widetilde{\gamma}_{ij}}{3}\right).
\eea
In particular, one can check that $M^A_{ij} M_{A'}^{ij} = \delta^A_{A'}$. Since $M_1$ is a pure trace, $\delta\gamma_1$ describes inhomogeneous but isotropic perturbations while $\delta\gamma_2$ characterises inhomogeneous and anisotropic perturbations.
Finally, the lapse and shift perturbations read
\bea
	&&\left\{\begin{array}{l}
		\delta N(\tau, \vec{x}) = \ds\int\frac{\dd^3k}{(2\pi)^{3/2}}\delta N(\tau,\vec{k})e^{i\vec{k}\cdot\vec{x}} , \\
		\delta N^i(\tau, \vec{x})  =  \ds\int\frac{\dd^3k}{(2\pi)^{3/2}}\delta N_1(\tau,\vec{k})\,i \frac{k^i}{k}e^{i\vec{k}\cdot\vec{x}}
		\,,
	\end{array}\right. 
\eea
where we again restricted the shift perturbation to its scalar component only.

\subsubsection{Hamiltonian}

Including the above expressions for the perturbed quantities into the total Hamiltonian, one can derive the perturbed Hamiltonian. A detailed study shows that the first order in perturbation vanishes identically when the background satisfies its equations of motion (see \eg \Refc{Artigas:2021zdk}). At quadratic order one then finds
\bea
	C^{(2)}[N+\delta N,\delta N^i]=\ds\int_{\mathbb{R}^{3+}}\dd^3\vec{k}\left[\left(\delta N^\star(\vec{k}) \,\mathcal{S}^{(1)}_{\vec{k}}+\mathrm{c.c.}\right)+k\left(\delta N_1^\star(\vec{k})\, \mathcal{D}^{(1)}_{\vec{k}}+\mathrm{c.c.}\right)+ 2N\,\mathcal{S}^{(2)}_{\vec{k}}\right], \nonumber \\
\label{eq:quadratic:action}	
\eea
where ``c.c.'' means ``complex conjugate'' of the previous term. We stress that integration is over $\mathbb{R}^{3+}=\mathbb{R}^2\times\mathbb{R}^+$ to avoid double-counting as required by the reality condition. In the above, $\mathcal{S}^{(1)}_{\vec{k}}$ and $\mathcal{D}^{(1)}_{\vec{k}}$ are the scalar constraint and the (scalar component of the) diffeomorphism constraint, linearised at first order in perturbative degrees of freedom. Differentiating the action with respect to the perturbed lapse and shift leads to the linear constraints $\mathcal{S}^{(1)}=0$ and $\mathcal{D}^{(1)}=0$. Their expression in Fourier space is given by\footnote{Although $\mathcal{S}^{(1)}$ and $\mathcal{D}^{(1)}$ are second-order constraints, they are linear in perturbative degrees of freedom, so they will be referred to as ``linear constraints'' to avoid any confusion with $\mathcal{S}^{(2)}$.}
\bea
	\mathcal{S}^{(1)}_{\vec{k}}&=&-\frac{\sqrt{3}}{\Mp^2}v^{2/3}\theta\,\delta\pi_1-\frac{v^{1/3}}{\sqrt{3}}\left(\frac{\pi_\phi^2}{v^2}-V+{\Mp^2}\frac{k^2}{v^{2/3}}\right)\,\delta\gamma_1+\frac{\Mp^2}{\sqrt{6}}\frac{k^2}{v^{1/3}}\,\delta\gamma_2+\frac{\pi_\phi}{v}\,\delta\pi_\phi+vV_{,\phi}\delta\phi, \nonumber \\ \label{eq:scal1gen}
\eea
and
\bea
	\mathcal{D}^{(1)}_{\vec{k}}&=&\pi_\phi\,\delta\phi+\frac{1}{\sqrt{3}}v^{1/3}\theta\,\left(\frac{1}{2}\delta\gamma_1-\sqrt{2}\delta\gamma_2\right)-\frac{2}{\sqrt{3}}v^{2/3}\,\left(\delta\pi_1+\sqrt{2}\delta\pi_2\right), \label{eq:diff1gen}
\eea
where we made the $\vec{k}$ and $\tau$ dependence of the phase-space variables implicit in order to lighten the notations. We stress that the background function multiplying $\delta\gamma_1$ in $\mathcal{S}^{(1)}_{\vec{k}}$ has been simplified thanks to the scalar constraint at the background level, \Eq{eq:Friedman}. 

Finally, when expanding the scalar constraint up to quadratic order in perturbations, one finds
\bea
\mathcal{S}^{(2)}_{\vec{k}}&=&\frac{v^{1/3}}{\Mp^2}\left(2\left|\delta\pi_2\right|^2-\left|\delta\pi_1\right|^2\right)+\frac{1}{2v}\left|\delta\pi_\phi\right|^2+\frac{v}{2}\left(\frac{k^2}{v^{2/3}}+V_{,\phi,\phi}\right)\left|\delta\phi\right|^2 \nonumber\\  
	&&+\frac{1}{3v^{1/3}}\left(\frac{\pi^2_\phi}{v^2}+\frac{V}{2} - \frac{\Mp^2 k^2}{4 v^{2/3}}\right)\left|\delta\gamma_1\right|^2 + \frac{1}{3v^{1/3}}\left(\frac{\pi^2_\phi}{v^2}+\frac{V}{2} - \frac{\Mp^2 k^2}{8 v^{2/3}}\right) \left|\delta\gamma_2\right|^2 \nonumber \\
	&&-\frac{\theta}{4\Mp^2}\left(\delta\pi_1\delta\gamma^\star_1+\mathrm{c.c.}\right)+\frac{\theta}{2\Mp^2}\left(\delta\pi_2 \delta\gamma^\star_2+\mathrm{c.c.}\right) 
	+\frac{\sqrt{2}\Mp^2}{24v}k^2\left(\delta\gamma_1\delta\gamma_2^\star+\mathrm{c.c.}\right)
	\nonumber \\
	&&-\frac{\sqrt{3}}{4}v^{1/3}\left[\left(\frac{\pi_\phi}{v^2}\delta\pi_\phi-V_{,\phi}\delta\phi\right)\delta\gamma_1^\star+\mathrm{c.c.}\right]. \label{eq:S2}
\eea
It is important to stress that technically, $\mathcal{S}^{(2)}$ is not a constraint since it is not multiplied by a Lagrange multiplier of the perturbative phase-space, hence it has no reason to vanish. For convenience however, we will refer to it as the ``quadratic constraint''.

\subsubsection{Equations of motion}

Since the background subtraction is a canonical transformation, and since the Fourier transform is a canonical transformation too, the phase space of the system is canonically described by the three perturbation variables $(\delta \phi(\vec{k}),\delta \pi_\phi(\vec{k}))$, $(\delta \gamma_1(\vec{k}),\delta \pi_1(\vec{k}))$ and $(\delta \gamma_2(\vec{k}),\delta \pi_2(\vec{k}))$. As a consequence, the Poisson brackets are preserved and act on the perturbation degrees of freedom as follow:
\bea
\label{eq:Poisson:CPT:gen}
	\left\{F,G\right\}&=&\ds\sum_{A=1}^3\int\dd^3\vec{k}\left(\frac{\delta F}{\delta \,\delta f_A}\frac{\delta G}{\delta\,\delta\pi^\star_A}-\frac{\delta G}{\delta \,\delta f^\star_A}\frac{\delta F}{\delta\,\delta\pi_A}\right),
\eea
where the index $A$ now runs up to three because of the three degrees of freedom listed above.

For conciseness of the equations, we choose to adopt a vector notation. We arrange the perturbative degrees of freedom into a vector
\bea
\label{eq:pert:vect:def}
\dz^{\mathrm{T}}=\left( \frac{\lambda^{3/2}}{m_\phi} \delta \phi(\vec{k})\, ,\, \frac{m_1^2}{\sqrt{\lambda}} \delta\gamma_1(\vec{k})\, ,\, \frac{m_2^2}{\sqrt{\lambda}} \delta\gamma_2(\vec{k})\, ,\,\frac{m_\phi}{\lambda^{3/2}} \delta\pi_\phi(\vec{k})\, ,\, \frac{\sqrt{\lambda}}{m_1^2}\delta\pi_1(\vec{k})\, ,\, \frac{\sqrt{\lambda}}{m_2^2}\delta\pi_2(\vec{k}) \right),
\label{eq:deltaz}
\eea
where ``$\mathrm{T}$'' stands for vector transpose. Note that the entries of $ \dz $ have been rescaled by the mass parameters $m_\phi$, $m_1$ and $m_2$ so they all have vanishing mass dimensions. Similarly, we have introduced a conformal parameter $\lambda$, which has the same conformal weight as the scale factor $a={v}^{1/3}$, such that all the components of $ \dz $ have the same conformal dimension $\lambda^{-3/2}$. The reason for introducing these additional parameters is two-fold. First, they will make all entries of vectorial objects of the same dimension, which is convenient when it comes to checking dimensional consistency of our expressions. Second, they will allow us to verify that our physical results do not depend on the units in which perturbations are being measured, which is an important sanity check. 

In vectorial notation, the Poisson brackets~\eqref{eq:Poisson:CPT:gen} give rise to
\bea
	\left\{ \dz ,\delta\vec{z}^{\,\dag}_{\vec{k}'}\right\}=\boldsymbol{\Omega}_6\,\delta^3(\vec{k}+\vec{k}'),
\eea
where $\dag$ stands for the complex conjugate transpose and 
\bea
\boldsymbol{\Omega}_6=\left(\begin{array}{cc}
		\boldsymbol{0}_3 & \boldsymbol{I}_3 \\
		-\boldsymbol{I}_3 & \boldsymbol{0}_3
	\end{array}\right)
\eea
is the $6\times 6$ symplectic matrix.\footnote{Hereafter, the $n\times n$ identity matrix is denoted $\boldsymbol{I}_n$, and 
\bea
	\boldsymbol{\Omega}_{2n}=\left(\begin{array}{cc}
		\boldsymbol{0} & \boldsymbol{I}_n \\
		-\boldsymbol{I}_n & \boldsymbol{0}
	\end{array}\right)
\eea
is the $2n\times2n$ symplectic matrix.} We note that from the reality condition, $ \dz ^{\,\dag}=\delta\vec{z}_{-\vec{k}}^{\,\mathrm{T}}$. Hence the above is equivalently rewritten as $\left\{ \dz ,\delta\vec{z}^{\mathrm{T}}_{\vec{k}'}\right\}=\boldsymbol{\Omega}_6\,\delta^3(\vec{k}-\vec{k}')$. Note also that at leading order in CPT, different Fourier modes decouple, given that the quadratic action~\eqref{eq:quadratic:action} does not involve mode mixing. As a consequence, the phase-space structure can be described Fourier mode by Fourier mode, and in the following we shall thus restrict to one particular Fourier mode $\vec{k}$. When doing so, it is convenient to rescale the phase-space vector by an (infinite) volume factor,
\bea
 \dz  \to \delta^{-3/2}(\vec{0})  \dz  \, ,
\eea
in order to absorb the Dirac distribution evaluated at $\vec{k}-\vec{k}=\vec{0}$ that appears in the above Poisson bracket. After this rescaling is performed, all entries of $ \dz $ have vanishing mass \emph{and} conformal dimensions, and the Poisson bracket can simply be written as
\bea
\label{eq:Poisson:simp}
\left\lbrace  \dz ,  \dz ^{\mathrm{T}}\right\rbrace = \boldmathsymbol{\Omega}_6\, . 
\eea

These definitions allow us to recast the linear constraints in a simple form,
\bea
	\mathcal{S}^{(1)}_{\vec{k}}:=\vec{S}_k\cdot \dz =0,& ~~~\mathrm{and}~~~& \mathcal{D}^{(1)}_{\vec{k}}:=\vec{D}_k\cdot \dz =0, \label{eq:const1vect}
\eea
and similarly for the quadratic constraint
\bea
\label{eq:quad:constraint:def:Hk}
\mathcal{S}^{(2)}_{\vec{k}}:= \dz ^{\,\dag}\,\boldsymbol{H}_{{k}}\, \dz .
\eea
The matrix $\boldsymbol{H}_k$ is a real-valued and symmetric $6\times6$-square matrix. The vectors $\vec{S}_k$ and $\vec{D}_k$, as well as the matrix $\boldsymbol{H}_k$, depend on the wavenumber (or the scale) $k$, and not on the direction of the wavevector $\vec{k}$, thanks to the isotropy of the background space-time. They are functions of the background phase-space variables and not of the background Lagrange multipliers, and detailed expressions are given in \App{app:vectornot}.

The evolution of the perturbations $\delta \vec{z}_{\vec{k}}$ is easily obtained by computing the Poisson bracket $\dot{\delta \vec{z}}_{\vec{k}}=\left\{\delta \vec{z}_{\vec{k}},C^{(2)}\right\}$, and one finds
\bea
	\dzdot=\delta N(\vec{k})\,\boldsymbol{\Omega}_6\vec{S}_k+k\delta N_1(\vec{k})\,\boldsymbol{\Omega}_6\vec{D}_k+2N\,\left(\boldsymbol{\Omega}_6\boldsymbol{H}_k\right) \dz , \label{eq:dynvect}
\eea
which has to be solved under the linear constraints, \Eqs{eq:const1vect}. 
For completeness, the detailed expressions of the linear constraints and of the equations of motion are given in \App{app:eqnsPert}.

\subsection{Constraints and gauge transformations}

The role played by the constraints in the above dynamical system is twofold. Firstly they define a surface in the phase space on which perturbations have to lie, \ie the so-called surface of constraints, which then has to be preserved throughout evolution. Secondly, they generate gauge transformations in the phase space, that is an equivalence class between different solutions of the dynamics. We now discuss each of these roles separately.

\subsubsection{Conservation of the constraints} \label{sssec:ConsConstraints}

Let us first note that the two linear constraints are first-class constraints, \ie their Poisson bracket vanishes
\bea
\label{eq:FirstClass:Strongly:Vanish}
	\Big\{\mathcal{S}^{(1)}_{\vec{k}},\mathcal{D}^{(1)}_{\vec{k}}\Big\}=\vec{S}_k^{{\mathrm{T}}} \boldsymbol{\Omega}_6\vec{D}_k=0.
\eea
The first equality is obtained by inserting the expression of the constraints as inner-dot products, \Eq{eq:const1vect}, and by using the Poisson bracket~\eqref{eq:Poisson:simp}. The second equality follows from the expressions of the constraint-vectors given in \App{app:vectornot}. One remarks that \Eq{eq:FirstClass:Strongly:Vanish} holds even when the linear constraints are not satisfied; in this sense, the Poisson bracket is said to  {\it{strongly}} vanish. However, the background constraint still needs to be imposed.

In addition, the constraints are preserved through the evolution (at least on the surface of constraints). Indeed, taking the time-derivative of $\vec{S}_k\cdot \dz $ and $\vec{D}_k\cdot \dz $ leads to
\bea
	\dot{\mathcal{S}}^{(1)}_{\vec{k}}=\left(\dot{\vec{S}}_k^{\,\mathrm{T}}+2N\vec{S}^{\,\mathrm{T}}_k\boldsymbol{\Omega}_6\boldsymbol{H}_k\right) \dz  & ~~~\mathrm{and}~~~&\dot{\mathcal{D}}^{(1)}_{\vec{k}}=\left(\dot{\vec{D}}_k^{\,\mathrm{T}}+2N\vec{D}^{\,\mathrm{T}}_k\boldsymbol{\Omega}_6\boldsymbol{H}_k\right)  \dz \, ,
	\label{eq:time:derivative:constraints}
\eea
where we have used the equation of motion~\eqref{eq:dynvect} for $ \dz $ together with the fact that $\vec{S}_k^{{\mathrm{T}}} \boldsymbol{\Omega}_6\vec{D}_k=\vec{D}_k^{{\mathrm{T}}} \boldsymbol{\Omega}_6\vec{S}_k= 0$ under the background constraint by virtue of \Eq{eq:FirstClass:Strongly:Vanish} and its transpose, and given that $\vec{V}^{\mathrm{T}} \boldsymbol{\Omega}_6\vec{V}=0$ for any vector $\vec{V}$ since $\boldsymbol{\Omega}_6$ is totally antisymmetric. It is interesting to note that the perturbed lapse and shift are absent from the right-hand side of these equations.\footnote{This is because the two constraints must hold irrespectively of the choice of the lapse function and the shift vector. In other words, the perturbed Lagrange multipliers cannot be fixed by using the conservation of the first-order constraints.} The vectors $\vec{S}_k$ and $\vec{D}_k$ depend on the background only, hence their time-derivative is given by the background Poisson bracket, \ie $\dot{\vec{S}}_k=\left\{\vec{S}_k,C^{(0)}[N]\right\}$ and similarly for $\vec{D}_k$. This leads to
\bea
	\dot{\vec{S}}_k&=&2N \boldsymbol{H}_k\boldsymbol{\Omega}_6 \vec{S}_k-N \frac{k^2}{v^{2/3}} \vec{D}_k\label{eq:dotS} \\
	\dot{\vec{D}}_k&=&2N \boldsymbol{H}_k\boldsymbol{\Omega}_6 \vec{D}_k \label{eq:dotD}
\eea
(see \App{app:GTHam} for an alternative derivation of these formulas). Inserting these relations in \Eq{eq:time:derivative:constraints} then yields
\bea
	\dot{\mathcal{S}}^{(1)}_{\vec{k}}=-N \frac{k^2}{v^{2/3}} \mathcal{D}^{(1)}_{\vec{k}} & ~~~\mathrm{and}~~~&\dot{\mathcal{D}}^{(1)}_{\vec{k}}=0,
\eea
where we have used $\boldsymbol{\Omega}_6^\mathrm{T}=-\boldsymbol{\Omega}_6$. Therefore, if initial conditions lie on the surface of constraints, these constraints are indeed preserved through the evolution.

\subsubsection{Gauge transformations}
\label{sssec:gtgen}

In the theory of constrained Hamiltonian systems (for a pedagogical review, see \eg \Refc{Wipf:1993xg}), the Lagrange multipliers $N(\vec{x},\tau)$ and $N^i(\vec{x},\tau)$ are auxiliary variables that are introduced to make the Legendre transform invertible. As a consequence, they do not describe physical parameters of the system. The presence of these arbitrary functions in the Hamiltonian implies that performing the evolution with one set of Lagrange multipliers or with another should lead to the same physical state, \ie described by the same physical observables. Let us consider the evolution of $ \dz $ over an infinitesimal time increment $\delta\tau$, with the lapse variable $N(\tau)$ on the one hand and $N(\tau)(1+\xi^0_{\vec{k}})$ on the other hand. From \Eq{eq:quadratic:action}, the difference between the evolved states is nothing but $\lbrace  \dz ,N(\tau) \xi^0_{\vec{k}} \mathcal{S}^{(1)}_{\vec{k}}\rbrace\delta\tau$. For this reason, we introduce 
\bea
\label{eq:Lie:def}
	\mathcal{L}_{\xi^0,0}\left( \dz \right)\equiv \left\{ \dz ,N\xi^0_{\vec{k}}\mathcal{S}^{(1)}_{\vec{k}}\right\}~~~&\mathrm{and}~~~&\mathcal{L}_{0,\xi}\left( \dz \right)\equiv \left\{ \dz ,\ds  \xi_{\vec{k}} k\mathcal{D}^{(1)}_{\vec{k}}\right\},
\eea
where $\mathcal{L}$ denotes the Lie derivative\footnote{Strictly speaking, \Eq{eq:Lie:def} does not involve the Lie derivative of the perturbed variables but instead the Lie derivative of their background counterpart \cite{Bojowald:2008jv}. This is because, in the full theory, gauge transformations act according to $X\to X+\mathcal{L}_{\xi^0,\xi}(X)$. Introducing a perturbative expansion and restricting it to first order leads to $\bar{X}+\delta X\to \bar{X}+\delta X+\mathcal{L}_{\xi^0,\xi}(\bar{X})$ where the term $\mathcal{L}_{\xi^0,\xi}(\delta{X})$ is discarded since it is second order in perturbation (given that both $\delta X$ and $\xi$ are order one). Hence in perturbation theory the gauge transformation is such that $\delta X\to\delta X+\mathcal{L}_{\xi^0,\xi}(\bar{X})$, and this is how the Lie derivative should be understood in \Eq{eq:Lie:def}.} and the second expression comes from considering the difference between the evolved states obtained with a vanishing shift and a shift equal to $\xi^i(\vec{k})=i\frac{k^i}{k}\xi_{\vec{k}}$ (keeping scalar degrees of freedom only).

The invariance of the theory under the transformations generated by \Eq{eq:Lie:def} is easy to interpret, since it merely corresponds to the invariance of General Relativity under the infinitesimal change of coordinates $x^\mu\to\widetilde{x}^\mu=x^\mu+\xi^\mu$ where $\xi^\mu=(\xi^0,\xi^i)$.\footnote{Note that $\xi^\mu$ should be decomposed on the direction orthogonal to the spatial hypersurfaces and the plane tangential to them since in the Hamiltonian language gauge transformations are interpreted as hypersurface deformations (see \Refc{Bojowald:2008jv} for details, see also footnote~\ref{footnote:projection}).} For this reason, hereafter they are referred to as ``gauge transformations'', and when restricted to the scalar type they are encoded by the two gauge parameters $\xi^0$ and $\xi$. 
Making use of the canonical Poisson bracket~\eqref{eq:Poisson:simp}, the gauge transformations induced by \Eq{eq:Lie:def} read
\bea
\label{eq:Lie:deltaz}
	\dztilde= \dz +N\xi^0_{\vec{k}} \boldsymbol{\Omega}_6\vec{S}_k+k\xi_{\vec{k}} \boldsymbol{\Omega}_6\vec{D}_k \, .
\label{eq:gtz}
\eea

Let us now consider the gauge transformation of the linear constraints,
\bea
\mathcal{L}_{\xi^\mu}\left(\mathcal{S}^{(1)}_{\vec{k}}\right)=\left\lbrace \mathcal{S}^{(1)}_{\vec{k}}, N \xi^0_{\vec{k}} \mathcal{S}^{(1)}_{\vec{k}} + \xi_{\vec{k}} k \mathcal{D}^{(1)}_{\vec{k}}\right\rbrace ,
\eea
and similarly for $\mathcal{D}^{(1)}_{\vec{k}}$. Given that the Poisson brackets between linear constraints strongly vanish, see \Eq{eq:FirstClass:Strongly:Vanish}, this implies that $\vec{S}_k\cdot \dztilde=\vec{S}_k\cdot \dz $ and $\vec{D}_k\cdot\dztilde=\vec{D}_k\cdot \dz $. Therefore, the surface of constraints is invariant under gauge transformations.

It is also important to note that gauge transformations preserve the Poisson brackets. The reason is that, at linear order in CPT, any phase-space function can be written as $F( \dz )=F_k+\vec{F}_{k}\cdot \dz $, where $F_k$ and $\vec{F}_{k}$ depend on the background only. Together with \Eq{eq:Lie:deltaz}, this implies that $\mathcal{L}_{\xi^\mu}[F( \dz )] = \vec{F}_k\cdot \mathcal{L}_{\xi^\mu}( \dz )$ only depends on $\xi^\mu$ and on the background, and not on the phase-space variables $ \dz $. As a consequence, for two phase-space functions $F( \dz )$ and $G( \dz )$, one has $\mathcal{L}_{\xi^\mu} [\lbrace F( \dz ), G( \dz ) \rbrace] = \lbrace \mathcal{L}_{\xi^\mu} [F( \dz )],G( \dz ) \rbrace  + \lbrace F( \dz ), \mathcal{L}_{\xi^\mu} [G( \dz )]\rbrace=0$, hence Poisson brackets are indeed invariant under gauge transformations.

This also implies that gauge transformations are canonical transformations. However, they are time-dependent canonical transformations, first through the two gauge parameters $N\xi^0$ and $\xi$, and second through the constraint vectors $\vec{S}_k$ and $\vec{D}_k$ [see Eq (\ref{eq:gtz})]. These two sources of time-dependence have, however, a different status: while the gauge parameters and their time-dependence are arbitrary (this is an ``explicit'' time-dependence), the time-dependence of the constraint vectors is determined by the background evolution (this is an ``implicit'' time-dependence, through the flow of the background phase-space variables). It is therefore useful to consider the equation of motion of the gauge-transformed variables,
\bea
	\frac{\dd \dztilde}{\dd\tau}&=&\left[\delta N(\vec{k})+\frac{\dd }{\dd\tau}\left(N\xi^0_{\vec{k}}\right)\right]\boldsymbol{\Omega}_6\vec{S}_k+\left[k\delta N_1(\vec{k})+k\dot{\xi}_{\vec{k}}\right]\boldsymbol{\Omega}_6\vec{D}_k+2N\boldsymbol{\Omega}_6\boldsymbol{H}_k \dz  \nonumber \\
	&&+N\xi^0_{\vec{k}}\boldsymbol{\Omega}_6 \dot{\vec{S}}_k+k\xi_{\vec{k}}\boldsymbol{\Omega}_6\dot{\vec{D}}_k\, ,
\eea
which is simply obtained by taking the time-derivative of \Eq{eq:gtz} and using the equations of motion~\eqref{eq:dynvect}. Inserting the equation of motion of the constraint vectors, \Eqs{eq:dotS}~and~\eqref{eq:dotD}, this leads to
\bea
	\frac{\dd \dztilde}{\dd\tau}&=&\left[\delta N(\vec{k})+\frac{\dd }{\dd\tau}\left(N\xi^0_{\vec{k}}\right)\right]\boldsymbol{\Omega}_6\vec{S}_k+k\left[\delta N_1(\vec{k})+\dot{\xi}_{\vec{k}}-\frac{N^2}{v^{2/3}}k\xi^0_{\vec{k}}\right]\boldsymbol{\Omega}_6\vec{D}_k \nonumber \\
	&&+2N\boldsymbol{\Omega}_6\boldsymbol{H}_k\left( \dz +N\xi^0_{\vec{k}}\boldsymbol{\Omega}_6\vec{S}_k+k\xi_{\vec{k}}\boldsymbol{\Omega}_6\vec{D}_k\right).
\eea
The gauge-transformed variables $\dztilde$ are easily recognised in the second line of the above, see \Eq{eq:gtz}, so one finds
\bea
	\frac{\dd \dztilde}{\dd\tau}&=&\widetilde{\delta N}(\vec{k})\,\boldsymbol{\Omega}_6\vec{S}_k+k\widetilde{\delta N_1}(\vec{k})\,\boldsymbol{\Omega}_6\vec{D}_k +2N\boldsymbol{\Omega}_6\boldsymbol{H}_k \dztilde ,
\eea
which has the same form as the equation of motion~\eqref{eq:dynvect} if the gauge-transformed Lagrange multipliers are given by
\bea
	\widetilde{\delta N}(\vec{k})&=&\delta N(\vec{k})+\dot{N}\xi^0_{\vec{k}}+N\dot{\xi}^0_{\vec{k}}\, , \label{eq:tildeN} \\
	\widetilde{\delta N_1}(\vec{k})&=&\delta N_1(\vec{k})-\frac{N^2}{v^{2/3}}\,k\xi^0_{\vec{k}}+\dot\xi_{\vec{k}}\, . \label{eq:tildeN1}
\eea
Therefore, in the Hamiltonian framework, gauge-transformations of the Lagrange multipliers are not given by Poisson brackets. The above procedure however shows how the dynamical equations can be used to obtain the gauge transform of the lapse and the shift vector \cite{Bojowald:2008jv} (see also \Refc{Pons:1995su,PhysRevD.55.658,Pons:1998ht} for in-depth studies of gauge transformations in constrained Hamiltonian systems with Lagrange multipliers). 

Let us finally stress that these transforms have been obtained by considering the time derivative of the gauge-transformed phase-space variables, which do not coincide with the gauge-transform of the time differentiated phase-space variables. This is because time derivation and gauge transformation do not commute when treating $\delta N$ and $\delta N_1$ as Lagrange multipliers, as shown in detail in \App{app:GTHam}.

\section{Physical and unphysical degrees of freedom}
\label{sec:PSgen}

In this section, we elaborate on the gauge transformations introduced above to discuss the concept of \emph{physical} and \emph{unphysical} degrees of freedom.
We start by introducing some notations and conventions. 
Any linear combination of the phase-space variables can be written as  $\vec{V}\cdot \dz $, where $\vec{V}$ is a vector of $\mathbb{R}^6$. In general, it is a function of the background variables, and it may exhibit explicit time-dependence. From \Eq{eq:Poisson:simp}, one has 
\bea
\left\{ \vec{V}_1\cdot \dz,\vec{V}_2\cdot \dz\right\}=\vec{V}_1^{\mathrm{T}} \boldsymbol{\Omega}_6\vec{V}_2\, ,
\eea 
hence two vectors $\vec{V}_1$ and $\vec{V}_2$ will be said to be {\it canonically conjugate} if $\vec{V}_1^{\mathrm{T}} \boldsymbol{\Omega}_6\vec{V}_2=1$. On the contrary, they will be said to be {\it canonically orthogonal} if $\vec{V}_1^{\mathrm{T}} \boldsymbol{\Omega}_6\vec{V}_2=0$.\footnote{If two vectors are not canonically orthogonal, they can always be renormalised to become canonically conjugate.}

In order to lighten some expressions, it will  also be useful to introduce the differential operator $\boldsymbol{\nabla}_\tau$, acting on 6-dimensional vectors as follows
\bea
	\boldsymbol{\nabla}_\tau\vec{F}=\left(\frac{\dd}{\dd\tau}-2N\boldsymbol{H}_k\boldsymbol{\Omega}_6\right)\vec{F}.
	\label{eq:nabla}
\eea
The transpose of the above simply reads $(\boldsymbol{\nabla}_\tau\vec{F})^\mathrm{T}=\dot{\vec{F}}^\mathrm{T}+2N\vec{F}^\mathrm{T}\boldsymbol{\Omega}_6\boldsymbol{H}_k$ where we have used that $\boldsymbol{H}_k^\mathrm{T}=\boldsymbol{H}_k$ and $\boldsymbol{\Omega}_6^\mathrm{T}=-\boldsymbol{\Omega}_6$. This operator is strongly related to the dynamics of cosmological perturbations since $-2N\boldsymbol{H}_k\boldsymbol{\Omega}_6$ is the operator adjoint to $2N\boldsymbol{\Omega}_6\boldsymbol{H}_k$ which generates the flow of $\dz$, see \Eq{eq:dynvect}. From the expression of $ \dz $ introduced in \Eq{eq:pert:vect:def}, one can define a natural basis of six vectors, each being associated to a single perturbed variable. We will call this basis the {\it natural basis} and we will denote it $\left\{\vec{e}^{\,\phi}_A,\vec{e}^{\,\pi}_A\right\}_{A\in\left\{0,1,2\right\}}$. Explicitly, we have 
\bea
	\vec{e}^{\,\phi}_0=\left(1,~0,~0;~0,~0,~0\right)^\mathrm{T}, &~~~~~~& \vec{e}^{\,\pi}_0=\left(0,~0,~0;~1,~0,~0\right)^\mathrm{T}, \nonumber \\
	\vec{e}^{\,\phi}_1=\left(0,~1,~0;~0,~0,~0\right)^\mathrm{T}, &~~~~~~& \vec{e}^{\,\pi}_1=\left(0,~0,~0;~0,~1,~0\right)^\mathrm{T}, \\
	\vec{e}^{\,\phi}_2=\left(0,~0,~1;~0,~0,~0\right)^\mathrm{T}, &~~~~~~&\vec{e}^{\,\pi}_2=\left(0,~0,~0;~0,~0,~1\right)^\mathrm{T}. \nonumber
\eea
This basis is orthonormal and is such that $\vec{e}^{\,\pi}_A\cdot  \dz $ is the canonical momentum of $\vec{e}^{\,\phi}_A\cdot  \dz $. This implies that
\bea
\vec{e}_A^{\,\varphi} \cdot \vec{e}_B^{\,\varphi'} =\delta^{\,\varphi,\varphi'} \delta_{A,B} &~~~\mathrm{and}~~~& \left\{ \vec{e}_A^{\,\varphi} \cdot \dz, \vec{e}_B^{\,\varphi'} \cdot \dz\right\}=\left[\boldsymbol{\Omega}_2\right]^{\varphi, \varphi'} \delta_{A,B} \,,
\eea
where $\varphi \equiv \phi,\pi$.

\subsection{Kinematical degrees of freedom} 
\label{ssec:gdofphysdof}

The natural basis does not seem particularly well suited when it comes to implementing constraints or gauge transformations. The former relies on the vectors $\vec{S}_k$ and $\vec{D}_k$, see \Eq{eq:const1vect}, while the later involve $\boldsymbol{\Omega}_6\vec{S}_k$ and $\boldsymbol{\Omega}_6\vec{D}_k$, see \Eq{eq:Lie:deltaz}. These vectors are not generically aligned with any of the vectors of the natural basis, and for this reason it is more convenient to consider the decomposition
\bea
	\mathbb{R}^6={\mathcal{C}}\otimes\mathcal{G}\otimes\mathcal{P}\, ,
\eea
where $\mathcal{C}$, $\mathcal{G}$ and $\mathcal{P}$ are three orthogonal planes that are defined as follows.

The first plane is generated by the two linear constraint-vectors, ${\mathcal{C}}=\spn{\vec{D}_k,\vec{S}_k}$, and is called {\it the plane of constraints}. An orthonormal basis can be constructed as $\vec{e}^{\,\mathcal{C}}_1=\lambda_D(\tau)\,\vec{D}_k$ and $\vec{e}^{\,\mathcal{C}}_2=\lambda_S(\tau)\,\vec{D}_k+\mu_S(\tau)\,\vec{S}_k$ where $\lambda_D$, $\lambda_S$ and $\mu_S$ are normalisation parameters.\footnote{Explicitly, they are given by $\lambda_D=\vert \vec{D}_k\vert^{-1}$, $\mu_S=\vert \vec{D}_k\vert/\sqrt{(\vert \vec{S}_k\vert \vert \vec{D}_k\vert)^2-(\vec{S}_k\cdot\vec{D}_k)^2}$ and $\lambda_S=-\mu_S \vec{S}_k\cdot\vec{D}_k /\vert\vec{D}_k\vert^2$, such that $\vec{e}^{\,\mathcal{C}}_\mu \cdot \vec{e}^{\,\mathcal{C}}_{\mu'}=\delta_{\mu,\mu'}$.\label{footnote:lambda:mu}}
The surface of constraints is orthogonal to the plane $\mathcal{C}$, \ie quantities like $\vec{e}^{\,\mathcal{C}} \cdot \dz $ vanish when the constraints are satisfied. Moreover,
since the linear constraints are strongly vanishing, see \Eq{eq:FirstClass:Strongly:Vanish}, one has
\bea
 \vec{e}^{\,\mathcal{C}}_\mu \cdot \left( \boldsymbol{\Omega}_6 \vec{e}^{\,\mathcal{C}}_{\mu'} \right) = 0\,,\quad \forall \mu,\mu'\equiv 1,\,2.
 \label{eq:eC:orthogonal}
\eea
Finally, from \Eqs{eq:dotS}-\eqref{eq:dotD}, for any vector $\vec{e}^{\,\mathcal{C}}\in\mathcal{C}$,  $\boldsymbol{\nabla}_\tau\vec{e}^{\,\mathcal{C}}$ lies in the plane of constraints. 

The second plane is  $\mathcal{G}=\spn{\boldsymbol{\Omega}_6\vec{D}_k,\boldsymbol{\Omega}_6\vec{S}_k}$ and is called {\it the plane of gauge degrees of freedom}. This is because, from \Eq{eq:gtz}, performing a gauge transformation corresponds to performing a translation in this plane. A basis can be obtained as $\{\vec{e}^{\,\mathcal{G}}_\mu\}_{\mu=1,2}:=\{-\boldsymbol{\Omega}_6 \vec{e}^{\,\mathcal{C}}_\mu\}_{\mu=1,2}$, and it is orthonormal by construction. Moreover, since the $\vec{e}^{\,\mathcal{C}}_\mu$ are canonically orthogonal, see \Eq{eq:eC:orthogonal}, so are the $\vec{e}^{\,\mathcal{G}}_\mu$. Finally, since $\boldsymbol{\Omega}_6^2=-\boldsymbol{I}_6$, the plane of gauge degrees of freedom is canonically conjugated to the plane of constraints, \ie
\bea
 \vec{e}^{\,\mathcal{C}}_\mu \cdot \left( \boldsymbol{\Omega}_6 \vec{e}^{\,\mathcal{G}}_{\mu'} \right) = \delta_{\mu,\mu'}
\eea
(hence the minus sign in the definition of the $\vec{e}^{\,\mathcal{G}}_\mu$). Therefore,
quantities like $\vec{e}^{\,\mathcal{G}} \cdot \dz $, \ie gauge degrees of freedom, are  canonically conjugated to the constraints~\cite{doi:10.1063/1.529065}.

The last plane, $\mathcal{P}$, is defined as the plane orthogonal to the two first ones.
Given that $ \vec{e}^{\,\mathcal{G}}_\mu=-\boldsymbol{\Omega}_6  \vec{e}^{\,\mathcal{C}}_\mu$ and $ \vec{e}^{\,\mathcal{C}}_\mu=\boldsymbol{\Omega}_6  \vec{e}^{\,\mathcal{G}}_\mu$, this plane is also canonically orthogonal to $\mathcal{C}\otimes\mathcal{G}$. Since it is both orthogonal and canonically orthogonal to $\mathcal{C}$ and $\mathcal{G}$, it carries the physical degrees of freedom, so we name it {\it the physical plane}. Once a vector $\vec{e}^{\,\mathcal{P}}_1$ of $\mathcal{P}$ is given, it is straightforward to show that $\vec{e}^{\,\mathcal{P}}_2=-\boldsymbol{\Omega}_6\vec{e}^{\,\mathcal{P}}_1$ also belongs to $\mathcal{P}$, and that $(\vec{e}^{\,\mathcal{P}}_1,\vec{e}^{\,\mathcal{P}}_2)$ forms an orthonormal basis if $\vec{e}^{\,\mathcal{P}}_1$ is normalised. Contrary to the two other planes, in $\mathcal{P}$ two basis vectors are canonically conjugated, \ie $\vec{e}^{\,\mathcal{P}}_1\cdot(\boldsymbol{\Omega}_6\vec{e}^{\,\mathcal{P}}_2)=\vec{e}^{\,\mathcal{P}}_1\cdot\vec{e}^{\,\mathcal{P}}_1=1$. Quantities like $\vec{e}^{\,\mathcal{P}}_1\cdot  \dz $ and $\vec{e}^{\,\mathcal{P}}_2\cdot  \dz $ constitute gauge-invariant canonical variables.

The above set of vectors forms an orthonormal basis that we call the {\it \Pbasis}. When ordered as follows,
\bea
\left\{\vec{e}_a\right\}_{a\in\{0,1,2,3,4,5\}}:=\left\{\vec{e}^{\,\mathcal{C}}_1,\vec{e}^{\,\mathcal{C}}_2,\vec{e}^{\,\mathcal{P}}_1;\vec{e}^{\,\mathcal{G}}_1,\vec{e}^{\,\mathcal{G}}_2,\vec{e}^{\,\mathcal{P}}_2\right\},
\eea
the three last vectors are respectively canonically conjugated to the three first ones (as in the natural basis). As a consequence, the scalar product and the Poisson bracket between two vectors reduce to the euclidean metric and to the symplectic 2-form
\bea
	\vec{e}_a \cdot \vec{e}_b=\delta_{a,b} &~~~\mathrm{and}~~~& \left\{\vec{e}_a\cdot\dz, \vec{e}_b\cdot\dz \right\}=\left[\boldsymbol{\Omega}_6\right]_{a,b} \,.
\eea

As an illustration, let us discuss the case of the Mukhanov-Sasaki variable~\cite{Mukhanov:1981xt,Kodama:1984ziu} 
\bea
\label{eq:MS:def}
Q_{{}_\mathrm{MS}} = \sqrt{\frac{v}{N}} \delta\phi + \frac{\Mp^2\pi_\phi}{\sqrt{6N}\theta  v^{7/6}}\left(\sqrt{2}\delta\gamma_1-\delta\gamma_2\right) 
\equiv \vec{B}\cdot \dz \, ,
\eea
which defines the vector $\vec{B}$. A direct calculation using \Eqs{eq:S1}-\eqref{eq:D1} shows that $\vec{B}$ is canonically orthogonal to $\vec{S}_k$ and $\vec{D}_k$, hence $\vec{B}$ is orthogonal to $\mathcal{G}$. This means that $Q_{{}_\mathrm{MS}}$ is a gauge-invariant combination, a well-known fact indeed. However, $\vec{B}$ is not orthogonal to the plane of constraints, and one may instead consider the (normalised) projection of $\vec{B}$ onto the physical plane $\mathcal{P}$. Hereafter, we will assume that $\vec{e}_1^{\, \mathcal{P}}$ coincides with that projection, which fixes the basis of $\mathcal{P}$, hence the whole \Pbasis.

\subsection{Dynamics}
\label{ssec:DynKin}

In the natural basis, the equations of motion were given by \Eq{eq:dynvect}. Let us now study the phase-space dynamics in the \Pbasis.

\subsubsection{Hamiltonian}

Let $\dq$ denote the phase-space vector in the \Pbasis. It is related to the phase-space vector in the natural basis, $\dz$, through a canonical transformation
\bea
\dq = \boldmathsymbol{P} \dz\, .
\eea 
Here, $\boldmathsymbol{P}=\mathrm{Row}(\vec{e}_a)$ is symplectic, \ie $\boldsymbol{P}^\mathrm{T}\boldsymbol{\Omega}_6\boldsymbol{P}=\boldsymbol{\Omega}_6$, and since it defines a transformation from an orthonormal basis to another orthonormal basis, it is  orthogonal, \ie $\boldsymbol{P}^\mathrm{T}\boldsymbol{P}=\boldsymbol{I}_6$ (these two relations also imply that $\boldsymbol{\Omega}_6\boldsymbol{P}=\boldsymbol{P}\boldsymbol{\Omega}_6$). The components of $\dq$ will be denoted as 
\bea
	\dq\equiv \left( Q_1(\vec{k}), \, Q_2(\vec{k}), \,Z_1(\vec{k}) ;\,P_1(\vec{k}),\,P_2(\vec{k}),\,Z_2(\vec{k})\right)^\mathrm{T}, \label{eq:deltaQ}
\eea
where $Q_1$ and $Q_2$ are the constraints (here viewed as configuration variables), $P_1$ and $P_2$ are the gauge degrees of freedom (here viewed as the canonical momenta conjugated to the constraints), and $(Z_1,Z_2)$ is the canonical pair of physical degrees of freedom. 

Since $ \boldmathsymbol{P} $ is a time-dependent canonical transformation, starting from \Eq{eq:quadratic:action} the new Hamiltonian reads \cite{goldstein2002classical,Grain:2019vnq}
\bea
	K&=&\ds\int_{\mathbb{R}^{3+}}\dd^3\vec{k}\left\{\left[\delta N^\star(\vec{k}) \,\left(\boldsymbol{P}\vec{S}_k\right)\cdot \dq +\mathrm{c.c.}\right]+k\left[\delta N_1^\star(\vec{k})\, \left(\boldsymbol{P}\vec{D}_{{k}}\right)\cdot \dq +\mathrm{c.c.}\right]\right\} \nonumber \\
	&&+ \ds\int_{\mathbb{R}^{3+}}\dd^3\vec{k}\,{ \dq }^{\,\dag}\,\underbrace{\left[2N\boldsymbol{P}\boldsymbol{H}_{{k}}\boldsymbol{P}^\mathrm{T}+\boldsymbol{P}\boldsymbol{\Omega}_6\frac{\dd\boldsymbol{P}^\mathrm{T}}{\dd\tau}\right]}_{\boldsymbol{K}_k}\, \dq , \label{eq:hamphys}
\eea
where the matrix $\boldsymbol{K}_k$ is easily shown to be symmetric since $\boldsymbol{H}_k$ is symmetric (see \Refc{Grain:2019vnq} for details). The first line is linear in the phase-space variables and corresponds to the linear constraints. It is given by
\bea
	K^{(1)}=\ds\int_{\mathbb{R}^{3+}}\dd^3\vec{k}\left[\Lambda^\star_1(\vec{k})\,Q_1(\vec{k})+\Lambda_2^\star(\vec{k})\,Q_2(\vec{k})+\mathrm{c.c.}\right], \label{eq:linearK}
\eea
where 
\bea
	\Lambda_1=\frac{1}{\lambda_D}\left(k\delta N_1-\frac{\lambda_S}{\mu_S}\,\delta N\right)
	\quad\quad\text{and}\quad\quad
	\Lambda_2=\frac{1}{\mu_S}\,\delta N
\label{eq:lapse:physical}
\eea
are the Lagrange multipliers now given by linear combinations of the perturbed lapse and shift. The first-order constraints are given by 
\bea
	Q_1(\vec{k})=0=Q_2(\vec{k})
\eea
and their (strong) first-class nature, $\{Q_1,Q_2\}=0$, is obvious from the fact that these are now two configuration variables. 

The second line of \Eq{eq:hamphys} is quadratic in the phase space. Since the matrix $\boldsymbol{P}$ is built from the \Pbasis vectors, the elements of $\boldsymbol{K}_k$ composing the quadratic part of the Hamiltonian are given by
\bea
	\left[\boldsymbol{K}_k\right]_{a,b}=2N\vec{e}_a\cdot\left(\boldsymbol{H}_k\vec{e}_b\right)+\vec{e}_a\cdot\left(\boldsymbol{\Omega}_6\dot{\vec{e}}_b\right) =\vec{e}_a\cdot\left[\boldsymbol{\nabla}_\tau\left(\boldsymbol{\Omega}_6\vec{e}_b\right)\right],
	\label{eq:Kab}
\eea
see \Eq{eq:nabla}. The entries of $\boldsymbol{K}_k$ involving the gauge degrees can be easily computed. They are obtained by setting $\vec{e}_{b=3,\,4}=\vec{e}^{\,\mathcal{G}}_{\mu=1,\,2}$. Using $\boldsymbol{\Omega}_6\vec{e}^{\,\mathcal{G}}_\mu=\vec{e}^{\,\mathcal{C}}_\mu$, and computing $\boldsymbol{\nabla}_\tau\vec{e}^{\,\mathcal{C}}_1$ and $\boldsymbol{\nabla}_\tau\vec{e}^{\,\mathcal{C}}_2$ with \Eqs{eq:dotS} and~\eqref{eq:dotD}, one obtains
\bea
	\left[\boldsymbol{K}_k\right]_{a,3}&=&\frac{\dot{\lambda}_D}{\lambda_D}\delta_{a,0}\, , \label{eq:Ka3} \\
	\left[\boldsymbol{K}_k\right]_{a,4}&=&\left(\frac{\dot{\lambda}_S}{\lambda_D}-\frac{\dot{\mu}_S}{\mu_S}\frac{\lambda_S}{\lambda_D}-N\frac{\mu_S}{\lambda_D} \frac{k^2}{v^{2/3}}\right)\delta_{a,0}+\frac{\dot{\mu}_S}{\mu_S}\delta_{a,1}\, . \label{eq:Ka4}
\eea
Hence the gauge degrees of freedom $P_1$ and $P_2$ are only coupled to the constraints $Q_1$ and $Q_2$. This is expected from the fact that the constraints are only sourced by themselves, \ie they are preserved on shell. In particular, there is no kinetic term of the form $P_\mu P_{\mu'}^\star+\mathrm{c.c.}$ nor couplings to the physical degrees of freedom.  

The other entries can also be derived by direct calculations. Their explicit expressions are not needed for the forthcoming discussion and we report them in \App{app:HamKinBasis}.

\subsubsection{Equations of motion}

Using Hamilton's equation with \Eq{eq:hamphys}, the equations of motion read\footnote{Alternatively, one can time differentiate $\dqi{a}=\vec{e}_a\cdot \dz $ and use \Eq{eq:dynvect}, leading to
\bea
    \frac{\dd}{\dd\tau}\left(\vec{e}_a\cdot \dz \right)=\Lambda_1(\tau)\,\left[\boldsymbol{\Omega}_6\right]_{a,0}+\Lambda_2(\tau)\,\left[\boldsymbol{\Omega}_6\right]_{a,1}+\left(\boldsymbol{\nabla}_\tau\vec{e}_a\right)\cdot \dz .
\eea
Comparing with \Eq{eq:EOMphysgen} leads to another expression for the Hamiltonian kernel in the \Pbasis, namely
\bea
    \left[\boldsymbol{K}_k\right]_{a,b}=-\ds\sum_{c}\left[\boldsymbol{\Omega}_6\right]_{a,c}\left(\vec{e}_b\cdot\boldsymbol{\nabla}_\tau\vec{e}_c\right).
\eea}
\bea
\frac{\dd{\dqi{a}}}{\dd\tau}=\Lambda_1(\tau)\,\left[\boldsymbol{\Omega}_6\right]_{a,0}+\Lambda_2(\tau)\,\left[\boldsymbol{\Omega}_6\right]_{a,1}+\left[\boldsymbol{\Omega}_6 \boldsymbol{K}_k \dq\right]_a \, . \label{eq:EOMphysgen}
\eea
The last term can be further simplified by noticing that since the Hamiltonian kernel is symmetric one has
\bea
\left[ \boldsymbol{K}_k \dq\right]_b & = &\left[  \boldsymbol{K}_k \right]_{b c} \dqi{c} = \left[  \boldsymbol{K}_k \right]_{c b} \dqi{c}\nonumber \\
& = & \vec{e}_c \cdot \left[ \boldmathsymbol{\nabla}_\tau \left(\boldmathsymbol{\Omega}_6 \vec{e}_b\right)\right] \left(\vec{e}_c \cdot \dz \right)\nonumber \\
& = & \left[\boldsymbol{\nabla}_\tau\left(\boldsymbol{\Omega}_6\vec{e}_b\right)\right]\cdot \dz\, ,
\eea 
where we have used \Eq{eq:Kab} in the second line, and in the third line the fact that $\sum_c \vec{e}_c \vec{e}_c^{\, \mathrm{T}}=\boldmathsymbol{I}_6$ since $\left\{\vec{e}_a\right\}$ forms an orthonormal basis. This gives rise to
\bea
 \frac{\dd\dqi{a}}{\dd\tau}=\Lambda_1(\tau)\,\left[\boldsymbol{\Omega}_6\right]_{a,0}+\Lambda_2(\tau)\,\left[\boldsymbol{\Omega}_6\right]_{a,1}+\ds\sum_{b}\left[\boldsymbol{\Omega}_6\right]_{a,b}\,\left[\boldsymbol{\nabla}_\tau\left(\boldsymbol{\Omega}_6\vec{e}_b\right)\right]\cdot \dz\, .
\eea
We stress that in the above, the symplectic matrix $\boldsymbol{\Omega}_6$ inside the operator $\boldsymbol{\nabla}_\tau$ acts on the components of the vector $\vec{e}_b$ in the natural basis, while the one preceding the operator $\boldsymbol{\nabla}_\tau$ acts across the basis vectors (in a way similar to the tetrad formalism in General Relativity).

Fourier modes can be treated separately and from now on, their dependence with $\vec{k}$ is omitted in order to keep equations as light as possible.

\subsubsection*{Unphysical degrees of freedom}

Let us now see what the equations of motion imply for the dynamics of the unphysical degrees of freedom, \ie the constraint and gauge degrees of freedom. As discussed below \Eqs{eq:Ka3} and~\eqref{eq:Ka4}, the time flow of the constraints is generated by the constraints only, hence the surface of constraints is invariant under dynamical evolution. 

For the gauge degrees of freedom, one has
\bea
	\dot{P}_1=-\frac{\partial K}{\partial Q_1}&=&-\Lambda_1(\tau)-\frac{\dot\lambda_D}{\lambda_D}\,P_1-\left[\frac{\dot{\lambda}_S}{\lambda_D}-\frac{\dot{\mu}_S}{\mu_S}\frac{\lambda_S}{\lambda_D}-N\frac{\mu_S}{\lambda_D}\left(\frac{k}{v^{1/3}}\right)^2\right]P_2 \nonumber  \\
	&&-\left[\boldsymbol{K}_k\right]_{00}Q_1-\left[\boldsymbol{K}_k\right]_{01}Q_2-\left[\boldsymbol{K}_k\right]_{02}Z_1-\left[\boldsymbol{K}_k\right]_{05}Z_2\, , \label{eq:P1dot}  \\
	\dot{P}_2=-\frac{\partial K}{\partial Q_2}&=&-\Lambda_2(\tau)-\frac{\dot{\mu}_S}{\mu_S}\,P_2 \nonumber \\
	&&-\left[\boldsymbol{K}_k\right]_{01}Q_1-\left[\boldsymbol{K}_k\right]_{11}Q_2-\left[\boldsymbol{K}_k\right]_{12}Z_1-\left[\boldsymbol{K}_k\right]_{15}Z_2\, . \label{eq:P2dot}
\eea
We stress that the $P_\mu$'s are the only variables whose dynamics is sourced by the two Lagrange multipliers, which  shows explicitly that they are gauge degrees of freedom, \ie they can be arbitrarily fixed by the choice of $\delta N$ and $\delta N_1$. On the surface of constraints where $Q_\mu=0$, these equations can be solved as
\bea
	P_1(\tau)&\approx&\frac{1}{\lambda_D(\tau)}\Bigg\{\alpha_1+\ds\int^\tau_{\tau_\uin}\dd\tau'\left[\frac{\lambda_S}{\mu_S}\,\delta N-k\,\delta N_1-\lambda_D\left[\boldsymbol{K}_k\right]_{02}Z_1-\lambda_D\left[\boldsymbol{K}_k\right]_{05}Z_2\right]_{\tau'}  \nonumber \\
	&&-\ds\int^\tau_{\tau_\uin}\dd\tau'\left[\dot\lambda_S-\frac{\dot\mu_S}{\mu_S}\lambda_S+N\mu_S\left(\frac{k}{v^{1/3}}\right)^2\right]_{\tau'}P_2(\tau')\Bigg\},  \label{eq:PDsol}  \\
	P_2(\tau)&\approx&\frac{1}{\mu_S(\tau)}\left\{\alpha_2-\ds\int^\tau_{\tau_\uin}\dd\tau'\left[\delta N-\mu_S\left[\boldsymbol{K}_k\right]_{12}Z_1-\mu_S\left[\boldsymbol{K}_k\right]_{15}Z_2\right]_{\tau'}\right\}, \label{eq:PSsol}
\eea
where $\alpha_1$ and $\alpha_2$ are two constants of time-integration (though they may depend on $k$), and where the notation $[\,\cdot\,]_{\tau'}$ means that all the time-dependent functions inside the bracket are evaluated at $\tau'$. We have also introduced the notation ``$\approx$'', which means that the equality holds on the surface of linear constraints (this is called a ``weak equality'' hereafter), and we have replaced $\Lambda_1$ and $\Lambda_2$ using \Eq{eq:lapse:physical}. As we will see below, the physical degrees of freedom are not coupled to the gauge degrees of freedom, hence $Z_1$ and $Z_2$ can be viewed as pure sources in the right-hand sides of the above. 

\subsubsection*{Physical degrees of freedom} 

As shown in \App{app:HamKinBasis}, $[\boldmathsymbol{K}_k]_{2,3}=[\boldmathsymbol{K}_k]_{2,4}=[\boldmathsymbol{K}_k]_{3,5}=[\boldmathsymbol{K}_k]_{4,5}=0$, hence the physical degrees of freedom $Z_1$ and $Z_2$ are decoupled from the gauge degrees of freedom. They are also decoupled from the Lagrange multipliers, which confirms that they can be interpreted as ``physical'' indeed. On the surface of constraints, their equations of motion reduce to
\bea
	\dot{Z}_1&\approx&\left[\boldsymbol{K}_{k}\right]_{5,5}\,Z_2+\left[\boldsymbol{K}_{k}\right]_{2,5}\,Z_1\, , \label{eq:eomQ} \\
	\dot{Z}_2&\approx&-\left[\boldsymbol{K}_{k}\right]_{2,2}\,Z_1-\left[\boldsymbol{K}_{k}\right]_{2,5}\,Z_2\, . \label{eq:eomP}
\eea
Although solutions to the above always exist, since the dynamics is generated by elements of the symplectic group \cite{Grain:2019vnq}, they cannot be written in a generic closed form, contrary to what was found for the unphysical degrees of freedom. This is precisely because $Z_1$ and $Z_2$ contain all the non-trivial part of the dynamics of cosmological perturbations. 

\subsection{Gauge transformations}
\label{ssec:GTransfophys}

As explained in \Sec{ssec:gdofphysdof}, gauge transformations consist in performing infinitesimal translations in the plane $\mathcal{G}$. This only affects the gauge degrees of freedom, which can be checked explicitly by writing the gauge transformation~\eqref{eq:Lie:deltaz} in the \Pbasis, leading to\footnote{These relations coincide with the formal solutions~\eqref{eq:PDsol}~and~\eqref{eq:PSsol} if $\delta N$ and $\delta N_1$ are replaced by their gauge-transformed expressions, \Eqs{eq:tildeN} and \eqref{eq:tildeN1} [in this replacement, expressions for $\dot{\lambda_S}$, $\dot{\lambda_D}$ and $\dot{\mu_S}$ have to be inserted, they can be obtained from combining footnote~\ref{footnote:lambda:mu} with \Eqs{eq:dotS} and~\eqref{eq:dotD}].}
\bea
	\widetilde{P}_1&=&P_1-\frac{1}{\lambda_D}\,k\xi_{\vec{k}}-\frac{\lambda_S}{\mu_S\lambda_D}\,N\xi^0_{\vec{k}}\, , \\
	\widetilde{P}_2&=&P_2-\frac{1}{\mu_S}\,N\xi^0_{\vec{k}},
\eea
while all the other degrees of freedom remain unchanged.

Usually, CPT is solved either by using gauge-invariant variables or by fixing the gauge. This ensures that the final solution is free from unphysical degrees of freedom prior to solving the equations of motion. However, the considerations presented above advocate for an alternative procedure, which is to use the \Pbasis to remove gauge degrees of freedom from solutions that neither need to be gauge-invariant, nor gauge-fixed. 
Indeed, suppose a solution ${\dz} _{\mathrm{sol}}$ of the constrained system, \Eqs{eq:const1vect} and~\eqref{eq:dynvect}, has been found. In full generality, this solution contains  gauge degrees of freedom. However, it is straightforward to separate these degrees of freedom from the physical ones by projecting the solution onto the \Pbasis, which is orthonormal, \ie $P_\mu= {\dz}_{\mathrm{sol}}\cdot\vec{e}^{\,\mathcal{G}}_\mu$, $Z_1= {\dz} _{\mathrm{sol}}\cdot\vec{e}^{\,\mathcal{P}}_1$, and $Z_2= {\dz}_{\mathrm{sol}}\cdot\vec{e}^{\,\mathcal{P}}_2$.

\subsection{Dynamical degrees of freedom}

Finally, let us mention the existence of another convenient basis that may be employed to separate the dynamics of the physical degrees of freedom from the unphysical ones. Although the \Pbasis follows naturally from kinematical considerations, its three sectors (constraints, gauge and physical) are not independent at the dynamical level: the constraints and the physical degrees of freedom source the evolution of the gauge degrees of freedom, see \Eqs{eq:PDsol} and~\eqref{eq:PSsol}, and the constraints also source the equation of motion of the physical degrees of freedom. On the surface of constraints, the physical degrees of freedom become fully independent, which is sufficient for most practical analyses. Nonetheless, through a canonical transformation it is formally possible to recast first-class constrained systems in a way that makes the Hamiltonian separable~\cite{doi:10.1063/1.529065}, \ie of the form
\bea
\label{eq:Hamiltonian:dynamical:basis}
	K(Q_\mu,P^\mu)=k(Q_i,P^i)+\Lambda^aQ_a \, .
\eea
In this expression, $Q_a$ are first-class constraints and their associated momenta $P^a$ correspond to the gauge degrees of freedom, $(Q_i,P^i)$ correspond to the physical degrees of freedom, and $\Lambda_a$'s are Lagrange multipliers whose specific values fix the gauge degrees of freedom $P^a$ (up to integration constants). In that setup, the constraints are strongly conserved (\ie not only on the surface of constraints), and the physical degrees of freedom are decoupled even off shell. 

In \App{ssec:dynPhys}, the canonical transformation leading to a Hamiltonian of the form~\eqref{eq:Hamiltonian:dynamical:basis} is constructed explicitly in the case of cosmological perturbations. It is shown that, in practice, its derivation requires to solve the dynamics in the \Pbasis, so it does not simplify the problem technically, but it might provide clearer interpretations in some cases.

\section{Gauge fixing}
\label{sec:GF}

Gauge-fixing exploits the freedom in selecting the perturbed lapse function and the perturbed shift vector in order to remove the gauge degrees of freedom from the final solution of the equation of motion. Hence, if the gauge is properly fixed, the final solution can be expressed using the physical degrees of freedom only (the constraint degrees of freedom being removed by solving on the surface of constraints). 

In the \Pbasis introduced above, the gauge-fixing procedure is trivial: it simply consists in setting the gauge degrees of freedom $P_1$ and $P_2$. In practice however, most gauges that have been proposed in the literature are defined through two algebraic conditions that may involve any of the phase-space variables and of the Lagrange multipliers. How such prescriptions may or may not fix the gauge degrees of freedom is in general not obvious, and it is the topic of this section.

\subsection{Gauge-fixed dynamics}
\label{sssec:diffgauge}

Since the dynamics is linear, the two gauge conditions are taken to be linear too. Their general expressions read  
\bea
	G_i\left( \dz ,\delta N,\delta N_1\right)=\vec{G}_i\cdot \dz +\lambda^{(N)}_i\,\delta N+\lambda^{(N_1)}_i\,k\delta N_1 \label{eq:gaugecond}
\eea
where $i$ runs from 1 to 2. Hereafter, the vectors $\vec{G}_i$  will be referred to as the gauge vectors and the set $(\lambda^{(N)}_i,\lambda^{(N_1)}_i)$ as the gauge multipliers. In general, they can all be time-dependent. Allowing the Lagrange multipliers to appear in the gauge conditions might seem at odds with the Hamiltonian perspective. However, there exists common gauges, introduced in the Lagrangian framework, in which the Lagrange multipliers are constrained. For instance, this is the case of the Newtonian gauge (see \Sec{sec:examples:gauges}) and this is why we adopt the general form given in \Eq{eq:gaugecond}. Eventually, solving the dynamics of cosmological perturbations in a gauge-fixed manner consists in solving the following system
\bea
	\left\{\begin{array}{ll}
		\ds \dzdot=\delta N\,\boldsymbol{\Omega}_6\vec{S}_k+k\delta N_1\,\boldsymbol{\Omega}_6\vec{D}_k+2N\,\left(\boldsymbol{\Omega}_6\boldsymbol{H}_k\right) \dz &~~~\text{[dynamical equation~\eqref{eq:dynvect}]} \\
		\ds\vec{S}_k\cdot \dz =0=\vec{D}_k\cdot \dz &~~~\text{[linear constraints~\eqref{eq:const1vect}]} \\
		G_1\left( \dz ,\delta N,\delta N_1\right)=0=G_2\left( \dz ,\delta N,\delta N_1\right)&~~~\text{[gauge conditions~\eqref{eq:gaugecond}]}
	\end{array}\right. \, .   \label{eq:GFsystem}
\eea

Though apparently algebraic, gauge conditions as defined in \Eq{eq:gaugecond} may in fact be differential in the phase-space variables, because of the Lagrange multipliers. Indeed, let us consider two arbitrary vectors $\vec{V}_\mu$, such that their projections onto the plane of gauge degrees of freedom $\mathcal{G}$ are linearly independent.\footnote{Examples of such pairs of vectors are $\vec{V}_\mu=\vec{e}_\mu^{\,\mathcal{G}}$, or $\vec{V}_1=\vec{e}^{\,\phi}_0$ and $\vec{V}_2=\vec{e}^{\,\phi}_2$, although the discussion does not depend on that choice.} Projecting the equations of motion~\eqref{eq:dynvect} onto these two vectors allows us to express the Lagrange multipliers as functions of $ \dz $ and $\dot{ \dz }$ . It leads to the following linear system
\bea
	\underbrace{\left(\begin{array}{ccc}
		\vec{V}_1\cdot(\boldsymbol{\Omega}_6\vec{S}_k) & \quad\quad& \vec{V}_1\cdot(\boldsymbol{\Omega}_6\vec{D}_k) \\
		\vec{V}_2\cdot(\boldsymbol{\Omega}_6\vec{S}_k) & \quad\quad& \vec{V}_2\cdot(\boldsymbol{\Omega}_6\vec{D}_k)
	\end{array}\right)}_{\boldsymbol{V}}\left(\begin{array}{c}
		\delta N \\
		k\delta N_1
	\end{array}\right)=\left(\begin{array}{c}
		\vec{V}_1\cdot\dzdot -2N\vec{V}_1\cdot(\boldsymbol{\Omega}_6\boldsymbol{H}_k \dz ) \\
		\vec{V}_2\cdot\dzdot -2N\vec{V}_2\cdot(\boldsymbol{\Omega}_6\boldsymbol{H}_k \dz )
	\end{array}\right),
\eea
which is invertible since the vectors $\vec{V}_\mu$ have two linearly independent projections onto the plane of gauge degrees of freedom. This leads to
\bea
	\delta N &=&\vec{W}_1\cdot\dzdot-2N\vec{W}_1\cdot(\boldsymbol{\Omega}_6\boldsymbol{H}_k \dz ), \\
	k\delta N_1&=&\vec{W}_2\cdot\dzdot -2N\vec{W}_2\cdot(\boldsymbol{\Omega}_6\boldsymbol{H}_k \dz ),
\eea	
where
\bea
	\vec{W}_\mu=\ds\sum_{\mu'=1}^2\left[\boldsymbol{V}^{-1}\right]_{\mu,\mu'}\vec{V}_{\mu'}.
\eea
As a consequence, the gauge-fixed dynamics~\eqref{eq:GFsystem} is equivalent to
\bea
	\left\{\begin{array}{ll}
		\ds\dzdot =\delta N\,\boldsymbol{\Omega}_6\vec{S}_k+k\delta N_1\,\boldsymbol{\Omega}_6\vec{D}_k+2N\,\left(\boldsymbol{\Omega}_6\boldsymbol{H}_k\right) \dz  \\
		\ds\vec{S}_k\cdot \dz =0=\vec{D}_k\cdot \dz  \\
		\vec{G'}_1\cdot \dz  +\vec{J}_1\cdot\dzdot \,=0=\vec{G'}_2\cdot \dz +\vec{J}_2\cdot\dzdot 
	\end{array}\right.  \, ,\label{eq:GFsystemdiff}
\eea
where the gauge conditions are now built from four vectors reading
\bea
\label{eq:Gp:def}
	\vec{G'}_i&=&\vec{G}_i+2N\boldsymbol{H}_k\boldsymbol{\Omega}_6\left[\lambda_i^{(N)}\,\vec{W}_1+\lambda^{(N_1)}_i\,\vec{W}_2\right], \\
	\vec{J}_i&=&\lambda_i^{(N)}\,\vec{W}_1+\lambda^{(N_1)}_i\,\vec{W}_2.
\label{eq:J:def}
\eea
Written in the form of \Eq{eq:GFsystemdiff}, it is now obvious that the gauge conditions are in full generality differential constraints on the phase space rather than algebraic ones. 

\subsection{Requirements for non-pathological gauges}
\label{ssec:RequNPgauge}

We now make use of the differential version of the gauge conditions to identify the requirements for removing the gauge degrees of freedom. To this end, we decompose any solution of \Eq{eq:GFsystemdiff}  on the \Pbasis as follows:
\bea	
	 \dz =\ds\sum_{\mu=1}^2\left(Q_\mu\,\vec{e}_\mu^{\,\mathcal{C}}+P_\mu\,\vec{e}_\mu^{\,\mathcal{G}}+Z_\mu\,\vec{e}_\mu^{\,\mathcal{P}}\right),
\eea
in accordance with the splitting introduced in \Eq{eq:deltaQ}. Solutions are on the surface of constraints, so $Q_\mu\approx0$ and $\dot{Q}_\mu\approx0$. Hence, the gauge conditions lead to two additional constraints mixing the gauge degrees of freedom and the physical degrees of freedom. They are given by
\bea
	\ds\sum_{\mu=1}^2\left[\left(\vec{G'}_i\cdot\vec{e}^{\,\mathcal{G}}_\mu+\vec{J}_i\cdot\dot{\vec{e}}^{\,\mathcal{G}}_\mu\right)P_\mu+\left(\vec{J}_i\cdot\vec{e}^{\,\mathcal{G}}_\mu\right)\dot{P}_\mu\right]\approx-\ds\sum_{\mu,\mu'=1}^2\left[\boldsymbol{J}_i\right]_{\mu\mu'}Z_{\mu'} \,,\label{eq:GCPhys}
\eea
where\footnote{For clarity, let us stress that we are not using the implicit summation notation here, hence $\mu$ is fixed  in equation \eqref{eq:Jmat}.}
\bea
	\left[\boldsymbol{J}_i\right]_{\mu\mu'}=\left(\vec{G'}_i\cdot\vec{e}^{\,\mathcal{P}}_\mu\right)\delta_{\mu,\mu'}+\left(\vec{J}_i\cdot\dot{\vec{e}}^{\,\mathcal{P}}_\mu\right)\delta_{\mu,\mu'}+\left(\vec{J}_i\cdot\vec{e}^{\,\mathcal{P}}_\mu\right)\left[\boldsymbol{\Omega}_2\boldsymbol{k}_{\mathrm{pp}}\right]_{\mu\mu'}. \label{eq:Jmat}
\eea
Note that we made use of the equations of motion of the physical degrees of freedom to express $\dot{Z}_\mu$ as a function of $Z_\mu$. 

The conditions for the gauge degrees of freedom $P_\mu$ to be unequivocally fixed by the gauge conditions are easily identified from \Eq{eq:GCPhys}. First, \Eq{eq:GCPhys} has to be algebraic in the $P_\mu$'s rather than differential, otherwise the gauge degrees of freedom would be fixed up to one (at least) integration constant. Remaining integration constants, however, can be arbitrarily fixed to fix the remaining gauge degrees of freedom (see \eg Sec. 5.2.3.3 of \Refc{Peter:2013avv}). Second, the resulting algebraic system has to be invertible, otherwise only one direction (at most) in the plane of gauge degrees of freedom is constrained. We now study these two requirements separately.

\subsubsection{Regaining algebraic conditions} 
\label{sssec:RegainingAlgConst}

Whether time derivatives can be removed or not depends on the projections of the vectors $\vec{J}_i$ onto the plane of gauge degrees of freedom that we now compute. 

To this end, we denote by $\vec{V}^{\,\mathcal{G}}_i\in\mathcal{G}$ the projections of $\vec{V}_1$ and $\vec{V}_2$ onto $\mathcal{G}$, and recall that they are linearly independent. These two vectors thus form a complete basis of the plane of gauge degrees of freedom. The matrix $\boldsymbol{V}$ is built from the component of the $\vec{V}^{\,\mathcal{G}}_i$ on the complete basis of $\mathcal{G}$ given by $(\boldsymbol{\Omega}_6\vec{S}_k)$ and $(\boldsymbol{\Omega}_6\vec{D}_k)$. Hence the projection of the $\vec{W}_i$ onto $\mathcal{G}$ also forms a complete basis of the plane of gauge degrees of freedom.\footnote{It can be shown that
\bea
	\vec{W}^{\,\mathcal{G}}_1&=&-\mu_S\,\vec{e}^{\,\mathcal{G}}_2, \\
	\vec{W}^{\,\mathcal{G}}_2&=&-\lambda_D\,\vec{e}^{\,\mathcal{G}}_1-\lambda_S\,\vec{e}^{\,\mathcal{G}}_2,
\eea
which is valid for any choice of the $\vec{V}_\mu$'s.}  As a consequence, the properties of $\vec{J}^{\,\mathcal{G}}_i=\lambda_i^{(N)}\,\vec{W}^{\,\mathcal{G}}_1+\lambda^{(N_1)}_i\,\vec{W}^{\,\mathcal{G}}_2$ are set by the gauge multipliers. Let us introduce the matrix
\bea
	\boldsymbol{\lambda}=\left(\begin{array}{cc}
		\lambda^{(N)}_1 & \lambda_1^{(N_1)} \\
		\lambda^{(N)}_2 & \lambda_2^{(N_1)}
	\end{array}\right) .
	\label{eq:lambda:def}
\eea
If $\boldsymbol{\lambda}$ is rank-2, then none of the $\vec{J}_i$'s have vanishing projection onto $\mathcal{G}$ and the $\vec{J}_i^{\,\mathcal{G}}$'s form a complete basis of the plane of gauge degrees of freedom.  If $\boldsymbol{\lambda}$ is rank-1, either one of the two $\vec{J}_i$'s has a vanishing projection onto $\mathcal{G}$ while the other has a non-zero projection, or the two $\vec{J}_i^{\,\mathcal{G}}$ are non-null but aligned one with the other. If $\boldsymbol{\lambda}$ is rank-0, then the two $\vec{J}_i$'s have vanishing projection onto the plane of gauge degrees of freedom (note that a vanishing rank is obtained if all the gauge multipliers are set to zero).

From these considerations, one can identify three cases for which time derivatives of the gauge degrees of freedom can be removed.

\subsubsection*{Case one} First, the most obvious case is when the two vectors $\vec{J}_i$ are both orthogonal to the plane of gauge degrees of freedom. This occurs when the whole set of gauge multipliers is set equal to zero, \ie $\mathrm{Rank}(\boldsymbol{\lambda})=0$. In this case, \Eq{eq:GCPhys} is free from any $\dot{P}_\mu$ and the gauge-fixed dynamics reduces to 
\bea
	\left\{\begin{array}{ll}
		\ds \dzdot=\delta N\,\boldsymbol{\Omega}_6\vec{S}_k+k\delta N_1\,\boldsymbol{\Omega}_6\vec{D}_k+2N\,\left(\boldsymbol{\Omega}_6\boldsymbol{H}_k\right) \dz  \\
		\ds\vec{S}_k\cdot \dz =0=\vec{D}_k\cdot \dz \\
		\vec{G}_1\cdot \dz =0=\vec{G}_2\cdot \dz 
	\end{array}\right.  \, , \label{eq:GFsystemSF}
\eea
where we readily see that gauge conditions are algebraic in the phase space.

\subsubsection*{Case two} Second, one can suppose that only one of the vectors $\vec{J}_i$ has a vanishing projection onto $\mathcal{G}$ [say $\vec{J}_1$ for instance, meaning that $\lambda^{(N)}_1=0=\lambda^{(N_1)}_1$], which corresponds to $\mathrm{Rank}(\boldsymbol{\lambda})=1$. Then the first gauge condition is written as $\vec{G}_1\cdot \dz =0$ while the second is $\vec{G'}_2\cdot \dz +\vec{J}_2\cdot\dzdot=0$. Our purpose is to recast the latter condition in a form that does not depend explicitly on derivatives of the phase-space variables. 

The first gauge condition has to be preserved throughout the evolution, which leads to $\vec{G}_1\cdot\dzdot +\dot{\vec{G}}_1\cdot \dz =0$. Decomposing the second gauge condition and the time-derivative of the first gauge condition on the \Pbasis leads to
\bea
	\ds\sum_{\mu=1}^2\left[\left(\vec{G'}_2\cdot\vec{e}^{\,\mathcal{G}}_\mu+\vec{J}_2\cdot\dot{\vec{e}}^{\,\mathcal{G}}_\mu\right)P_\mu+\left(\vec{J}_2\cdot\vec{e}^{\,\mathcal{G}}_\mu\right)\dot{P}_\mu\right]\approx-\ds\sum_{\mu,\mu'=1}^2\left[\boldsymbol{J}_2\right]_{\mu\mu'}Z_{\mu'}, \\
	\ds\sum_{\mu=1}^2\left[\left(\dot{\vec{G}}_1\cdot\vec{e}^{\,\mathcal{G}}_\mu+{\vec{G}}_1\cdot\dot{\vec{e}}^{\,\mathcal{G}}_\mu\right)P_\mu+\left({\vec{G}}_1\cdot\vec{e}^{\,\mathcal{G}}_\mu\right)\dot{P}_\mu\right]\approx-\ds\sum_{\mu,\mu'=1}^2\left[\widetilde{\boldsymbol{J}}_1\right]_{\mu\mu'}Z_{\mu'},
\eea
where $\widetilde{\boldsymbol{J}}_1$ is given by \Eq{eq:Jmat} by substituting $\vec{G'}_i$ by $\dot{\vec{G}}_1$ and $\vec{J}_i$ by $\vec{G}_1$. Hence if the projection of $\vec{J}_2$ onto the plane of gauge degrees of freedom is aligned with the projection of $\vec{G}_1$ onto that same plane [\ie if $\vec{J}_2\cdot\vec{e}^{\,\mathcal{G}}_\mu=\alpha(\tau)\,(\vec{G}_1\cdot\vec{e}^{\,\mathcal{G}}_\mu)$ for all $\mu=1,\,2$ with $\alpha$ any function of time], then the time-derivative of the first gauge condition can be used to eliminate $\dot{P}_\mu$ in the second gauge condition, which then reads
\bea
	\ds\sum_{\mu=1}^2\left[\left(\vec{G'}_2-\alpha\dot{\vec{G}}_1\right)\cdot\vec{e}^{\,\mathcal{G}}_\mu + \left(\vec{J}_2  - \alpha \vec{G}_1 \right)\cdot \dot{\vec{e}}_\mu^{\,\mathcal{G}}\right]\,P_\mu\approx-\ds\sum_{\mu,\mu'=1}^2\left[\boldsymbol{J}_2-\alpha\widetilde{\boldsymbol{J}}_1\right]_{\mu\mu'}Z_{\mu'}\, .
\eea

Let us now introduce the gauge vector $\vec{G''}_2$ defined as
\bea
	\vec{G''}_2=\ds\sum_{\mu=1}^2\left[\left(\vec{G'}_2-\alpha\dot{\vec{G}}_1\right)\cdot\vec{e}^{\,\mathcal{G}}_\mu + \left(\vec{J}_2  - \alpha \vec{G}_1 \right)\cdot \dot{\vec{e}}_\mu^{\,\mathcal{G}}\right]\vec{e}^{\,\mathcal{G}}_\mu+\ds\sum_{\mu,\mu'=1}^2\left[\boldsymbol{J}_2-\alpha\widetilde{\boldsymbol{J}}_1\right]_{\mu\mu'}\vec{e}^{\,\mathcal{P}}_{\mu'}\,. \label{eq:Gsecond2''}
\eea
By differentiating $(\vec{J}_2  - \alpha \vec{G}_1 )\cdot \vec{e}_\mu^{\,\mathcal{G}}=0$ with respect to time, one finds $(\vec{J}_2  - \alpha \vec{G}_1 )\cdot \dot{\vec{e}}_\mu^{\,\mathcal{G}}=(-\dot{\vec{J}}_2+\alpha\dot{\vec{G}}_1+\dot{\alpha}\vec{G}_1)\cdot\vec{e}_\mu^{\,\mathcal{G}}$. Using this identity, the gauge vector simplifies to 
\bea
	\vec{G''}_2&=&\ds\sum_{\mu=1}^2\left[\left(\vec{G'}_2-\dot{\vec{J}}_2 + \dot{\alpha} \vec{G}_1 \right)\cdot\vec{e}^{\,\mathcal{G}}_\mu \right]\vec{e}^{\,\mathcal{G}}_\mu+\ds\sum_{\mu,\mu'=1}^2\left[\boldsymbol{J}_2-\alpha\widetilde{\boldsymbol{J}}_1\right]_{\mu\mu'}\vec{e}^{\,\mathcal{P}}_{\mu'}\,, \label{eq:Gsecond2} \\
 &=& \ds\sum_{\mu=1}^2\left[\left(\vec{G}_2 -\boldsymbol{\nabla}_\tau \vec{J}_2 + \dot{\alpha} \vec{G}_1 \right)\cdot\vec{e}^{\,\mathcal{G}}_\mu \right]\vec{e}^{\,\mathcal{G}}_\mu+\ds\sum_{\mu,\mu'=1}^2\left[\boldsymbol{J}_2-\alpha\widetilde{\boldsymbol{J}}_1\right]_{\mu\mu'}\vec{e}^{\,\mathcal{P}}_{\mu'}\,, \nonumber
\eea
where we have introduced the $\boldsymbol{\nabla}_\tau$ operator defined in \Eq{eq:nabla} and used \Eqs{eq:Gp:def} and~\eqref{eq:J:def} to express $\vec{G}'_2$ in terms of $\vec{G}_2$ and $\vec{J}_2$. Finally, the gauge-fixed dynamical system reduces to
\bea
	\left\{\begin{array}{ll}
		\ds\dzdot=\delta N\,\boldsymbol{\Omega}_6\vec{S}_k+k\delta N_1\,\boldsymbol{\Omega}_6\vec{D}_k+2N\,\left(\boldsymbol{\Omega}_6\boldsymbol{H}_k\right) \dz  \\
		\ds\vec{S}_k\cdot \dz =0=\vec{D}_k\cdot \dz  \\
		\vec{G}_1\cdot \dz =0=\vec{G''}_2\cdot \dz 
	\end{array}\right. \, . \label{eq:GFsystemN}
\eea
Both gauge conditions are now free from any time derivative providing a redefinition of the second gauge vector, and the gauge conditions are in a form such that $\mathrm{Rank}(\boldsymbol{\lambda})=0$.
 
\subsubsection*{Case three} Third, we suppose that none of the $\vec{J}_i$'s are orthogonal to the plane of gauge degrees of freedom. We decompose them as $\vec{J}_i=\vec{J}_i^{\,\mathcal{G}}+\vec{J}_i^{\,\perp}$ where $\vec{J}_i^{\,\mathcal{G}}\in\mathcal{G}$ and $\vec{J}_i^{\,\perp}\in\mathcal{C}\otimes\mathcal{P}$. By inserting this decomposition into the gauge conditions~\eqref{eq:GFsystemdiff} and by further using the equations of motion for $\vec{J}_i^{\,\perp}\cdot\dzdot $, we arrive at
\bea
	\left(\vec{G'}_i-2N\boldsymbol{H}_k\boldsymbol{\Omega}_6\vec{J}_i^{\,\perp}\right)\cdot \dz +\vec{J}_i^{\,\mathcal{G}}\cdot\dzdot=0.
\eea
We now assume that the two $\vec{J}_i^{\,\mathcal{G}}$'s are aligned, \ie $\vec{J}_2^{\,\mathcal{G}}=\alpha(\tau)\vec{J}_1^{\,\mathcal{G}}$, which means that $\mathrm{Rank}(\boldsymbol{\lambda})=1$. One can recast the two gauge conditions as
\bea
	{\vec{G''}_1}\cdot \dz =0
\quad\quad\text{and}\quad\quad
	\vec{G'}_2\cdot \dz +\vec{J}_2\cdot\dzdot =0 \, ,
\eea
where
\bea
	\vec{G''}_1=\alpha\left(\vec{G'}_1-2N\boldsymbol{H}_k\boldsymbol{\Omega}_6\vec{J}_1^{\,\perp}\right)-\vec{G'}_2+2N\boldsymbol{H}_k\boldsymbol{\Omega}_6\vec{J}_2^{\,\perp}\, .
\eea
The same situation as in ``case two'' is recovered and one can perform the same analysis to define the conditions for the time-derivatives of the gauge degrees of freedom to be removed. It requires $\vec{J}_2^{\,\mathcal{G}}$ to be aligned with the projection of $\vec{G''}_1$ onto the plane of gauge degrees of freedom. This third case is thus equivalent to the second one and the gauge conditions can be written in a form such that $\mathrm{Rank}(\boldsymbol{\lambda})=0$. 

\subsubsection*{Other cases} Up to our investigations, there is no other way to remove the time-derivatives of the gauge degrees of freedom. 

In particular, this is not possible if $\mathrm{Rank}(\boldsymbol{\lambda})=2$. Indeed, if both $\vec{J}_i$'s have non-zero projections onto the plane of the gauge degrees of freedom, which are linearly independent, one can either use the equations of motion or the time-derivative of the linear constraints to try to remove $\dot{P}_\mu$ (note that using the time-derivative of the gauge conditions is useless here since one would end up with second time-derivatives of the gauge degrees of freedom). The latter are useless since the time-derivatives of the linear constraints  are confined to the plane of constraints. Using the former will lead to injecting the Lagrange multipliers, hence recovering the original form of the gauge conditions, \Eq{eq:gaugecond}. 

If $\mathrm{Rank}(\boldsymbol{\lambda})=1$ but $\vec{J}^{\,\mathcal{G}}_2$ is not aligned with $\vec{G}^{\,\mathcal{G}}_1$, one can use the time-derivative of the first gauge condition to get the set $\dot{\vec{G}}_1\cdot \dz +\vec{G}_1\cdot\dzdot=0$ and $\vec{G'}_2\cdot \dz +\vec{J}_2\cdot\dzdot=0$ as the new gauge conditions. Since $\vec{J}^{\,\mathcal{G}}_2$ is not aligned with $\vec{G}^{\,\mathcal{G}}_1$, we recover the situation in which $\mathrm{Rank}(\boldsymbol{\lambda})=2$, hence the time-derivative of the gauge degrees of freedom cannot be removed.\footnote{Note that here we have an implication instead of an equivalence, \ie
\bea
\begin{cases}
\ds\vec{G}_1\cdot \dz =0 \\
		\ds\vec{G'}_2\cdot \dz +\vec{J}_2\cdot\dzdot=0
\end{cases}
\Rightarrow\quad 
\begin{cases}
\dot{\vec{G}}_1\cdot \dz +\vec{G}_1\cdot\dzdot=0 \\
		\ds\vec{G'}_2\cdot \dz +\vec{J}_2\cdot\dzdot=0
\end{cases}\, ,
\eea
which is however sufficient to prove that time-derivatives of the gauge degrees of freedom cannot be removed.}

\subsubsection{Solving the algebraic constraints} 

Gauge conditions which are free of time-derivatives of the gauge degrees of freedom can be systematically rewritten as $\vec{G}_i\cdot \dz =0$. One should now identify the conditions under which these two algebraic constraints in the phase space lead to a unique expression of the gauge degrees of freedom. By decomposing the solution in the \Pbasis and working on the surface of constraints, the two gauge conditions can be casted  in the following linear system
\bea
	\underbrace{\left(\begin{array}{ccc}
		\vec{G}_1\cdot\vec{e}^{\,\mathcal{G}}_1 &  &\vec{G}_1\cdot\vec{e}^{\,\mathcal{G}}_2 \\
		\vec{G}_2\cdot\vec{e}^{\,\mathcal{G}}_1 & & \vec{G}_2\cdot\vec{e}^{\,\mathcal{G}}_2
	\end{array}\right)}_{\boldsymbol{G}}\left(\begin{array}{c}
		P_1 \\
		P_2 
	\end{array}\right)\approx\ds\sum_{\mu=1}^2\left(\begin{array}{c}
		\vec{G}_1\cdot\vec{e}^{\,\mathcal{P}}_\mu \\
		\vec{G}_2\cdot\vec{e}^{\,\mathcal{P}}_\mu
	\end{array}\right)Z_\mu. \label{eq:gaugephyssystem}
\eea
This system is invertible providing that $\det(\boldsymbol{G})\neq0$, \ie providing that the projections of the vectors $\vec{G}_i$ onto the plane of gauge degrees of freedom lead to two linearly independent vectors of $\mathcal{G}$. 

Then, it is straightforward to unequivocally express the gauge degrees of freedom in terms of the physical degrees of freedom. This can be inserted in the final solution, which is thus solely determined by the physical degrees of freedom, as required for a gauge to be non-pathological. It reads
\bea
	 \dz \approx\ds\sum_{\mu=1}^2\left[\vec{e}^{\,\mathcal{P}}_\mu\,Z_\mu+\vec{e}^{\,\mathcal{G}}_\mu\ds\sum_{\mu',\mu''}\left[\boldsymbol{G}^{-1}\right]_{\mu\mu'}\left(\vec{G}_{\mu'}\cdot\vec{e}^{\,\mathcal{P}}_{\mu''}\right)\,Z_{\mu''}\right],
\eea
where the second term in the square bracket lies in the plane of gauge degrees of freedom and is explicitly set by the physical degrees of freedom.

\subsection{Lagrange multipliers} 
\label{sec:LagMult}

As explained above, physically, fixing a gauge amounts to working with a specific set of space-time coordinates, hence with a specific perturbed lapse function and shift vector. However, so far the gauge-fixing procedure has been described as one leading to the gauge degrees of freedom to be removed from the final solution. We now explain how this implies that, indeed,  the Lagrange multipliers can be expressed as functions of the phase-space variables.

In non-pathological gauges, the gauge conditions can be cast in the form $\vec{G}_i\cdot \dz =0$. The corresponding expression of the Lagrange multipliers is obtained requiring the gauge conditions to be preserved in time, \ie $\dd (\vec{G}_i\cdot \dz )/\dd\tau=0$. Taking the time-derivative of the gauge conditions and making use of the equations of motion of $ \dz $, \Eq{eq:GFsystemdiff}, we arrive at the linear system
\bea
	\underbrace{\left(\begin{array}{ccc}
		\vec{G}_1\cdot(\boldsymbol{\Omega}_6\vec{S}_k) & & \vec{G}_1\cdot(\boldsymbol{\Omega}_6\vec{D}_k) \\
		\vec{G}_2\cdot(\boldsymbol{\Omega}_6\vec{S}_k) & & \vec{G}_2\cdot(\boldsymbol{\Omega}_6\vec{D}_k)
	\end{array}\right)}_{\boldsymbol{G}_{\mathrm{LM}}}\left(\begin{array}{c}
		\delta N \\
		k\delta N_1
	\end{array}\right)=-\left(\begin{array}{c}
		\boldsymbol{\nabla}_\tau\vec{G}_1\,\cdot \dz  \\
		\boldsymbol{\nabla}_\tau\vec{G}_2\,\cdot \dz 
	\end{array}\right), \label{eq:LMfix}
\eea
where the equality holds for gauge-fixed solutions. The above system is invertible providing that $\det(\boldsymbol{G}_\mathrm{LM})\neq0$. The vectors  $(\boldsymbol{\Omega}_6\vec{S}_k)$ and $(\boldsymbol{\Omega}_6\vec{D}_k)$ form a complete basis of the plane of gauge degrees of freedom. Since the projections of the vectors $\vec{G}_i$'s onto $\mathcal{G}$ have to form a complete basis of $\mathcal{G}$ for the gauge to be non-pathological, the matrix $\boldsymbol{G}_\mathrm{LM}$ is necessarily non-singular.\footnote{In fact, $\det(\boldsymbol{G}_\mathrm{LM})\neq0$ if and only if $\det(\boldsymbol{G})\neq0$. One easily finds that $\boldsymbol{G}_\mathrm{LM}=\boldsymbol{G \,L}$ where 
\bea
\boldsymbol{L}= \left(\begin{array}{cc}
        \lambda_S\lambda^{-1}_D\mu_S^{-1} & -\lambda_D^{-1} \\
        -\mu_S^{-1} & 0
    \end{array}\right)\,. \label{eq:GLM-G}
\eea
This matrix is non-singular and is built from the projections of $(\boldsymbol{\Omega}_6\vec{S}_k)$ and $(\boldsymbol{\Omega}_6\vec{D}_k)$ onto the basis $\lbrace \vec{e}^{\,\mathcal{G}}_\mu \rbrace$.\label{foot:GLMG}} As a consequence, the Lagrange multipliers can be uniquely determined as functions of the phase-space variables. 

Let us note that, since the gauge degrees of freedom are uniquely expressed as functions of the physical degrees of freedom in non-pathological gauges, the above implies that the Lagrange multipliers are also fixed entirely by the physical degrees of freedom. This is done in detail in \App{app:LagMultKin}, where the constraints and the gauge conditions are inserted in the right-hand side of \Eq{eq:LMfix}, and the contributions from the two constraint vectors and the two gauge vectors in $\boldsymbol{\nabla}_\tau\vec{G}_i$ are shown to vanish in $(\boldsymbol{\nabla}_\tau\vec{G}_i)\cdot \dz $. By decomposing the gauge conditions, $\vec{G}_i\cdot \dz =0$, and the right-hand side of \Eq{eq:LMfix}, on the \Pbasis, one finds
\bea
    \boldsymbol{G}_{\mathrm{LM}}\left(\begin{array}{c}
		\delta N \\
		k\delta N_1
	\end{array}\right)\approx\boldsymbol{N}\left(\begin{array}{c}
	    Z_1 \\
	    Z_2
	 \end{array}\right), \label{eq:LMphysdof}
\eea
where the weak equality holds for gauge-fixed solutions.\footnote{This equation is equivalent to \Eqs{eq:P1dot} and \eqref{eq:P2dot} in which the gauge conditions and the constraints have been applied.} The matrix $\boldsymbol{N}$ reads
\bea
    \boldsymbol{N}=\left(\dot{\boldsymbol{G}}-\boldsymbol{G}\boldsymbol{k}_{\mathrm{cg}}\right)\boldsymbol{G}^{-1}\boldsymbol{G}_{\mathrm{p}}-\left(\dot{\boldsymbol{G}}_{\mathrm{p}}+\boldsymbol{G}_{\mathrm{p}}\boldsymbol{\Omega}_2 \boldsymbol{k}_{\mathrm{pp}}\right)+\boldsymbol{G k}_{\mathrm{cg}} \, , \label{eq:LMmatN}
\eea
where the $2\times2$ matrix $\boldsymbol{G}_{\mathrm{p}}$ is built from the components of the gauge vector in the plane of physical degrees of freedom, \ie $[\boldsymbol{G}_{\mathrm{p}}]_{i,\mu}=\vec{G}_i\cdot\vec{e}^{\,\mathcal{P}}_\mu$, and where the matrices $\boldsymbol{k}_i$'s are given in \App{ssec:dynPhys}.

It is worth stressing that for non-pathological gauges, the Lagrange multipliers are obtained {\it on-shell}, since using the equations of motion is needed to derive expressions for $\delta N$ and $\delta N_1$ [\ie $ \dz $ in the right-hand side of \Eq{eq:LMfix} has to satisfy the dynamical equations]. This highlights the role played by Lagrange multipliers, which are \emph{derived} quantities ensuring the gauge conditions to be preserved through evolution, rather than quantities that are chosen {\it a priori} to fix the gauge. In other words, they are consequences of the gauge conditions, not the gauge conditions themselves. On the contrary, for gauges that are pathological because of the presence of time-derivatives of the gauge degrees of freedom, if $\mathrm{Rank}(\boldsymbol{\lambda})=2$ the Lagrange multipliers are given by
\bea
	\left(\begin{array}{c}
		\delta N \\
		k\delta N_1
	\end{array}\right)=\boldsymbol{\lambda}^{-1}\left(\begin{array}{c}
		\vec{G}_1\cdot \dz  \\
		\vec{G}_2\cdot \dz 
	\end{array}\right). \label{eq:LMpath}
\eea
Here, the perturbed lapse and shift are derived {\it off-shell} [\ie $ \dz $ in the right-hand side of \Eq{eq:LMpath} does not have to satisfy the dynamical equations]. We finally note that if the gauge is pathological because $\det(\boldsymbol{G})=0$ and $\mathrm{Rank}(\boldsymbol{\lambda})=0$, then the Lagrange multipliers cannot be uniquely fixed since $\boldsymbol{G}_\mathrm{LM}$ is singular (at least one linear combination of them remains fully undetermined).

\subsection{Unicity of gauges}
\label{sssec:gaugeuni}

A criterion that is usually invoked to determine if a gauge is pathological or not is the unicity of the gauge-fixing procedure. Here, we show that this criterion is equivalent to the definition introduced above, where non-pathological gauges are those where the gauge degrees of freedom are uniquely fixed by the physical ones. 
A gauge $G$ is said to be uniquely determined if the gauge parameters $\xi^0$ and $\xi$ that generate the gauge transformation to go from any other gauges $G'$ to the considered gauge $G$ are unequivocally fixed. Let us denote by an overtilde the phase-space variables and the perturbed Lagrange multipliers in the gauge $G$. The gauge conditions associated to $G$ are 
\bea
\label{eq:gauge:cond:GT:1}
	\vec{G}_1\cdot\widetilde{ \dz }+\lambda^{(N)}_1\,\widetilde{\delta N}+\lambda^{(N_1)}_1\,k\widetilde{\delta N_1}&=&0\, , \\
	\vec{G}_2\cdot\widetilde{ \dz }+\lambda^{(N)}_2\,\widetilde{\delta N}+\lambda^{(N_1)}_2\,k\widetilde{\delta N_1}&=&0\, .
\label{eq:gauge:cond:GT:2}
\eea
We now rewrite $\widetilde{ \dz }$, $\widetilde{\delta N}$, and $k\widetilde{\delta N_1}$ in the gauge $G$ as functions of the same variables in another gauge $G'$, that we denote ${ \dz }$, ${\delta N}$, and $k{\delta N_1}$. This can be done using the Lie derivatives~\eqref{eq:Lie:def}, and one obtains \Eqs{eq:gtz}, \eqref{eq:tildeN} and~\eqref{eq:tildeN1}. By inserting these relations into the gauge conditions~\eqref{eq:gauge:cond:GT:1} and~\eqref{eq:gauge:cond:GT:2}, one finds
\bea
	\boldsymbol{\lambda}\left(\boldsymbol{I}_2\frac{\dd}{\dd\tau}+\boldsymbol{T}\right)\left(\begin{array}{c}
		N\xi^0 \\
		k\xi
	\end{array}\right)+\boldsymbol{G}_\mathrm{LM}\left(\begin{array}{c}
		N\xi^0 \\
		k\xi
	\end{array}\right)=-\boldsymbol{\lambda}\left(\begin{array}{c}
		\delta N \\
		k\delta N_1
	\end{array}\right)-
	\left(\begin{array}{c}
		\vec{G}_{1}\cdot \dz  \\
		\vec{G}_{2}\cdot \dz 
	\end{array}\right), \label{eq:unisys}
\eea
where $\boldsymbol{T}$ is a $2\times2$ matrix given by
\bea
	\boldsymbol{T}=\left(\begin{array}{cc}
		0 & ~~~0 \\
		-{N} k^2/{v^{2/3}} & ~~~0
	\end{array}\right).
\eea

In full generality then, the gauge parameters $\xi^0$ and $\xi$ are determined by a 2-dimensional system of ordinary differential equations. This leads to two possible sources of underdetermination of the gauge parameters. First, because time-differentiation is involved, $\xi^0$ and $\xi$ might be fixed up to some arbitrary initial conditions only (that may depend on space). Second, even if time derivatives can be removed and the system of differential equations is turned into an ordinary algebraic system, another source of underdetermination appears if the resulting algebraic system is not invertible. 
Removing time-derivatives of the gauge parameters in \Eq{eq:unisys} requires to set $\mathrm{Rank}(\boldsymbol{\lambda})=0$.\footnote{More precisely, according to the terminology of \Sec{sssec:RegainingAlgConst}, this corresponds to ``case one'', while ``case two'' and case ``case three'' can be shown to follow similar arguments.} We are then left with an algebraic system which is invertible if $\det(\boldsymbol{G}_\mathrm{LM})\neq0$, which is equivalent to requiring $\det(\boldsymbol{G})\neq 0 $, see footnote~\ref{foot:GLMG}. 
Hence the gauge $G$ is unique if $\mathrm{Rank}(\boldsymbol{\lambda})=0$ {\it and} $\det(\boldsymbol{G})\neq0$, which coincides with the two conditions obtained above for $G$ to be non-pathological, \ie for the final solution to be free of any gauge degrees of freedom. The reverse is yet to be confirmed and would require to show that $\mathrm{Rank}(\boldsymbol{\lambda})=0$ and $\det(\boldsymbol{G})\neq0$ is the unique way to solve \Eq{eq:unisys}.

\subsection{Gauge classification} 
\label{sssec:gaugeclass}

\begin{figure}
\begin{center}
	\includegraphics[scale=0.5]{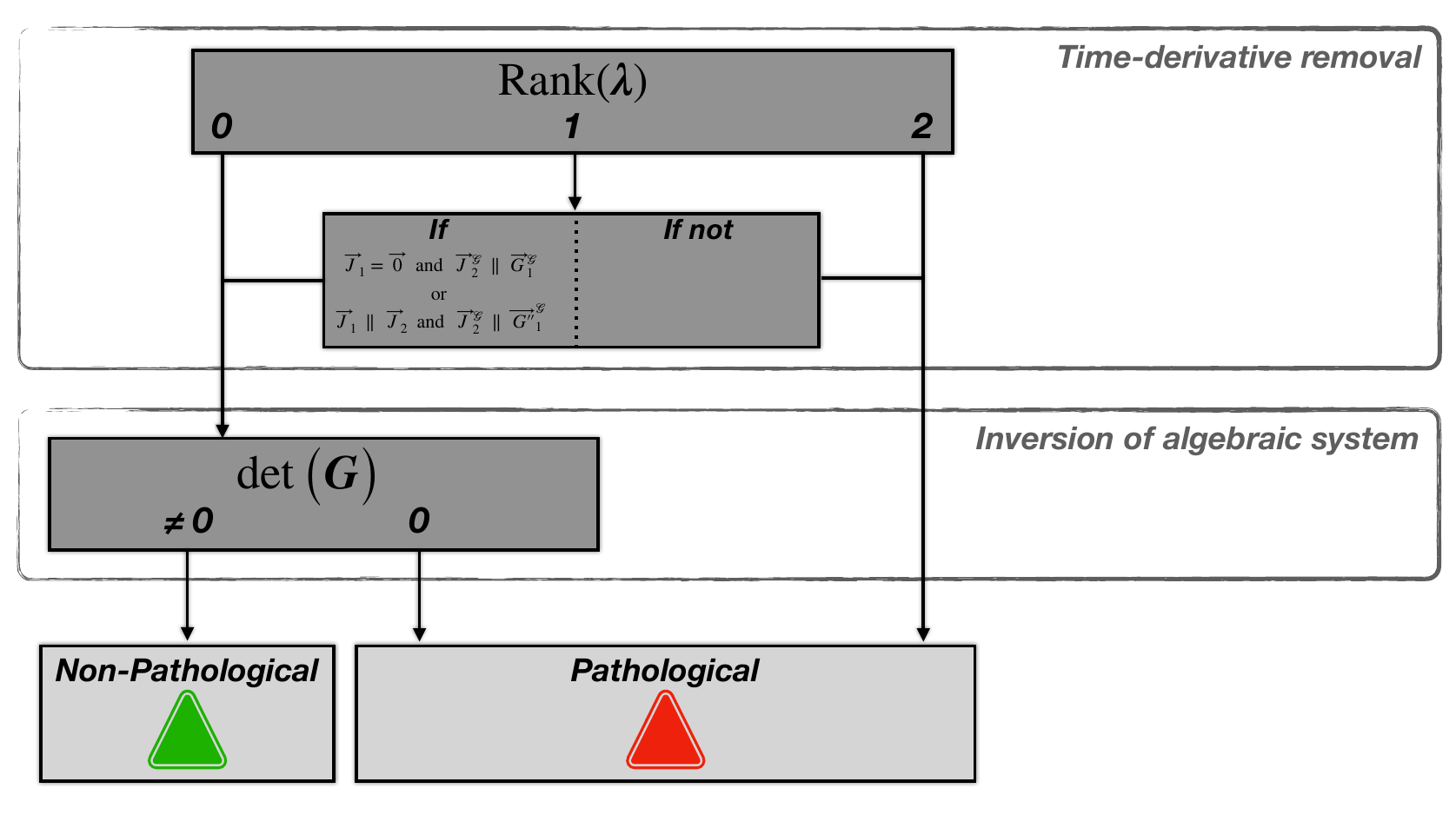}
	\caption{Classification algorithm to determine if a gauge is pathological or not. Non-pathological gauges are such that the gauge degrees of freedom are fully determined by the physical ones, hence they drop out of the final solution for the dynamics. In contrast, in pathological gauges, the gauge degrees of freedom are not fully eliminated. These properties are determined by the rank of the matrix $\boldsymbol{\lambda}$ and by the determinant of the matrix $\boldsymbol{G}$. The notation $\vec{V}_1\,\parallel\,\vec{V}_2$ means that the two vectors are aligned one with each other and the superscript $\mathcal{G}$ denotes the projection onto the plane of gauge degrees of freedom (see main text for details).}
	\label{fig:gauge}
\end{center}
\end{figure}

The results derived previously are summarised in \Fig{fig:gauge} and are detailed hereafter. We first recall that the gauge-fixed dynamics~\eqref{eq:GFsystem} can be written as
\bea
	\left\{\begin{array}{ll}
		\ds\dzdot=\delta N\,\boldsymbol{\Omega}_6\vec{S}_k+k\delta N_1\,\boldsymbol{\Omega}_6\vec{D}_k+2N\,\left(\boldsymbol{\Omega}_6\boldsymbol{H}_k\right) \dz  \\
		\ds\vec{S}_k\cdot \dz =0=\vec{D}_k\cdot \dz  \\
		\vec{G}_1\cdot \dz +\lambda^{(N)}_1\,\delta N+\lambda^{(N_1)}_1\,k\delta N_1=0=\vec{G}_2\cdot \dz +\lambda^{(N)}_2\,\delta N+\lambda^{(N_1)}_2\,k\delta N_1
	\end{array}\right. \, .  \label{eq:GFsystemGen}
\eea
The gauge multipliers can be arranged into the matrix $\boldsymbol{\lambda}$ defined in \Eq{eq:lambda:def}, and in \Sec{sssec:RegainingAlgConst} it was shown that if $\mathrm{Rank}(\boldsymbol{\lambda})=1$, after a suitable redefinition of the gauge vectors and of the gauge multipliers one can rewrite the gauge conditions in a form such that either $\mathrm{Rank}(\boldsymbol{\lambda})=0$ (\ie all the gauge multipliers are set equal to zero) or $\mathrm{Rank}(\boldsymbol{\lambda})=2$. 

\subsubsection*{Non-pathological gauges} A gauge is non-pathological if the dynamical system can be equivalently rewritten as
\bea
	\left\{\begin{array}{ll}
		\ds\dzdot=\delta N\,\boldsymbol{\Omega}_6\vec{S}_k+k\delta N_1\,\boldsymbol{\Omega}_6\vec{D}_k+2N\,\left(\boldsymbol{\Omega}_6\boldsymbol{H}_k\right) \dz  \\
		\ds\vec{S}_k\cdot \dz =0=\vec{D}_k\cdot \dz  \\
		\vec{G}_1\cdot \dz =0=\vec{G}_2\cdot \dz 
	\end{array}\right. \, ,  \label{eq:GFsystemNP}
\eea
such that the two vectors $\vec{G}_i$ have projections onto the plane of gauge degrees of freedom that are linearly independent (\ie their projections form a complete basis of $\mathcal{G}$). This is equivalent to requiring $\mathrm{Rank}(\boldsymbol{\lambda})=0$ and $\mathrm{Rank}(\boldsymbol{G})=2$, where the last condition is equivalent to $\det(\boldsymbol{G})\neq0$.

Rewriting the gauge conditions as algebraic constraints in the phase space is mandatory to remove time-derivatives of the gauge degrees of freedom from the gauge conditions. When doing so, the equations of motion for $\dz$ are used, hence the equivalence between \Eq{eq:GFsystemGen} and \Eq{eq:GFsystemNP} is at the level of the whole systems, it does not hold equation by equation. The condition on the projection of the gauge vectors onto the plane $\mathcal{G}$ is needed to further cast the gauge conditions as expressions for the gauge degrees of freedom in terms of the physical ones.

We note that a non-pathological gauge is not necessarily one where the gauge degrees of freedom are equal to zero in the final solution. Instead, it is such that the gauge degrees of freedom are uniquely fixed by the physical ones. Indeed, fixing the gauge consists in introducing two variables in the phase space, $G_i=\vec{G}_i\cdot \dz $, which are constants of motion set equal to zero, \ie ${G}_i=0$ and $\dot{G}_i=0$ for $i=1$ and $2$. For the gauge to be non-pathological, these two constants of motion should bear two linearly independent combinations of the gauge degrees of freedom. Hence, they necessarily have non-zero Poisson brackets with the constraints.\footnote{\label{foot:16}The Poisson bracket between the two gauge conditions is $\{G_1,G_2\}=\vec{G}_1\cdot(\boldsymbol{\Omega}_6\vec{G}_2)$ and can assume any value. 
When their Poisson bracket vanishes, up to a canonical transformation the gauge conditions can either be seen as two configuration variables or two momentum variables. 
If the bracket is non-zero, the two gauge conditions can be rescaled to form a pair of canonically conjugate variables without affecting the gauge choice.\label{footnote:G1G2PoissonBracket}} In this picture, the Lagrange multipliers given in \Eq{eq:LMfix} are dynamical \emph{consequences} of the gauge conditions ensuring their preservation throughout evolution. They are not part of the gauge definition itself.

\subsubsection*{Pathological gauges} 
A gauge choice is pathological if either $\mathrm{Rank}(\boldsymbol{\lambda})=2$ {\it or} $\mathrm{Rank}(\boldsymbol{G})\neq2$. 
If $\mathrm{Rank}(\boldsymbol{\lambda})=2$, the gauge conditions are not free from time-derivatives of the gauge degrees of freedom, which are thus still present in the final solution via their arbitrary initial conditions. If $\mathrm{Rank}(\boldsymbol{\lambda})=0$ but $\mathrm{Rank}(\boldsymbol{G})\neq2$, the two gauge vectors have aligned projections onto $\mathcal{G}$, or at least one of the two gauge vectors has a vanishing projection onto $\mathcal{G}$.  Hence the linear system relating the gauge degrees of freedom to the physical ones is under-constrained, making the gauge degrees of freedom not uniquely determined by the physical degrees of freedom. 

\section{Applications}
\label{sec:WorkEx}

In order to illustrate the formalism developed so far, in this section we apply it to famous gauges that have been proposed in the literature, for which it is well known whether they are pathological or not. In practice, the gauges we consider have been selected to cover the whole set of branches displayed in \Fig{fig:gauge}. In a second part, we further make use of our formalism to construct new types of gauges. 

In this section, the mass parameters and the conformal parameter introduced in \Eq{eq:deltaz} are set equal to one for convenience.\footnote{This corresponds to working with the rescaled vectors $\dz\to\boldsymbol{M}\dz$ and $\vec{V}\to\boldsymbol{M}^{-1}\vec{V}$ where $\vec{V}$ stands for the gauge vectors or the vectors of constraints, and where $\boldsymbol{M}$
is a time-independent and symplectic matrix given by
\bea                \boldsymbol{M}=\mathrm{diag}\left[m_\phi/\lambda^{3/2},\,\sqrt{\lambda}/m^2_1,\,\sqrt{\lambda}/m^2_2,\,\lambda^{3/2}/m_\phi,\,m^2_1/\sqrt{\lambda},\,m^2_2/\sqrt{\lambda}\right].
\eea
As a consequence, inner-dot products $\vec{V}\cdot\dz$ and symplectic products $\vec{V}_1\cdot(\boldsymbol{\Omega}_6\vec{V}_2)$ are invariant under such a rescaling.}

\subsection{Examples of known gauges}
\label{sec:examples:gauges}

\subsubsection*{Spatially-flat gauge}

This gauge is defined by $\delta\gamma_i=0$ for $i=1,\,2$. This means that the scalar field perturbations directly give the gauge-invariant Mukhanov-Sasaki variables, see \Eq{eq:MS:def}. In our language, it corresponds to setting $\boldsymbol{\lambda}$ to the null matrix and to introducing the gauge vectors $\vec{G}_{1}=\vec{e}^{\,\phi}_1$ and $\vec{G}_{2}=\vec{e}^{\,\phi}_2$ (this corresponds to the case where the two gauge conditions are two configuration variables, see footnote~\ref{footnote:G1G2PoissonBracket}). Time-derivatives of the gauge degrees of freedom are absent since $\mathrm{Rank}(\boldsymbol{\lambda})=0$. Then, the elements of the matrix $\boldsymbol{G}_\mathrm{LM}$ defined in \Eq{eq:LMfix}  are
\bea
	\vec{G}_1\cdot\left(\boldsymbol{\Omega}_6\vec{S}_k\right)=-\frac{\sqrt{3}}{\Mp^2}v^{2/3}\theta, \,\,\,\,\,\, \vec{G}_1\cdot\left(\boldsymbol{\Omega}_6\vec{D}_k\right)=-\frac{2}{\sqrt{3}}v^{2/3},
\eea
as well as
\bea
	\vec{G}_2\cdot\left(\boldsymbol{\Omega}_6\vec{S}_k\right)=0, \,\,\,\,\,\, \vec{G}_2\cdot\left(\boldsymbol{\Omega}_6\vec{D}_k\right)=-2\sqrt{\frac{2}{3}}v^{2/3}. \label{eq:Ggauge2SF}
\eea	
It is easily checked that $\det(\boldsymbol{G}_\mathrm{LM})\neq0$ and so is $\det(\boldsymbol{G})$ (see footnote \ref{foot:GLMG}). Hence, we recover that the spatially-flat gauge is non-pathological. The Lagrange multipliers are subsequently computed using \Eq{eq:LMfix}. Its right-hand side reads
\bea
	\boldsymbol{\nabla}_\tau\vec{{G}}_1&=&-\frac{N\theta}{2\Mp^2}\,\vec{e}^{\,\phi}_1-\frac{2Nv^{1/3}}{\Mp^2}\,\vec{e}^{\,\pi}_1, \\ 
	\boldsymbol{\nabla}_\tau\vec{{G}}_2&=&\frac{N\theta}{\Mp^2}\,\vec{e}^{\,\phi}_2+\frac{4Nv^{1/3}}{\Mp^2}\,\vec{e}^{\,\pi}_2. \label{eq:NG2SF} 
\eea
Finally, the Lagrange multipliers are derived by inverting the linear system and by making use of the gauge conditions. It boils down to
\bea
	\delta N&=&-\frac{2}{\sqrt{3}}\frac{N}{v^{1/3}\theta}\left(\delta\pi_1+\sqrt{2}\delta\pi_2\right), \\
	k\delta N_1&=&\frac{\sqrt{6}}{\Mp^2}\frac{N}{v^{1/3}}\,\delta\pi_2.
\eea
We recover the standard expressions in the spatially-flat gauge (see \eg Sec. 3.3 in \Refc{Artigas:2021zdk}).

\subsubsection*{Unitary gauge}
This gauge is defined by imposing $\delta\phi=0$ and $\delta\gamma_2=0$, which implies that the curvature perturbation is given by $\delta\gamma_1$. The corresponding gauge vectors are $\vec{G}_1=\vec{e}^{\,\phi}_0$ and $\vec{G}_2=\vec{e}^{\,\phi}_2$, while the matrix $\boldsymbol{\lambda}$ vanishes. Projecting the first gauge vector onto the plane of gauge degrees of freedom yields
\bea
    	\vec{G}_1\cdot\left(\boldsymbol{\Omega}_6\vec{S}_k\right)=\frac{\pi_\phi}{v} \,\,\,\mathrm{and}\,\,\, \vec{G}_1\cdot\left(\boldsymbol{\Omega}_6\vec{D}_k\right)=0.
\eea
The projection onto $\mathcal{G}$ of the second gauge vector is given in \Eq{eq:Ggauge2SF}. This leads to $\det(\boldsymbol{G}_\mathrm{LM})\neq0$, hence the unitary gauge is non-pathological. The right-hand side of \Eq{eq:LMfix} is derived from
\bea
    \boldsymbol{\nabla}_\tau\vec{G}_1=-\frac{\sqrt{3}N\pi_\phi}{2v^{5/3}}\,\vec{e}^{\,\phi}_1+\frac{N}{v}\,\vec{e}^{\,\pi}_0,
\eea
and from \Eq{eq:NG2SF} for $\boldsymbol{\nabla}_\tau\vec{G}_2$. Eventually, this gives for the Lagrange multipliers
\bea
    \delta N&=&\frac{\sqrt{3}N}{2v^{2/3}}\,\delta\gamma_1-\frac{N}{\pi_\phi}\,\delta\pi_\phi, \\
    k\delta N_1&=&\frac{\sqrt{6}}{\Mp^2}\frac{N}{v^{1/3}}\,\delta\pi_2,
\eea
where the gauge conditions have been used as well.

\subsubsection*{Newtonian gauge}
This gauge is defined by $\delta\gamma_2=0$ and $\delta N_1=0$ and it is often used since it gives simple expressions for the Bardeen potentials. In our language, the Newtonian gauge is fixed by introducing $\vec{G}_1=\vec{e}^{\,\phi}_2$ and $\vec{G}_{2}=\vec{0}$, as well as the matrix 
\bea
\boldsymbol{\lambda}=\left(\begin{array}{cc} 0 & 0 \\ 0 & 1 \end{array}\right)
\eea
whose rank equals 1. 

Following \Sec{sssec:diffgauge}, we first rewrite the second gauge conditions as a derivative condition on the phase space. To this end, we use $\vec{V}_1=\vec{e}^{\,\phi}_0$ and $\vec{V}_2=\vec{e}^{\,\phi}_2$ to project the equations of motion yielding $\vec{W}_1=(v/\pi_\phi)\,\vec{e}^{\,\phi}_0$ and $\vec{W}_2=- \frac{1}{2}\sqrt{\frac{3}{2}} \frac{1}{v^{2/3}} \,\vec{e}^{\,\phi}_2$. Then, the derivative version of $\delta N_1=0$ reads 
\bea
	\vec{G'}_2\cdot \dz +\vec{J}_2\cdot\dzdot=0,
\eea
where
\bea
	&&\vec{G'}_2= \sqrt{\frac{3}{2}} \frac{N}{v^{2/3}} \left(\frac{\theta}{2M^2_\mathrm{Pl}}\,\vec{e}^{\,\phi}_2+\frac{2v^{1/3}}{M^2_\mathrm{Pl}}\,\vec{e}^{\,\pi}_2\right), \\
	&&\vec{J}_2=-\frac{1}{2}\sqrt{\frac{3}{2}} \frac{1}{v^{2/3}} \,\vec{e}^{\,\phi}_2. \label{eq:J2newt}
\eea 
This shows that $\vec{J}_2$ is aligned with $\vec{G}_1$, and so are their projections onto $\mathcal{G}$. Hence the gauge conditions can be rewritten such that $\mathrm{Rank}(\boldsymbol{\lambda})=0$ and it is free of time-derivatives of the gauge degrees of freedom. The simplest rewriting of the Newtonian gauge is to keep $\vec{G}_1$ equal to $\vec{e}^{\,\phi}_2$ and to set $\vec{G}_2=\vec{e}^{\,\pi}_2$ (see also \Refc{Artigas:2021zdk}).\footnote{Note that $ \vec{G}_2\cdot \dz +\vec{J}_2\cdot\dzdot=0$ can be rewritten as $(\alpha \vec{e}^{\,\phi}_2+\beta\vec{e}^{\,\pi}_2)\cdot \dz =0$. Since the first condition is $ \vec{e}^{\,\phi}_2\cdot \dz =0$, it is equivalent to rewrite the set of gauge conditions as $ \vec{e}^{\,\phi}_2\cdot \dz =0$ and $ \vec{e}^{\,\pi}_2\cdot \dz =0$.} The projection of $\vec{G}_1$ onto the plane of gauge degrees of freedom is given in \Eq{eq:Ggauge2SF}. Projecting the second vector onto $\mathcal{G}$ gives
\bea
	&&\vec{G}_2\cdot\left(\boldsymbol{\Omega}_6\vec{S}_k\right)=-\frac{\Mp^2}{\sqrt{6}}\frac{k^2}{v^{1/3}} \,\,\,\,\,\,\mathrm{and}\,\,\,\,\,\, \vec{G}_2\cdot\left(\boldsymbol{\Omega}_6\vec{D}_k\right)=\sqrt{\frac{2}{3}}v^{1/3}\theta.
\eea 
Then, it is straightforward to show that $\det(\boldsymbol{G}_\mathrm{LM})\neq0$, hence that the Newtonian gauge is non-pathological.

To derive the expressions of the Lagrange multipliers, we first note that
\bea
	\boldsymbol{\nabla}_\tau\vec{{G}}_2&=&- \frac{\sqrt{2} N}{12 v} \Mp^2 k^2 \,\vec{e}^{\,\phi}_1 - \frac{2N}{3v^{1/3}}\left(\frac{\pi^2_\phi}{v^2}+ \frac{V}{2}-\frac{\Mp^2}{8}\frac{k^2}{v^{2/3}}\right)\,\vec{e}^{\,\phi}_2 -  \frac{N \theta}{\Mp^2}\,\vec{e}^{\,\pi}_2,
\eea
while the expression of $\boldsymbol{\nabla}_\tau\vec{{G}}_2$ is given in \Eq{eq:NG2SF}. Plugging this into \Eq{eq:LMfix} and further using the gauge conditions, \ie $\delta\gamma_2=0=\delta\pi_2$, we arrive at
\bea
	\delta N&=& -\frac{N}{2\sqrt{3}v^{2/3}}\delta\gamma_1, \label{eq:LapseNewt} \\
	k\delta N_1&=&0,
\eea
which exactly match the results obtained in \Refc{Artigas:2021zdk}. We note that \Eq{eq:LapseNewt} simply states that the two Bardeen's potentials are equal. It is also worth stressing that $\delta N_1=0$ now appears as a consequence of the gauge conditions instead of a gauge condition {\it per se}, in agreement with the discussion of \Sec{sec:GF}.

\subsubsection*{Comoving gauge}
The comoving gauge is defined by imposing $\delta \phi=0$ and $\delta N_1 =0$ which corresponds to setting $\vec{G}_1 = \vec{e}_0^{\,\phi}$, $\vec{G}_2 = \vec{0}$, and 
\bea
\boldsymbol{\lambda}=\left(\begin{array}{cc} 0 & 0 \\ 0 & 1 \end{array}\right).
\eea
Thus, it differs from the Newtonian gauge by the first gauge vector. Following the calculation carried out for the Newtonian gauge, the condition $\delta N_1=0$ leads to a derivative condition on the phase space with a vector $\vec{J}_2$ given by \Eq{eq:J2newt}. Its projection onto the plane of gauge degrees of freedom is $\vec{J}^{\,\mathcal{G}}_2=-\lambda_D \vec{e}_1^{\,\mathcal{G}} - \lambda_S \vec{e}_2^{\,\mathcal{G}}$, while projecting $\vec{G}_1$ onto $\mathcal{G}$ leads to $\vec{G}^{\,\mathcal{G}}_1=-\mu_S\pi_\phi v^{-1}\,\vec{e}^{\,\mathcal{G}}_1$. This shows that $\vec{J}_2^{\,\mathcal{G}}$ and $\vec{G}_1^{\,\mathcal{G}}$ are not aligned and time-derivatives of gauge degrees of freedom are not fully removed. Hence the comoving gauge is pathological.

\subsubsection*{Synchronous gauge}

This gauge is defined by $\delta N=0$ and $\delta N_1=0$, which ensures the foliation of the perturbed FLRW space-time to be synchronised with the homogeneous and isotropic background. In our formalism, it is fixed by setting the two gauge vectors to zero and by using the matrix
\bea
\boldsymbol{\lambda}=\left(\begin{array}{cc} 1 & 0 \\ 0 & 1 \end{array}\right).
\eea
Its rank equals two hence we straightforwardly find that the synchronous gauge is pathological since time-derivative of the gauge degrees of freedom cannot be removed. 

We note that it remains possible to define a non-pathological gauge such that $\delta N$ and $\delta N_1$ are both vanishing. This is explicitly done in \App{app:synchronous}. In this alternative construction, the Lagrange multiplier are not imposed to vanish independently of the constraints and of the equations of motion. Instead, we show that there exist two gauge vectors $\vec{G}_i$ such that a) they have linearly independent projections onto the plane of gauge degrees of freedom, and b) the resulting $\boldsymbol{\nabla}_\tau\vec{G}_i$'s are linear combinations of the gauge vectors themselves. As a consequence, imposing the gauge conditions $\vec{G}_i\cdot\dz=0$ leads to a non-pathological gauge. Moreover, the right-hand side of \Eq{eq:LMfix} is vanishing once the gauge conditions are imposed, hence leading to $\delta N=0=\delta N_1$. Constructing the synchronous gauge that way makes use of the equations of motion through \Eq{eq:LMfix}. Therefore, the fact that the Lagrange multipliers vanish is derived from the gauge conditions and the equations of motion, \ie it is only valid on-shell. This is in contrast with the standard definition of the synchronous gauge, in which $\delta N=0$ and $\delta N_1=0$ are imposed off-shell as gauge conditions, \ie they hold independently of the equations of motion.

\subsubsection*{Uniform-expansion gauge}
\label{ssec:unifexp}
The uniform-expansion gauge is built to ensure vanishing perturbations of the integrated expansion rate, that we dub $\delta\mathcal{N}_\mathrm{int}$ hereafter. It is commonly used in the framework of the stochastic $\delta N$-formalism \cite{Fujita:2013cna,Vennin:2015hra,Pattison:2019hef}. However, different sets of gauge conditions can yield $\delta\mathcal{N}_\mathrm{int}=0$ and it is important to determine which of these (if any) define non-pathological gauges.

In the Hamiltonian formalism, the perturbation of the expansion rate is given by
\bea
	\delta\mathcal{N}_\mathrm{int}=-\frac{1}{3}\ds\int\dd\tau\left(N\,\delta\Theta+\Theta\,\delta N\right), \label{eq:NintHam}
\eea
where $\Theta=3\theta/(2\Mp^2)$ is the background expansion rate and
\bea
	\delta\Theta=\frac{\sqrt{3}}{v^{2/3}\Mp^2}\left(v^{1/3}\delta\pi_1-\frac{\theta}{4}\delta\gamma_1\right)
\eea
is the perturbed expansion rate. Alternatively, one can use the equation of motion to recast the integrated expansion rate as \cite{Artigas:2021zdk}
\bea
	\delta\mathcal{N}_\mathrm{int}=\frac{\delta\gamma_1}{2\sqrt{3}v^{2/3}}+\frac{k}{3}\ds\int\dd\tau\,\delta N_1\,. \label{eq:NintStandard}
\eea
These expressions offer two different options to impose the uniform-expansion gauge, either $\delta N=0=\delta\Theta$ or $\delta N_1=0=\delta\gamma_1$ (note that the second option is the one most commonly used in the literature). However, we show in \App{app:UnifExpGauge} that these two ways of fixing the uniform-expansion gauge are pathological because the matrix $\boldsymbol{G}$ is singular for $\delta N=0=\delta\Theta$, and because time-derivatives of gauge degrees of freedom cannot be fully removed for $\delta N_1=0=\delta\gamma_1$. 

Another possibility consists in imposing $N\,\delta\Theta+\Theta\,\delta N=0$ as the first gauge condition, and to complement with a second gauge condition of the form $\vec{G}_2\cdot\dz=0$. Nevertheless, it is proved in \App{app:UnifExpGauge} that any choice for that second gauge condition leads to a pathological gauge too. The reason is that the second gauge condition should be designed to remove time-derivatives of the gauge degrees of freedom and to yield a non-singular matrix $\boldsymbol{G}$. However, these two conditions cannot be satisfied together because of the peculiar form of the relation $N\,\delta\Theta+\Theta\,\delta N=0$.

The above attempts are based on gauge conditions that involve one of the Lagrange multiplier. Instead, one could directly start from gauge conditions restricted to the phase space. This strategy demands to find two gauge vectors $\vec{G}_i$ such that a) their projections onto $\mathcal{G}$ are linearly independent and b) the perturbation of the expansion rate reads
\bea   
    \frac{\dd \delta\mathcal{N}_{\mathrm{int}}}{\dd\tau}=\ds\sum_{i=1}^2\left[\alpha_i(\tau)G_i+\beta_i(\tau)\dot{G}_i\right], \label{eq:UEGgen}
\eea
where $G_i\equiv\vec{G}_i\cdot\dz$ and where the $\alpha_i$'s and $\beta_i$'s are four arbitrary functions of time. The first requirement guarantees healthy gauges. The second yields vanishing perturbations of the expansion rate since $G_i=0=\dot{G}_i$ upon imposing gauge conditions. We stress that the expansion rate involves the perturbed lapse function [or the perturbed shift, depending on which of the forms~\eqref{eq:NintHam} and~\eqref{eq:NintStandard} is adopted], hence the time-derivative of the gauge conditions must necessarily appear in \Eq{eq:UEGgen}. Otherwise, its right-hand side would be free of any Lagrange multipliers. This implies that a uniform expansion rate necessarily appears as a dynamical consequence of the gauge conditions. In this approach, two gauge vectors can be varied to try to fulfil the above requirements. This contrasts with the previous approach where only one gauge vector could be varied, hence it should  accommodate for more possibilities. We try to implement this strategy in \App{app:UnifExpGauge} where, nonetheless, we can only find a set of gauge vectors that have the same projections onto $\mathcal{G}$, hence they define a pathological gauge. 

Up to our investigations, we have not found a set of gauge vectors that meets the two requirements identified above, and all implementations of the uniform-expansion gauge that have been proposed so far are pathological. Two words of caution are however in order. First, although we have been able to rule out a large class of possible implementations of the uniform-expansion gauge, we have not exhausted all possibilities and this does not prove that a gauge in which $\delta\mathcal{N}_{\mathrm{int}}=0$ on-shell does not exist. Further investigation is required, in the spirit of the program proposed below in \Sec{sec:GeneralStrategy}. Second, the uniform-expansion gauge is usually employed in the context of the $\delta N$ formalism, where only the leading-order terms in the gradient expansion are retained. For this purpose, it might be sufficient to find a gauge in which the above requirements are valid only at leading order in $k^2$. This would require to develop the present formalism in the reduced phase space of the separate universe~\cite{Artigas:2021zdk}, which we plan to address in a future work.

\subsection{Gauges with vanishing Lagrange multipliers}

Our formalism not only allows us to readily diagnose well-known gauges but also provides an efficient framework in which new classes of gauges can be constructed. We start by considering gauges where one of the two Lagrange multipliers vanishes.

\subsubsection*{Vanishing shift vector}
Gauges with a vanishing shift vector play a key role in the context of the separate universe approximation. Indeed, this approximation holds providing that the gauge conditions imposed at the level of CPT leads to a perturbed lapse function and a perturbed shift vector which, at large scales, match the ones of the independent FLRW patches \cite{Pattison:2019hef, Artigas:2021zdk}. Since homogeneity and isotropy of the FLRW patches impose a vanishing shift vector, gauges with $\delta N_1=0$ are expected to be particularly appropriate for using the separate universe approximation. However, the examples presented previously show that this class of gauges contains pathological elements, and it is important to circumscribe the sub-class that is non-pathological.

The easiest way to impose a vanishing shift vector is to consider the gauge vector $\vec{G}_2=\vec{0}$ and the matrix
\bea
	\boldsymbol{\lambda}=\left(\begin{array}{cc}
		0 & 0 \\
		0 & 1
	\end{array}\right)
\eea
whose rank equals 1. The first gauge vector, $\vec{G}_1$, is left unspecified but it should be such that the gauge is non-pathological. Following \Sec{sssec:diffgauge} and the calculations performed for the Newtonian gauge, the vector $\vec{J}_2$ is given in \Eq{eq:J2newt} and its projection onto the plane of gauge degrees of freedom reads $\vec{J}^\mathcal{G}_2=-\lambda_D\, \vec{e}_1^{\,\mathcal{G}}-\lambda_S\, \vec{e}_2^{\,\mathcal{G}}$. For time-derivatives of the gauge degrees of freedom to be removed, the vector $\vec{G}_1$ must be of the form
\bea
	\vec{G}_1=\frac{-1}{\alpha_k(\tau)}\left(\lambda_D\,\vec{e}_1^{\,\mathcal{G}}+\lambda_S\, \vec{e}_2^{\,\mathcal{G}}\right)+\vec{G}_1^{\,\mathcal{P}}+\vec{G}_1^{\,\mathcal{C}}, \label{eq:G1nullshift}
\eea
where $\alpha_k(\tau)$ is an arbitrary function of time and scale, and where $\vec{G}_1^{\,\mathcal{P}}$ and $\vec{G}_1^{\,\mathcal{C}}$ can be any vectors lying in the physical plane and in the plane of constraints respectively. One now needs to check that the matrix $\boldsymbol{G}$ is not singular. Its elements are given by the projections of $\vec{G}_1$ and $\vec{G}_2-\boldsymbol{\nabla}_\tau{\vec{J}}_2+\dot{\alpha}_k\vec{G}_1$ onto the plane of gauge degrees of freedom. Making use of $\vec{G}_2=\vec{0}$ and of $\vec{J}_2^{\,\mathcal{G}}=\alpha_k(\tau)\,\vec{G}^{\,\mathcal{G}}_1$, the resulting determinant reads
\bea
	\det\left(\boldsymbol{G}\right)=\frac{1}{\alpha_k(\tau)}\left[\left(\vec{J}_2\cdot\vec{e}^{\,\mathcal{G}}_1\right)\left(\boldsymbol{\nabla}_\tau\vec{J}_2\cdot\vec{e}^{\,\mathcal{G}}_2\right)-\left(\vec{J}_2\cdot\vec{e}^{\,\mathcal{G}}_2\right)\left(\boldsymbol{\nabla}_\tau\vec{J}_2\cdot\vec{e}^{\,\mathcal{G}}_1\right)\right] ,
\eea
which reduces to
\bea
	\det\left(\boldsymbol{G}\right)=-N\,\frac{\mu_S\,\lambda_D}{\alpha_k(\tau)}\,\left(\frac{k}{v^{1/3}}\right)^2.
\eea
This is non-vanishing and the gauges defined by $\delta N_1=0$ and $\vec{G}_1\cdot\dz=0$ with $\vec{G}_1$ given in \Eq{eq:G1nullshift} are therefore non-pathological. 

It is worth stressing that the above class of gauges does not necessarily exhaust all the possible non-pathological gauges with a vanishing shift vector. Indeed, we started from a specific way of writing the gauge, which introduces a matrix $\boldsymbol{\lambda}$ with a rank equal to 1. We showed that upon an appropriate choice of the gauge vector $\vec{G}_1$, such gauges can be recast into gauges with a vanishing matrix $\boldsymbol{\lambda}$. However, the reverse is not necessarily true. In other words, there exists gauges where $\boldsymbol{\lambda}=\boldsymbol{0}$, which nevertheless lead to $\delta N_1=0$ as a dynamical consequence. The non-pathological implementation of the synchronous gauge is an example of such type of gauges (see \App{app:synchronous}). Starting from \Eq{eq:LMfix}, the shift vector is given by $k\delta N_1=\vec{N}_1\cdot\dz$ where
\bea
	\vec{N}_1=\frac{1}{\det(\boldsymbol{G}_\mathrm{LM})}\left\{\left[\vec{G}_2\cdot\left(\boldsymbol{\Omega}_6\vec{S}_k\right)\right]\,\boldsymbol{\nabla}_\tau\vec{G}_1-\left[\vec{G}_1\cdot\left(\boldsymbol{\Omega}_6\vec{S}_k\right)\right]\,\boldsymbol{\nabla}_\tau\vec{G}_2\right\}.
\eea
Hence, $k\delta N_1$ vanishes if the vector $\vec{N}_1$ is a linear combination of the two gauge vectors and of the vectors of constraints. Finding the pairs of gauge vectors for which such a condition holds (in addition to the condition $\det(\boldsymbol{G}_\mathrm{LM})\neq0$ for the gauge to be non-pathological) requires to explore the space of eigenvectors of $\boldsymbol{\nabla}_\tau$, and this task is beyond the scope of this paper. However, this might offer another route towards gauges with vanishing perturbations of the shift vector in situations where the construction presented above fails to provide appropriate gauges. As a concrete example, let us consider the two following gauge vectors $\vec{G}_1=\boldsymbol{\nabla}_\tau\vec{G}_2$ and $\vec{G}_2=\vec{e}^{\,\mathcal{G}}_1+(\lambda_S/\lambda_D)\,\vec{e}^{\,\mathcal{G}}_2$. By construction, this gauge is free of time-derivatives of the gauge degrees of freedom. The elements of the matrix $\boldsymbol{G}_\mathrm{LM}$ are\footnote{Note that the projection of $\vec{G}_1$ onto $\mathcal{G}$ is expressed as a function of the couplings between the gauge degrees of freedom and the constraints (see Apps. \ref{app:HamKinBasis}, \ref{ssec:dynPhys} and \ref{app:LagMultKin}).}
\bea
	\vec{G}_1\cdot\left(\boldsymbol{\Omega}_6\vec{S}_k\right)=N\,\frac{\mu_S}{\lambda_D}\left(\frac{k}{v^{1/3}}\right)^2 &\,\,\,\mathrm{and}\,\,\,&\vec{G}_1\cdot\left(\boldsymbol{\Omega}_6\vec{D}_k\right)=\frac{\lambda_S}{\mu_S}\,\frac{\dot{\lambda}_D}{\lambda^2_D},
\eea
as well as
\bea
	\vec{G}_2\cdot\left(\boldsymbol{\Omega}_6\vec{S}_k\right)=0 &\,\,\,\mathrm{and}\,\,\,&\vec{G}_1\cdot\left(\boldsymbol{\Omega}_6\vec{D}_k\right)=\frac{1}{\lambda_D}.
\eea
This leads to $\det(\boldsymbol{G}_\mathrm{LM})\neq0$ and the gauge is non-pathological. Then, it is straightforward to show that $\delta N_1\propto \vec{G}_1\cdot\dz$, which vanishes upon imposing the first gauge condition.

\subsubsection*{Vanishing lapse function}
The construction developed above can also be applied to explore gauges with a vanishing lapse function. Let us thus introduce the vector $\vec{G}_2=\vec{0}$ and the rank-1 matrix 
\bea
	\boldsymbol{\lambda}=\left(\begin{array}{cc}
		0 & 0 \\
		1 & 0
	\end{array}\right).
\eea
The first gauge vector $\vec{G}_1$ is left unspecified. Following the calculations done in \App{app:UnifExpGauge}, the gauge condition $\delta N=0$ yields a vector $\vec{J}_2$ given in \Eq{eq:J2unif}. Its projection onto the plane of gauge degrees of freedom is $\vec{J}^{\,\mathcal{G}}_2=-\mu_S\,\vec{e}^{\,\mathcal{G}}_2$. Thus, we constrain the first gauge vector to be of the form
\bea
	\vec{G}_1=\frac{-\mu_S}{\alpha_k(\tau)}\,\vec{e}^{\,\mathcal{G}}_2+\vec{G}_1^{\,\mathcal{P}}+\vec{G}_1^{\,\mathcal{C}}. \label{eq:G1nulllapse}
\eea
Since the vector $\vec{G}_1$ has a vanishing projection onto $\vec{e}^{\,\mathcal{G}}_1$, the matrix $\boldsymbol{G}$ is non-singular if the vector $\vec{G}_2-\boldsymbol{\nabla}_\tau{\vec{J}}_2+\dot{\alpha}_k\vec{G}_1$ has a non-zero projection onto $\vec{e}^{\,\mathcal{G}}_1$. This projection reduces to $\boldsymbol{\nabla}_\tau\vec{J}_2\cdot(\boldsymbol{\Omega}_6\vec{D}_k)$, which equals zero (see \App{app:UnifExpGauge}). Thus, the gauge is pathological since $\det(\boldsymbol{G})=0$. Note that allowing for $\vec{G}_1$ to have a non-zero projection onto $\vec{e}^{\,\mathcal{G}}_1$ is mandatory for the matrix $\boldsymbol{G}$ to be non-singular. However, time-derivatives of the gauge degrees of freedom are not removed anymore in that case. As a consequence, all the gauges fixed by imposing $\delta N=0$ and $\vec{G}_1\cdot\dz=0$ are pathological, irrespectively of the choice of $\vec{G}_1$. This generalises our finding about the uniform-expansion gauge detailed in \App{app:UnifExpGauge}.

We stress that these considerations do not mean that non-pathological gauges leading to $\delta N=0$ do not exist, but rather imply that such gauges cannot be constructed by directly imposing $\delta N=0$. And indeed, we found a non-pathological implementation of the synchronous gauge in which $\delta N=0$. This highlights that the space of gauges with $\mathrm{Rank}(\boldsymbol{\lambda})=0$ cannot be entirely mapped to the space of gauges with $\mathrm{Rank}(\boldsymbol{\lambda})=1$ complemented with an appropriate choice of the first gauge vector. From \Eq{eq:LMfix}, the perturbed lapse function reads $\delta N=\vec{N}\cdot\dz$ where
\bea
	\vec{N}=\frac{1}{\det(\boldsymbol{G}_\mathrm{LM})}\left\{\left[\vec{G}_2\cdot\left(\boldsymbol{\Omega}_6\vec{D}_k\right)\right]\,\boldsymbol{\nabla}_\tau\vec{G}_1-\left[\vec{G}_1\cdot\left(\boldsymbol{\Omega}_6\vec{D}_k\right)\right]\,\boldsymbol{\nabla}_\tau\vec{G}_2\right\}.
\eea
Any gauge choice which leads to $\vec{N}$ given by a linear combination of the gauge vectors would thus yield vanishing perturbations of the lapse function. As is the case for gauges with vanishing perturbations of the shift vector, deriving the set of gauge vectors fulfilling the above condition is beyond the scope of this study.

\subsection{Gauges with vanishing gauge degrees of freedom}

We end this section by introducing the set of gauges for which the gauge degrees of freedom are simply set to zero. From the perspective of the \Pbasis introduced above, these are the most natural gauges to consider. They correspond to setting the gauge vectors $\vec{G}_\mu=\vec{e}^{\,\mathcal{G}}_\mu$ and $\boldsymbol{\lambda}=\boldsymbol{0}$. These gauges are manifestly non-pathological since the two gauge vectors form a complete basis of the plane of gauge degrees of freedom. Moreover, the two gauge vectors are orthogonal to the plane of physical degrees of freedom and to the plane of constraints. Hence, the right-hand side of \Eq{eq:gaugephyssystem} vanishes and one recovers that the gauge degrees of freedom are fixed to zero. We note that any set of gauge vectors that forms a complete basis of $\mathcal{G}$ will lead to the same properties. For instance, we can equivalently fix the gauge by imposing $(\boldsymbol{\Omega}_6\vec{D}_k)\cdot\dz =0$ and $(\boldsymbol{\Omega}_6\vec{S}_k)\cdot\dz=0$.

The expression of the Lagrange multipliers are derived using \Eqs{eq:LMphysdof} and \eqref{eq:LMmatN} in which $\boldsymbol{G}_{\mathrm{p}}$ is now vanishing. Using footnote \ref{foot:GLMG}, it boils down to
\bea
	\left(\begin{array}{c}
		\delta N \\
		k\delta N_1
	\end{array}\right)=\left(\begin{array}{cc}
		0 & {\mu}_S \\
		{\lambda}_D & {\lambda}_S
	\end{array}\right)\,\boldsymbol{k}_\mathrm{cp}\,\left(\begin{array}{c}
		Z_1 \\
		Z_2
	\end{array}\right),
\eea
where $\boldsymbol{k}_\mathrm{cp}$ is built from the couplings between the constraints and the physical degrees of freedom (see Apps. \ref{app:HamKinBasis}, \ref{ssec:dynPhys} and \ref{app:LagMultKin}). This formula holds irrespectively of the choice of the two gauge vectors as long as they form a complete basis of the plane of gauge degrees of freedom. 

Let us briefly comment on that gauge. Here, the two constants of motion are the two gauge degrees of freedom, hence they have vanishing Poisson bracket since they correspond to two momenta (see footnote \ref{foot:16}). It is also worth noticing that removing the gauge degrees of freedom by projecting the solutions of the equations of motion onto the physical plane (as proposed in \Sec{ssec:GTransfophys}) is equivalent to working in the gauge with vanishing gauge degrees of freedom.

\subsection{General strategy for defining gauges}
\label{sec:GeneralStrategy}

The above investigations lay the ground for developing a general strategy to construct gauges. As explained previously, gauges are usually chosen in order to adjust the foliation on a quantity of interest. For instance, in the unitary gauge, spatial hypersurfaces have uniform inflaton-field values. Hence, in full generality, one often searches for a gauge such that $\delta A=0$ where $ A$ is a quantity of interest. A general methodology to build a non-pathological gauge in which $\delta A=0$ is presented below. It depends on whether $\delta A$ contains Lagrange multipliers or not. Here, we only consider perturbations of physical quantities at linear order.

The easiest case is when $\delta A$ involves phase-space variables only, thus reading $\delta A=\vec{A}\cdot\dz$. A first possibility is that $\vec{A}$ has a vanishing projection onto $\mathcal{G}$ and $\delta A$ does not involve gauge degrees of freedom. Hence $\delta A$ is either vanishing if it is a linear combination of the constraints or non-vanishing if it bears physical degrees of freedom, but in both situations, its value cannot be changed by fixing a gauge. A second possibility is that $\vec{A}$ has a non-zero projection onto $\mathcal{G}$. Then, one simply impose $\delta A=0$ as a first gauge condition, which is easily complemented with a second condition in which the projection of the second gauge vector onto $\mathcal{G}$ is selected to be linearly independent of the one of $\vec{A}$.

More involved is the case where $\delta A$ contains Lagrange multipliers, \ie $\delta A=A^{(N)}\delta N+A^{(N_1)}\delta N_1+\vec{A}\cdot\dz$. A first method consists in imposing $\delta A=0$ and to complement it with $\vec{G}_2\cdot\dz=0$. The second gauge vector then needs to be designed to remove time-derivatives of gauge degrees of freedom and to lead to $\det(\boldsymbol{G})\neq0$. However, the example of the uniform-expansion gauge shows that this approach can fail in providing a non-pathological gauge. If it is so, one needs to resort to gauges with $\mathrm{Rank}(\boldsymbol{\lambda})=0$ and find two gauge vectors such that
\bea    
    &&\det(\boldsymbol{G})\neq0, \label{eq:detG} \\
    &&\delta A=\ds\sum_{i=1}^2\left[\alpha_i(\tau)\,\vec{G}_i\cdot\dz+\beta_i(\tau)\,\frac{\dd }{\dd\tau}\left(\vec{G}_i\cdot\dz\right)\right]. \label{eq:dAgauge}
\eea
The second requirement ensures that $\delta A$ is vanishing by imposing the gauge condition. Using the equations of motion in \Eq{eq:dAgauge}, we arrive at
\bea    
\ds\sum_{i}\beta_i\,\vec{G}_i\cdot\left(\boldsymbol{\Omega}_6\vec{S}_k\right)&=&A^{(N)}\, , \\
\ds\sum_{i}\beta_i\,\vec{G}_i\cdot\left(\boldsymbol{\Omega}_6\vec{D}_k\right)&=&A^{(N_1)}\, , \\
\ds\sum_{i=1}^2\left[\alpha_i(\tau)\,\vec{G}_i+\beta_i(\tau)\,\boldsymbol{\nabla}_\tau\vec{G}_i\right]&\approx&\vec{A}\, ,
\eea
where it is sufficient that the last equality holds on the surface of constraints.\footnote{The two first equalities are conditions in the plane of gauge degrees of freedom, hence weak equality and strong equality are identical.} The two first conditions are algebraic constraints in the plane of gauge degrees of freedom and complement \Eq{eq:detG}. The last condition is an inhomogeneous differential system and solving it demands to identify the eigenvectors of the operator $\boldsymbol{\nabla}_\tau$. This study is beyond the scope of this article but it would allow us to construct any gauge in a non-pathological way (or at least to show that gauges cancelling $\delta A$ do not exist).

\section{Gauge-invariant variables}
\label{sec:GIvar}

Gauge transformations are infinitesimal translations in the plane of gauge degrees of freedom and gauge-invariant quantities are given by the set of variables orthogonal to that plane. Hence, canonical pairs of gauge-invariant variables are defined by $Q_\mathrm{GI}=\vec{I}_Q\cdot \dz $ and $P_\mathrm{GI}=\vec{I}_P\cdot \dz $ where $\vec{I}_{Q/P}\in\mathcal{C}\otimes\mathcal{P}$. The two variables are canonically conjugated providing that $\left\{Q_\mathrm{GI},P_\mathrm{GI}\right\}=1$, which leads to 
\bea
    \vec{I}_Q\cdot(\boldsymbol{\Omega}_6\vec{I}_P)=1. \label{eq:GIbracket}
\eea
Decomposing the vectors on the plane of constraints and the physical plane as $\vec{I}_{Q/P}=\vec{I}^{\,\mathcal{C}}_{Q/P}+\vec{I}_{Q/P}^{\,\mathcal{P}}$, and inserting this decomposition into \Eq{eq:GIbracket}, one finds
\bea
	\vec{I}_{Q}^{\,\mathcal{P}}\cdot\left(\boldsymbol{\Omega}_6\vec{I}^{\,\mathcal{P}}_P\right)=1, \label{eq:GIvect}
\eea
while the vectors $\vec{I}^{\,\mathcal{C}}_{Q/P}$ remain unconstrained. The above relation is easily inverted to give
\bea
	\vec{I}^{\,\mathcal{P}}_P=-\left|\vec{I}_{Q}^{\,\mathcal{P}}\right|^{-2}\,\left(\boldsymbol{\Omega}_6\vec{I}_{Q}^{\,\mathcal{P}}\right)+\lambda_k(\tau)\,\vec{I}_{Q}^{\,\mathcal{P}},
\eea
where $\lambda_k(\tau)$ is an arbitrary function of time and scale. Thus, the set of canonical pairs of gauge-invariant variables read
\bea
	Q_\mathrm{GI}&=&\vec{I}_{Q}^{\,\mathcal{P}}\cdot \dz +\vec{I}_{Q}^{\,\mathcal{C}}\cdot \dz , \\
	P_\mathrm{GI}&=&-{\left|\vec{I}_{Q}^{\,\mathcal{P}}\right|^{-2}}\,\left(\boldsymbol{\Omega}_6\vec{I}_{Q}^{\,\mathcal{P}}\right)\cdot \dz +\lambda_k(\tau)\,\vec{I}_{Q}^{\,\mathcal{P}}\cdot \dz +\vec{I}_{P}^{\,\mathcal{C}}\cdot \dz .
\eea
We stress that \Eq{eq:GIvect} imposes the vectors $\vec{I}_{Q/P}^{\,\mathcal{P}}$ to be non-aligned, ensuring the physical degrees of freedom to be entirely captured. 

At the quantum level, defining gauge-invariant states is straightforward using the \Pbasis. To this end, we promote the variables to operators $\widehat{\dq}$ acting on quantum states that we denote $\left|\Psi\right>$. If we were to upgrade Poisson brackets to commutators without introducing Dirac brackets, then gauge-fixed quantum states would necessarily violate the constraints. Indeed, let us consider the gauge-fixed quantum states $\left|\Psi_{\mathrm{GF}}\right>$, satisfying $\widehat{G}_i(\vec{k})\left|\Psi_{\mathrm{GF}}\right>=0$ where the classical gauge conditions have been promoted to operators, \ie $G_i(\vec{k})=\vec{G}_i\cdot\dz\to\widehat{G}_i(\vec{k})=\vec{G}_i\cdot\widehat{\dz}$. The Heisenberg uncertainty principle leads to two obstructions for building such states. First, the commutator between the two gauge-fixing operators reads
\bea
    \left[\widehat{G}_1(\vec{k}),\widehat{G}_2(\vec{k})\right]=\vec{G}_1\cdot\left(\boldsymbol{\Omega}_6\vec{G}_2\right),
\eea
which does not vanish in general, since the gauge vectors mix the constraints, the gauge degrees of freedom and the physical degrees of freedom, see footnote~\ref{footnote:G1G2PoissonBracket}. This means that it is not always possible to find states for which both gauge-conditions are fulfilled, unless additional prescriptions on the vectors $\vec{G}_i$ are imposed to ensure $[\widehat{G}_1(\vec{k}),\widehat{G}_2(\vec{k})]=0$.\footnote{The example of the Newtonian gauge perfectly illustrates this point. Classically, the Newtonian gauge amounts to imposing $\delta\gamma_2(\vec{k})=0=\delta\pi_2(\vec{k})$ for all $\vec{k}$. However, at the quantum level, the gauge-fixed states would be defined as $\widehat{\delta\gamma}_2(\vec{k})\left|\Psi_\mathrm{GF}\right>=0=\widehat{\delta\pi}_2(\vec{k})\left|\Psi_\mathrm{GF}\right>$ which is impossible since $[\widehat{\delta\gamma}_2(\vec{k}),\widehat{\delta\pi}_2(\vec{k})]=i$ by construction.} Second, fixing a gauge consists in imposing conditions on the plane of gauge degrees of freedom and these conditions necessarily have a non-zero Poisson bracket with the constraints since those are canonically conjugated to the gauge degrees of freedom. At the quantum level, this means that gauge-fixing operators do not commute with constraints operators, \ie $[\widehat{G}_i(\vec{k}),\widehat{Q}_\mu(\vec{k})]\neq0$. Hence, quantum states which cancel both gauge conditions and the constraints cannot exist, and gauge-fixed quantum states necessarily have non-vanishing constraints (see also \Refc{Markkanen:2014dba}).

To avoid this, Poisson brackets can be upgrated to Dirac brackets,
\bea
\left\{\cdot , \cdot \right\}_D &=& \left\{\cdot , \cdot \right\} +  \left\{\cdot , Q_\mu \right\} \mathcal{M}^{-1} \left\{ P_\nu , \cdot \right\}  \,,
\eea
before promoting them to commutators~\cite{Boldrin:2021xrm}. The matrix $\mathcal{M}$ depends on constraints and gauge conditions as follows:
\begin{equation}
\mathcal{M} := \begin{pmatrix}
\left\{Q_\mu , Q_\nu\right\} &  \boldsymbol{G} \\
 -\boldsymbol{G} & \left\{G_i , G_j \right\} 
\end{pmatrix} \,.
\end{equation}
Therefore the introduction of Dirac brackets requires first to gauge fix the system such that $\mathcal{M}$ is well defined. Furthermore, since the constraints are first class, see \Eq{eq:FirstClass:Strongly:Vanish}, and since the gauge conditions can always be rewritten as conditions belonging to $\mathcal{P}$, the matrix is antidiagonal and $\det\left(\mathcal{M}\right) = \det\left(\boldsymbol{G}\right)^2$. Our non-pathology criterion therefore ensures that Dirac brackets are well defined. After promoting Dirac brackets to commutators, the only non-vanishing commutator is therefore $[\widehat{Z}_1(\vec{k}),\widehat{Z}_2(\vec{k})]=i$. Gauge-invariant states $\left|\Psi_\mathrm{GI}\right>$ are defined as the set of states on which the constraints vanish, \ie $\widehat{Q}_1(\vec{k})\left|\Psi_\mathrm{GI}\right>=0=\widehat{Q}_2(\vec{k})\left|\Psi_\mathrm{GI}\right>$ for all $\vec{k}$. Both constraints can be jointly realised since  $\widehat{Q}_1(\vec{k})$ commute with $\widehat{Q}_2(\vec{k}')$. Gauge transformations are given by the quantum operator generating translations in the plane of gauge degrees of freedom, which reads
\bea
	\widehat{\mathcal{L}}_{\xi^\mu}=\ds\exp\left[i\int\dd^3\vec{k}\sum_{\mu=1}^2\xi^\mu_{\vec{k}} \, \widehat{Q}_\mu(\vec{k})\right],
\eea
where the $\xi^\mu$'s are two infinitesimal parameters. The exponential is easily expanded since $[\widehat{Q}_1(\vec{k}),\widehat{Q}_2(\vec{k}')]=0$. By further using that $\widehat{Q}_\mu(\vec{k})\left|\Psi_\mathrm{GI}\right>=0$, it is straightforward to show that $\widehat{\mathcal{L}}_{\xi^\mu}\left|\Psi_\mathrm{GI}\right>=\left|\Psi_\mathrm{GI}\right>$. Hence the states $\left|\Psi_\mathrm{GI}\right>$ are indeed gauge-invariant.

\section{Conclusion}
\label{sec:Conclusion}

In this paper, we have investigated the gauge-fixing procedure in a Hamiltonian formalism for the case of (linear) cosmological perturbation theory (CPT), restricting ourselves to its scalar sector with a scalar field as matter content. Although the background was fixed to a Friedmann-Lema\^itre-Robertson-Walker (FLRW) universe for concreteness, the discussion remained general as the gauge fixing was performed on the six-dimensional phase space described by perturbed variables only. 

We have introduced a decomposition of this six-dimensional phase space into three two-dimensional planes: a) the plane of constraints $\mathcal{C}$ generated by the scalar and the momentum constraints, b) the plane of gauges $\mathcal{G}$ generated by the momenta of these constraints, which role is to generate gauge transformations and c) the physical plane $\mathcal{P}$ that contains all the remaining quantities, which are {\it de facto} physical. This splitting amounts to the so-called Kucha\v{r} decomposition \cite{Boldrin:2023jqc,doi:10.1063/1.1666050}. By construction, $\mathcal{C}$, $\mathcal{G}$ and $\mathcal{P}$ are orthogonal to each other and obey certain relations under the action of the symplectic structure $\boldsymbol{\Omega}_6$, see \Sec{ssec:gdofphysdof}. Such a geometrical approach makes straightforward \eg the extraction of gauge-invariant quantities or the application of constraints by simply projecting vectors onto the plane of interest. 

In this \Pbasis, the Hamiltonian can be obtained by performing a time-dependent canonical transformation, see \App{app:HamKinBasis}. Although the resulting expressions may be cumbersome to use in practice, they highlight that the gauge degrees of freedom in the total Hamiltonian are solely coupled to the constraints. This has to be the case in order to ensure that the constraints are conserved over time, if they are imposed initially. Equivalently, the dynamics of the gauge degrees of freedom is generated by the physical degrees of freedom and by the Lagrange multipliers only (when applying the constraints). As a consequence, once the gauge is fixed and is imposed to be conserved over time, the Lagrange multipliers are directly parametrised in terms of the physical degrees of freedom.

We have then brought our attention to the gauge-fixing procedure. Gauge prescriptions in the literature appear in two ways: a) healthy gauges in which all the gauge degrees of freedom are fixed and b) {pathological gauges} where some gauge degrees of freedom remain unfixed. When using pathological gauges, it is commonly assumed that the remaining gauge degrees of freedom can be fixed somehow, eventually leading to a healthy gauge. However, we have shown here that this is not necessarily the case.

We first presented a formal description of healthy and pathological gauges. On the one hand, a healthy gauge is a gauge that can be recast as a set of conditions on the phase-space variables only (\eg $\delta\gamma_2=\delta\pi_2=0$), that generate the entire gauge plane $\mathcal{G}$ ($\det(\boldsymbol{G})\neq 0$ in the language of \Sec{sec:LagMult}). In healthy gauges, the Lagrange multipliers are derived on-shell, by demanding that the gauge conditions are preserved over time. They are not involved in the off-shell definition of the gauge, they are on-shell consequences of that definition. 
On the other hand, when the gauge conditions directly fix the Lagrange multipliers or the derivatives of the phase-space variables (\eg $\delta N=0$ or $\dot{\delta\gamma_1}=0$ respectively), a dedicated analysis is required. The procedure presented in \Sec{sssec:RegainingAlgConst} shows how to recast these conditions into a form that makes the gauge explicitly healthy, if this is possible, see \Fig{fig:gauge}. Otherwise, the gauge is pathological. A healthy reformulation of the gauge can then be looked for using the generic method presented in \Sec{sec:GeneralStrategy}, although it is not clear whether or not such a reformulation always exists, and we leave this question for future work.

We then illustrated this methodology with some concrete examples. For instance, while the Newtonian gauge ($\delta\gamma_2 = \delta N_1=0$) may appear pathological at first sight as it fixes a Lagrange multiplier prior to solving the equations of motion, the methodology presented in this paper shows that this gauge prescription can equivalently be recast under the form $\delta\gamma_2 =\delta\pi_2=0$, which clearly fixes all the gauge degrees of freedom. The condition $\delta N_1=0$ then appears as a dynamical consequence of this gauge fixing, instead of an off-shell condition. We also studied the example of the uniform-expansion gauge in which the integrated-expansion rate vanishes, $\delta \mathcal{N}_\mathrm{int}=0$. This gauge plays a key role in the stochastic-inflation formalism, in which the modelling of quantum-backreaction effects is crucial to describe non-perturbative non-Gaussianities and the statistics or rare fluctuations, relevant to the formation of primordial black holes for instance~\cite{Pattison:2019hef, Escriva:2022duf}. Applying our method, we have investigated several ways to implement such a gauge that nonetheless all turned out to be pathological. This does not prove that a healthy gauge where $\delta \mathcal{N}_\mathrm{int}\approx 0$ does not exist, but if it does it has to lie outside the large class of gauges we have tested. Note also that, in the context of the (stochastic or classical) $\delta N$ formalism, physical quantities are computed at leading order in the gradient expansion, hence it may be enough to impose that $\delta \mathcal{N}_\mathrm{int}$ is $k^2$ suppressed. This would require to develop the present formalism for the reduced phase space of the separate-universe approach, which we plan to carry out in a future work.

Moreover, we made use of our formalism to explore new classes of gauges. For example, we proposed a generic class of non-pathological gauges that yield vanishing perturbations of the shift vector $\delta N_1=0$. This type of gauges could be of particular interest in the context of the separate-universe picture (and consequently of stochastic inflation) since this approximation scheme assumes the shift vector to be zero. We also considered the class of gauges where the gauge degrees of freedom vanish, which appears as the most natural prescription from the geometrical perspective we have been using.

The methodology developed in \Sec{ssec:RequNPgauge} is prescriptive: given two gauge conditions, it is straightforward to follow the steps summarised in \Fig{fig:gauge} and assess whether the gauge is healthy or not. Adopting a constructive approach instead, we showed that building a gauge with a desired property eventually amounts to searching for the eigenvectors of a linear differential operator. This operator is adjoint to the one generating the dynamics of cosmological perturbations, see \Sec{sec:GeneralStrategy}.

Finally, we highlighted that this formalism allows us to define gauge-invariant variables easily, since it simply amounts to projecting solutions onto the plane of physical degrees of freedom. This approach is straightforwardly extended to the construction of gauge-invariant quantum states.

To put this work in the broader context, we would like to emphasise that, although we focused on cosmological perturbations, our approach is generic and could be extended to other perturbation theories, for instance to the post-Newtonian theory \cite{Clifton:2020oqx} or to any gauge theory. In the cosmological context, our formalism can be adapted to the separate-universe picture. This would allow us to build a well-defined prescription for matching this picture with CPT at large scales in a gauge-fixed and/or a gauge-invariant manner. Another natural extension of this methodology would consist in including higher orders in CPT \cite{Malik_2004,Finelli:2006wk,Domenech:2017ems}. While we showed that gauges defined by conditions that do not contain the Lagrange multipliers are always well defined, it is yet to be proven whether this property is still satisfied at higher order or in other gauge theories. However, the vectorial formalism developed in this article relies on the linear behaviour of cosmological perturbations. A non-linear formulation would therefore request to rethink our methodology. Finally, these investigations may provide new understanding at the quantum level. We show that the non-pathology criteria established in this article ensure that Dirac brackets are naturally well defined. The methodology developed here may also inspire new ways to treat constraints and the gauge-fixing process in a Hamiltonian setting after quantising the theory. While it is expected that reduction of gauge system and quantisation are two commuting procedures, no such proof has been provided so far. Such investigations would be of great interest in particular in the context of quantum gravity \cite{Dirac:1958sq,Amorim:1994ua,Kleinert:1997em}.

\acknowledgments
It is a pleasure to thank Przemys\l aw Ma\l kiewicz and Alice Boldrin for interesting comments and discussions. D.~A is supported by JSPS Grant-in-Aid for Scientific Research No. JP23KF0247.

\appendix
\addtocontents{toc}{\protect\setcounter{tocdepth}{1}}

\section{Constraints in vector notations}
\label{app:vectornot}
The vectors generating the first-order constraints and introduced in \Eq{eq:const1vect} are given by
\bea
\vec{S}_k^{\,\mathrm{T}} &=& \Bigg(\frac{m_\phi}{\lambda^{3/2}} v V_{,\phi}\, ,\, -\frac{\sqrt{\lambda}}{m_1^2} \frac{v^{1/3}}{\sqrt{3}}\left(\frac{\pi_\phi^2}{v^2} - V + \Mp^2 \frac{k^2}{v^{2/3}} \right) \, ,\,  \frac{\sqrt{\lambda}}{m_2^2} \frac{\Mp^2}{\sqrt{6}}\frac{k^2}{v^{1/3}}\, , \label{eq:S1} \\
&\, & \frac{\lambda^{3/2}}{m_\phi}\frac{\pi_\phi}{v}\, ,\, -\frac{m_1^2}{\sqrt{\lambda}} \frac{\sqrt{3}}{\Mp^2} v^{2/3} \theta\, ,\, 0\Bigg) ,  \nonumber \\
\vec{D}_k^{\,\mathrm{T}} &=& \left(\frac{m_\phi}{\lambda^{3/2}} \pi_\phi\, ,\, \frac{\sqrt{\lambda}}{m_1^2}\frac{1}{2\sqrt{3}} v^{1/3}\theta \, ,\, -\frac{\sqrt{\lambda}}{m_2^2}\sqrt{\frac{2}{3}} v^{1/3} \theta\, ,\, 0\, ,\, -\frac{m_1^2}{\sqrt{\lambda}} \frac{2}{\sqrt{3}} v^{2/3}\, ,\, -2 \frac{m_2^2}{\sqrt{\lambda}} \sqrt{\frac{2}{3}} v^{2/3} \right) . \kern2em
\label{eq:D1}
\eea 
The matrix $\boldsymbol{H}_k$ appearing in the quadratic constraint~\eqref{eq:quad:constraint:def:Hk} can be decomposed into $3\times3$ blocks in the configuration/momentum decomposition as follows:
\bea
	\boldsymbol{H}_k=\left(\begin{array}{cc}
		\boldsymbol{H}_{\phi\phi}(k) & \boldsymbol{H}_{\phi\pi} \\
		\boldsymbol{H}^{\,\mathrm{T}}_{\phi\pi} & \boldsymbol{H}_{\pi\pi}
	\end{array}\right),
\eea
where only the configuration block is scale-dependent (note that $\boldsymbol{H}_k$ is symmetric hence the two off-diagonal blocks are the transpose of each-other.) Each block can be read off from \Eq{eq:S2}. The two diagonal blocks are symmetric and they are given by
\bea
	\boldsymbol{H}_{\phi\phi}(k)=\left(\begin{array}{ccc}
		\frac{m_\phi^2}{\lambda^3} \frac{v}{2}\left(\frac{k^2}{v^{2/3}}+V_{,\phi,\phi}\right) & \frac{m_\phi}{\lambda m_1^2} \frac{\sqrt{3}}{4}v^{1/3}V_{,\phi} & 0 \\
		\frac{m_\phi}{\lambda m_1^2} \frac{\sqrt{3}}{4}v^{1/3}V_{,\phi} & \frac{\lambda}{m_1^4} \frac{1}{3v^{1/3}}\left(\frac{\pi^2_\phi}{v^2}+\frac{V}{2}-\frac{\Mp^2}{4}\frac{k^2}{v^{2/3}}\right) & \frac{\lambda}{m_1^2 m_2^2} \frac{\sqrt{2}}{24v}\Mp^2k^2 \\
		0 & \frac{\lambda}{m_1^2 m_2^2} \frac{\sqrt{2}}{24v}\Mp^2k^2 &  \frac{\lambda}{m_2^4} \frac{1}{3v^{1/3}}\left(\frac{\pi^2_\phi}{v^2}+\frac{V}{2}-\frac{\Mp^2}{8}\frac{k^2}{v^{2/3}}\right)
	\end{array}\right), \nonumber \\
	\label{eq:hphiphi}
\eea
and
\bea
	\boldsymbol{H}_{\pi\pi}=\left(\begin{array}{ccc}
		\frac{\lambda^3}{m_\phi^2}\frac{1}{2v} & 0 & 0 \\
		0 & -\frac{m_1^4}{\lambda} \frac{v^{1/3}}{\Mp^2} & 0 \\
		0 & 0 & \frac{m_2^4}{\lambda} \frac{2v^{1/3}}{\Mp^2}
	\end{array}\right). \label{eq:hpipi}
\eea
The off-diagonal block contains the couplings between configuration and momentum variables. It is given by 
\bea
	\boldsymbol{H}_{\phi\pi} =\left(\begin{array}{ccc}
		0 & 0 & 0 \\
		-\frac{\lambda^2}{m_\phi m_1^2}\frac{\sqrt{3}}{4}\frac{\pi_\phi}{v^{5/3}} & -\frac{\theta}{4\Mp^2} & 0 \\
		0 & 0 & \frac{\theta}{2\Mp^2}
	\end{array}\right). \label{eq:hphipi}
\eea

\section{Equations of motion for the perturbations}
\label{app:eqnsPert}
Vector notations allow us to write the dynamics of perturbations in a compact way. It should however not hide its complexity. We thus here provide the explicit expressions of the set of eight equations (2 constraints plus 6 Hamilton equations) compactly encoded in \Eqs{eq:const1vect} and~\eqref{eq:dynvect}. The linear constraint equations read
\bea
	-\frac{\sqrt{3}}{\Mp^2}v^{2/3}\theta\,\delta\pi_1-\frac{v^{1/3}}{\sqrt{3}}\left[\frac{\pi_\phi^2}{v^2}-V+{\Mp^2}\left(\frac{k^2}{v^{2/3}}\right)\right]\,\delta\gamma_1+\frac{\Mp^2}{\sqrt{6}}\frac{k^2}{v^{1/3}}\,\delta\gamma_2+\frac{\pi_\phi}{v}\,\delta\pi_\phi+vV_{,\phi}\delta\phi=0 \nonumber \\ \label{eq:scalconst=0}
\eea
and
\bea
	&&\pi_\phi\,\delta\phi+\frac{1}{\sqrt{3}}v^{1/3}\theta\,\left(\frac{1}{2}\delta\gamma_1-\sqrt{2}\delta\gamma_2\right)-\frac{2}{\sqrt{3}}v^{2/3}\,\left(\delta\pi_1+\sqrt{2}\delta\pi_2\right)=0\, . \label{eq:diffconst=0}
\eea
The Hamilton equations for the scalar-field perturbations are
\bea
	\dot{\delta\phi}&=&\frac{\pi_\phi}{v}\delta N+N\left(\frac{1}{v}\delta\pi_\phi-\frac{\sqrt{3}}{2}\frac{\pi_\phi}{v^{5/3}}\delta\gamma_1\right), \label{eq:dotPhiGen} \\
	\dot{\delta\pi}_\phi&=&-vV_{,\phi}\delta N-\pi_\phi k\delta N_1-N\left[v\left(\frac{k^2}{v^{2/3}}+V_{,\phi,\phi}\right)\delta\phi+\frac{\sqrt{3}}{2}v^{1/3}V_{,\phi}\delta\gamma_1\right]. \label{eq:dotPiphiGen}
\eea
For the isotropic gravitational perturbations, they are given by
\bea
	\dot{\delta\gamma}_1&=&-\frac{\sqrt{3}}{\Mp^2}v^{2/3}\theta\delta N-\frac{2}{\sqrt{3}}v^{2/3} k\delta N_1-\frac{N}{\Mp^2}\left(2v^{1/3}\delta\pi_1+\frac{\theta}{2}\delta\gamma_1\right), \label{eq:dotGamma1Gen} \\
	\dot{\delta\pi}_1&=&\frac{v^{1/3}}{\sqrt{3}}\left(\frac{\pi_\phi^2}{v^2}-V+{\Mp^2} \frac{k^2}{v^{2/3}}\right)\delta N-\frac{1}{2\sqrt{3}}v^{1/3}\theta k\delta N_1  \label{eq:dotPi1Gen} \\
	&&+N\left[-\frac{2}{3v^{1/3}}\left(\frac{\pi^2_\phi}{v^2}+\frac{V}{2}-\frac{\Mp^2}{4}\frac{k^2}{v^{2/3}}\right)\delta\gamma_1+\frac{\theta}{2\Mp^2}\delta\pi_1-\frac{\sqrt{2}}{12v} \Mp^2 k^2 \delta\gamma_2\right] \nonumber \\
	&&+N\frac{\sqrt{3}}{2}v^{1/3}\left(\frac{\pi_\phi}{v^2}\delta\pi_\phi-V_{,\phi}\delta\phi\right). \nonumber
\eea
Finally, for the anisotropic degrees of freedom, the Hamilton equations are
\bea
	\dot{\delta\gamma}_2&=&-2\sqrt{\frac{2}{3}}v^{2/3} k\delta N_1+\frac{N}{\Mp^2}\left(4v^{1/3}\delta\pi_2+\theta\delta\gamma_2\right), \label{eq:dotGamma2Gen} \\
	\dot{\delta\pi}_2&=&-\frac{\Mp^2}{\sqrt{6}}\frac{k^2}{v^{1/3}}\delta N+\sqrt{\frac{2}{3}}v^{1/3}\theta k\delta N_1 \label{eq:dotPi2Gen} \\
	&&+N\left[-\frac{2}{3v^{1/3}}\left(\frac{\pi^2_\phi}{v^2}+\frac{V}{2}-\frac{\Mp^2}{8}\frac{k^2}{v^{2/3}}\right)\delta\gamma_2-\frac{\theta}{\Mp^2}\delta\pi_2-\frac{\Mp^2}{6\sqrt{2}}\frac{k^2}{v}\delta\gamma_1\right]. \nonumber
\eea

\section{Gauge transformation of Hamilton equations}
\label{app:GTHam}

In \Eq{eq:deltaz} we have treated $\delta N$ and $\delta N_1$ as Lagrange multipliers, and as such they do not appear in the reduced phase space considered in this work.  Alternatively, the lapse $N$ and shift $N^i$ can also be seen as belonging to the phase space. However, since the Hamiltonian~\eqref{HamGR} does not involve their time derivative, their momenta identically vanish. Following Dirac's method, one can introduce Lagrange multipliers and primary constraints in the action before performing the Legendre transformation. One can then compute the equations of motion for all the fields (including $N$ and $N^i$) and impose the primary constraints to be conserved over time. This directly leads to the appearance of the secondary constraints $\mathcal{S}^{(G)}+\mathcal{S}^{(\phi)}=0$ and $\mathcal{D}_i^{(G)} + \mathcal{D}_i^{(\phi)}=0$. Since secondary constraints are automatically conserved, one can consider them as primary constraints from the beginning with $N$ and $N^i$ their associated Lagrange multipliers. In this paper, we therefore follow this point of view and $N$ and $N^i$ are not treated as phase-space variables \cite{thieman_book,Pons:1998ht}.

Under this setting, time-evolution and gauge transformations do not commute. The reason is that both are representations of the algebra of hypersurface deformation, \ie time-evolution of any phase-space function $F$ and gauge transformation of that function are both generated by the scalar and diffeomorphism constraints which form an algebroid \cite{Hojman:1976vp}. Because gauge transformation and time-evolution are generated by the same set of non-commuting constraints, this leads to the fact that the gauge transformation of the time-derivative of ${F}$ is not given by the time-derivative of the gauge transformation of $F$. This was shown for the specific case of cosmological perturbations in \Refc{Bojowald:2008jv}, and here we prove this result in full generality. We then apply it to cosmological perturbations, where we show that this non-commutativity is related to the time evolution of the constraint vectors.

Let $\Gamma\equiv(\phi_\mathrm{m},\gamma_{ij};\pi_\mathrm{m},\pi^{ij})$ denote any point in the phase space where $(\phi_\mathrm{m},\pi_\mathrm{m})$ describes any matter content (not necessarily a scalar field). The full Hamiltonian, generating both evolution and gauge-transformations, is 
\bea
	C[N,N^i]=\ds\int\dd^3x\left[N\,\mathcal{S}(\Gamma)+N^i\,\mathcal{D}_i(\Gamma)\right],
\eea
where $\mathcal{S}(\Gamma)$ and $\mathcal{D}(\Gamma)$ are the scalar and diffeomorphism constraints for gravity and matter. Their expressions are not needed here. We denote the ``smeared constraints'' by $S[\Lambda]\equiv\int\dd^3x\Lambda \,\mathcal{S}(\Gamma)$ and $D[V^i]\equiv\int\dd^3xV^i\,\mathcal{D}_i(\Gamma)$, with $\Lambda$ and $V^i$ any scalar and vector space-time functions. The smeared constraints obey the following algebra of constraints:
\bea
\label{eq:constraint:smeared:albegra:start}
	\left\{S[\Lambda_1],S[\Lambda_2]\right\}&=&D\left[\gamma^{ij}(\Lambda_1\partial_j\Lambda_2-\Lambda_2\partial_j\Lambda_1)\right], \\
	\left\{D[V^i_1],D[V^j_2]\right\}&=&D\left[V^j_1\partial_jV^i_2-V^j_2\partial_jV^i_1\right],\\
	\left\{D[V^i],S[\Lambda]\right\}&=&S\left[V^i\partial_i\Lambda\right],
\label{eq:constraint:smeared:albegra:end}
\eea
 which is no more than the representation of the algebra of hypersurface deformation (see \eg \Refs{Hojman:1976vp,thieman_book}).\footnote{Note that in \Refc{thieman_book}, the Poisson bracket is defined as
 \bea
 \left\{F,G\right\}=\Mp^{-2}\ds\int\dd^3x\left(\frac{\delta F}{\delta \pi^{ij}}\frac{\delta G}{\delta \gamma_{ij}}-\frac{\delta G}{\delta \pi^{ij}}\frac{\delta F}{\delta \gamma_{ij}}\right),
\eea
while here we define it as 
\bea
 \left\{F,G\right\}=\ds\int\dd^3x\left(\frac{\delta F}{\delta \gamma_{ij}}\frac{\delta G}{\delta \pi^{ij}}-\frac{\delta G}{\delta \gamma_{ij}}\frac{\delta F}{\delta \pi^{ij}}\right).
\eea
This explain the slight difference in the algebra of constraints as compared to Eq. (1.2.15) of \Refc{thieman_book}.} Time-evolution of any function defined on the phase space with no explicit time dependence is given by its Poisson bracket with the full Hamiltonian, \ie
\bea
	\dot{F}\left(\Gamma\right)=\left\{F\left(\Gamma\right),S[N]+D[N^i]\right\}.
\eea
Consider now a gauge transformation $x^\mu\to x^\mu+\xi^\mu$. In the Hamiltonian framework it is generated by\footnote{Note that in the Hamiltonian framework $\xi^\mu$ has to be projected onto the vector orthogonal to the hypersurface, \ie $n^\mu=(1/N,-N^i/N)$, leading to $\Lambda_{\xi^\mu}=n_\mu\xi^\mu=N\xi^0$, and on the plane tangential to the hypersurface using the projector $\perp^\nu_\mu=\delta^\nu_\mu+n^\nu n_\mu$, leading to $V^i_{\xi^\mu}=\xi^i$.\label{footnote:projection}}
\bea
\label{eq:gauge:transform:generic}
	\mathcal{L}_{\xi^0,\xi}(F)= \left\{F,S[N\xi^0]+D[\xi^i]\right\}.
\eea
We now consider the gauge transformation of $\dot{F}$ as being defined by the Poisson bracket
\bea
	\mathcal{L}_{\xi^0,\xi}(\dot{F})=\Big\{\big\{F,S[N]+D[N^i]\big\},S[N\xi^0]+D[\xi^i]\Big\}.
\eea
We note that $\dot{F}$ is not strictly speaking a function on the phase space, since it may depend on time explicitly, through the lapse and shift functions, hence the interpretation of the above Poisson bracket as a Lie derivative should be taken with a grain of salt. Nevertheless, we shall keep using the notation $\mathcal{L}_{\xi^0,\xi}(\dot{F})$ for the above Poisson bracket. Using the Jacobi identities, it can be written as
\bea
	\mathcal{L}_{\xi^0,\xi}(\dot{F})&=&\Big\{F,\big\{S[N]+D[N^i],S[N\xi^0]+D[\xi^i]\big\}\Big\} \nonumber \\ 
	&&+\Big\{\big\{F,S[N\xi^0]+D[\xi^i]\big\},S[N]+D[N^i]\Big\}.
\label{eq:gaugeTransform:Fdot:interm1}
\eea
The first Poisson bracket can be simplified using the algebra of constraints~\eqref{eq:constraint:smeared:albegra:start}-\eqref{eq:constraint:smeared:albegra:end}. It boils down to
\bea
	\left\{S[N]+D[N^i],S[N\xi^0]+D[\xi^i]\right\}&=&S\left[N^i\partial_i(N\xi^0)-\xi^i\partial_iN\right] \nonumber \\ 
	&&+D\left[N^j\partial_j\xi^i-\xi^j\partial_jN^i+N^2\gamma^{ij}\partial_j\xi^0\right].
\eea
The first term in \Eq{eq:gaugeTransform:Fdot:interm1} thus reads as the gauge transformation of $F$ along the vector field $\widetilde{\xi}^\mu$, \ie $\big\{F,\big\{S[N]+D[N^i],S[N\xi^0]+D[\xi^i]\big\}\big\}=\mathcal{L}_{\widetilde{\xi}^0,\widetilde{\xi}^i}(F)$, with $N\widetilde{\xi}^{\,0}=N^i\partial_i(N\xi^0)-\xi^i\partial_iN$ and $\widetilde{\xi}^{\,i}=N^j\partial_j\xi^i-\xi^j\partial_jN^i+N^2\gamma^{ij}\partial_j\xi^0$, see \Eq{eq:gauge:transform:generic}. The second term in \Eq{eq:gaugeTransform:Fdot:interm1} is the Poisson bracket of the Lie derivative of $F$ with the total Hamiltonian, \ie $\{\mathcal{L}_{\xi^0,\xi}({F}),S[N]+D[N^i]\}$. It should however not be interpreted as the total time-derivative of $\mathcal{L}_{\xi^0,\xi}({F})$. Indeed, the two gauge-parameters are time-dependent in full-generality but they are not functions of the phase-space variables. Hence their time evolution is totally arbitrary, \ie it is not generated by taking their Poisson bracket with the Hamiltonian, and $\mathcal{L}_{\xi^0,\xi}(F)$ is also explicitly time-dependent. This Poisson bracket can still be related to $\frac{\dd}{\dd\tau}\mathcal{L}_{\xi^0,\xi}({F})$ by noting that
\bea
	\frac{\dd}{\dd\tau}\mathcal{L}_{\xi^0,\xi}({F})=\frac{\partial}{\partial\tau}\mathcal{L}_{\xi^0,\xi}({F})+\left\{\mathcal{L}_{\xi^0,\xi}({F}),S[N]+D[N^i]\right\} ,
\eea
where
\bea
	\frac{\partial}{\partial\tau}\mathcal{L}_{\xi^0,\xi}({F})=\left\{F,S\left[\frac{\partial }{\partial\tau}\left(\xi^0 N\right)\right]+D\left[\frac{\partial}{\partial\tau}\left(\xi^i\right)\right]\right\}=\mathcal{L}_{\frac{1}{N}\frac{\partial (N\xi^0)}{\partial\tau},\frac{\partial \xi}{\partial\tau}}\left(F\right).
\eea

Combining the above results, the commutator between gauge transformations and total time-derivatives is thus given by one single Lie derivative, \ie
\bea
	\mathcal{L}_{\xi^0,\xi}\left(\frac{\dd F}{\dd\tau}\right)-\frac{\dd}{\dd\tau}\mathcal{L}_{\xi^0,\xi}({F})=\mathcal{L}_{\zeta^0,\zeta}(F), \label{eq:GTtimediff}
\eea
where the vector field $(\zeta^0,\zeta^i)$ is
\bea
\label{eq:zeta0:def}
	\zeta^0&=& \widetilde{\xi}^{\,0} - \frac{1}{N}\frac{\partial N\xi^0}{\partial\tau} = \frac{1}{N}\left[N^i\partial_i(N\xi^0)-\xi^i\partial_iN\right]-\frac{1}{N}\frac{\partial N\xi^0}{\partial\tau}, \\
	\zeta^i&=&  \widetilde{\xi}^{\,i} - \frac{\partial\xi^i}{\partial\tau} =  N^j\partial_j\xi^i-\xi^j\partial_jN^i+N^2\gamma^{ij}\partial_j\xi^0-\frac{\partial\xi^i}{\partial\tau}. 
\label{eq:zetai:def}
\eea
We stress that the above holds under the condition that $F$ is a function of the phase-space variables only, and does not have any explicit time dependence. Otherwise, the total time-derivatives should be replaced with Poisson brackets in \Eq{eq:GTtimediff}. This concludes the proof that, in general, gauge transformation and time differentiation do not commute.

We now apply the above result to the case of cosmological perturbations, \ie we replace $F$ by $ \dz $ in \Eq{eq:GTtimediff},
\bea
\label{eq:GTtimediff:CPT}
\mathcal{L}_{\xi^0,\xi}\left( \dzdot\right)-\frac{\dd}{\dd\tau}\mathcal{L}_{\xi^0,\xi}( \dz )=\mathcal{L}_{\zeta^0,\zeta}( \dz )\, .
\eea
Let us evaluate each of the terms appearing in that expression separately. We first recall that, within each Fourier subspace,
\bea	
	\mathcal{L}_{\xi^0,\xi}\left(F\right)=\left\{F,N\xi^0_{\vec{k}}\mathcal{S}^{(1)}_{\vec{k}}+i \xi^i k_i \mathcal{D}^{(1)}_{\vec{k}}\right\} ,
\eea
see the discussion around \Eq{eq:Lie:def}. When specified to $F= \dz $, this reduces to
\bea	
\label{eq:Lie:deltaz:app}
	\mathcal{L}_{\xi^0,\xi}( \dz )=N\xi^0_{\vec{k}}\boldsymbol{\Omega}_6\vec{S}_k+k\xi_{\vec{k}}\boldsymbol{\Omega}_6\vec{D}_k\, ,
\eea
see \Eq{eq:Lie:deltaz}. This allows us to evaluate the second term in \Eq{eq:GTtimediff:CPT}. For the first term, we can make use of the equations of motion~\eqref{eq:dynvect} to replace $ \dzdot$ by $\delta N(\vec{k})\boldsymbol{\Omega}_6\vec{S}_k+k\delta N_1(\vec{k})\boldsymbol{\Omega}_6\vec{D}_k+2N\boldsymbol{\Omega}_6\boldsymbol{H}_k \dz $. Its Poisson bracket with the constraints at first order is easily obtained using \Eq{eq:Lie:deltaz:app} and yields\footnote{If $\delta N$ and $\delta N_1$ were to be considered as phase-space variables, one would have to take into account their gauge transformation in the Lie derivative, leading to
\bea
	\mathcal{L}_{\xi^0,\xi}\left( \dzdot\right)&=&2N\boldsymbol{\Omega}_6\boldsymbol{H}_k\left(N\xi^0_{\vec{k}}\boldsymbol{\Omega}_6\vec{S}_k+k\xi_{\vec{k}}\boldsymbol{\Omega}_6\vec{D}_k\right) \\
	&&+ \left(\widetilde{\delta N}-\delta N\right)\boldsymbol{\Omega}_6\vec{S}_k + k \left(\widetilde{\delta N_1} - \delta N_1\right)\boldsymbol{\Omega}_6\vec{D}_k\, , \nonumber
\eea
which matches the equation above \eqref{eq:tildeN}. So time derivative and gauge transformation commute if one works in the extended phase space which comprises $\delta N$ and $\delta N_1$ \cite{Pons:1995su,PhysRevD.55.658,Pons:1998ht}.}
\bea
	\mathcal{L}_{\xi^0,\xi}\left( \dzdot\right)=2N\boldsymbol{\Omega}_6\boldsymbol{H}_k\left(N\xi^0_{\vec{k}}\boldsymbol{\Omega}_6\vec{S}_k+k\xi_{\vec{k}}\boldsymbol{\Omega}_6\vec{D}_k\right).
\eea
Finally, the third term in \Eq{eq:GTtimediff:CPT} follows from \Eq{eq:Lie:deltaz:app} where $\xi$ is replaced with $\zeta$, leading to
\bea
\mathcal{L}_{\xi^0,\xi}( \dz )&=&N\zeta^0_{\vec{k}}\boldsymbol{\Omega}_6\vec{S}_k+k\zeta_{\vec{k}}\boldsymbol{\Omega}_6\vec{D}_k\\
&=& \left[N\widetilde{\xi}^0_{\vec{k}}-\frac{\partial}{\partial\tau}\left(N\xi^0_{\vec{k}}\right)\right]\boldsymbol{\Omega}_6\vec{S}_k+k \left(\widetilde{\xi}_{\vec{k}}-\frac{\partial\xi_{\vec{k}}}{\partial\tau}\right)\boldsymbol{\Omega}_6\vec{D}_k
\eea
where we have used \Eqs{eq:zeta0:def}-\eqref{eq:zetai:def}. The above expression needs to be evaluated at leading order in CPT. Upon replacing $N$ by $N(\tau)+\delta N(\vec{k},\tau)$ and $N^i$ by $\delta N^i(\vec{k},\tau)$, one finds that $\widetilde{\xi}^0$ is quadratic in perturbations and can thus be discarded, while $\widetilde{\xi}= N2 k v^{-2/3}\xi^0-\dot{\xi}$ at leading order. Combining the above results, \Eq{eq:GTtimediff:CPT} boils down to
\bea
	\xi^0_{\vec{k}} N \boldsymbol{\Omega}_6 \left(\frac{\dd \vec{S}_k}{\dd\tau}-2N\boldsymbol{H}_k\boldsymbol{\Omega}_6\vec{S}_k+N\frac{k^2}{v^{2/3}}\vec{D}_k\right)+\xi_{\vec{k}} k \boldsymbol{\Omega}_6 \left(\frac{\dd \vec{D}_k}{\dd\tau}-2N\boldsymbol{H}_k\boldsymbol{\Omega}_6\vec{D}_k\right)=0.
\eea
This must hold for all gauge parameters $\xi^0_{\vec{k}}$ and $\xi_{\vec{k}}$, so 
\bea
	\frac{\dd \vec{S}_k}{\dd\tau}&=&2N\boldsymbol{H}_k\boldsymbol{\Omega}_6\vec{S}_k-N\frac{k^2}{v^{2/3}}\vec{D}_k\, , \\
	\frac{\dd \vec{D}_k}{\dd\tau}&=&2N\boldsymbol{H}_k\boldsymbol{\Omega}_6\vec{D}_k\, .
\eea
These two identities are consistent with those obtained in \Eqs{eq:dotS}-\eqref{eq:dotD}, from the explicit expression of $\vec{S}_k$, $\vec{D}_k$, and $\boldsymbol{H}_k$ given in \App{app:vectornot}. Here we see that they are related to the non-commutativity between gauge transformations and time differentiation.

\section{Hamiltonian in the \Pbasis} 
\label{app:HamKinBasis}

In this appendix, we derive the components of the Hamiltonian in the \Pbasis, denoted $\left[\boldsymbol{K}_k\right]_{a,b}$. The final results are gathered at the end of this appendix. For simplicity, we take all the mass fiducial parameters to be equal to the Planck mass $m_1=m_2=m_\phi=\Mp$, but we keep the conformal parameter $\lambda$ unspecified.

\subsection*{Useful properties}
We write here again \Eq{eq:Kab},
\bea
	\left[\boldsymbol{K}_k\right]_{a,b}=2N\vec{e}_a\cdot\left(\boldsymbol{H}_k\vec{e}_b\right)+\vec{e}_a\cdot\left(\boldsymbol{\Omega}_6\dot{\vec{e}}_b\right) =\vec{e}_a\cdot\left[\boldsymbol{\nabla}_\tau\left(\boldsymbol{\Omega}_6\vec{e}_b\right)\right]\,,
 \label{eq:Kab-app}
\eea
and recall that the Hamiltonian is symmetric $\left[\boldsymbol{K}_k\right]_{a,b}=\left[\boldsymbol{K}_k\right]_{b,a}$.
We give some additional properties that may be useful in the following. For any vectors $\vec{X}$, $\vec{Y}$, one has
\bea
\vec{X} \cdot  \left[\boldsymbol{H}_k\, , \, \boldsymbol{\Omega}_6 \right] \vec{Y} &=& \vec{Y} \cdot  \left[\boldsymbol{H}_k\, , \, \boldsymbol{\Omega}_6 \right] \vec{X} \,,\\
\vec{X} \cdot  \left[\boldsymbol{H}_k\, , \, \boldsymbol{\Omega}_6 \right] \vec{Y} &=& -\boldsymbol{\Omega}_6 \vec{X} \cdot  \left[\boldsymbol{H}_k\, , \, \boldsymbol{\Omega}_6 \right] \boldsymbol{\Omega}_6 \vec{Y} \,, \\
\vec{X} \cdot  \left[\boldsymbol{H}_k\, , \, \boldsymbol{\Omega}_6 \right] \boldsymbol{\Omega}_6 \vec{Y} &=& \vec{Y} \cdot  \left[\boldsymbol{H}_k\, , \, \boldsymbol{\Omega}_6 \right] \boldsymbol{\Omega}_6 \vec{X} \,.
\eea
From \Eq{eq:Kab-app}, one can see that 
\bea
	\left[\boldsymbol{K}_k\right]_{a,0}&=& 2 N \vec{e}_a \cdot \left( \left[\boldsymbol{H}_k\, , \, \boldsymbol{\Omega}_6 \right]  \vec{e}^{\,\mathcal{G}}_1 \right) - \frac{\dot{\lambda}_D}{\lambda_D}\,\underbrace{\vec{e}_a\cdot\vec{e}^{\,\mathcal{G}}_1}_{=\delta_{a,3}}\,, \\
	\left[\boldsymbol{K}_k\right]_{a,1}&=& 2 N \vec{e}_a \cdot \left( \left[\boldsymbol{H}_k\, , \, \boldsymbol{\Omega}_6 \right]  \vec{e}^{\,\mathcal{G}}_2 \right) - \frac{\dot{\mu}_S}{\mu_S}\,\underbrace{\vec{e}_a\cdot\vec{e}^{\,\mathcal{G}}_2}_{=\delta_{a,4}} \nonumber \\
 && -\left[\frac{\dot{\lambda}_S}{\lambda_D}-\frac{\dot{\mu}_S}{\mu_S}\frac{\lambda_S}{\lambda_D}-N\frac{\mu_S}{\lambda_D}\left(\frac{k}{v^{1/3}}\right)^2\right]\,\underbrace{\vec{e}_a\cdot\vec{e}^{\,\mathcal{G}}_1}_{=\delta_{a,3}}\,. 
\eea
Because of the results already obtained for $\left[\boldsymbol{K}_k\right]_{a,3}$ and $\left[\boldsymbol{K}_k\right]_{a,4}$ in \Eqs{eq:Ka3} and~\eqref{eq:Ka4}, and using the fact that the Hamiltonian is symmetric, one can infer the following relations
\bea
 N \vec{e}^{\,\mathcal{G}}_1 \cdot \left( \left[\boldsymbol{H}_k\, , \, \boldsymbol{\Omega}_6 \right]  \vec{e}^{\,\mathcal{G}}_1 \right) &=&  \frac{\dot{\lambda}_D}{\lambda_D}\,, \\
  N \vec{e}^{\,\mathcal{G}}_2 \cdot \left( \left[\boldsymbol{H}_k\, , \, \boldsymbol{\Omega}_6 \right]  \vec{e}^{\,\mathcal{G}}_2 \right) &=&  \frac{\dot{\mu}_S}{\mu_S}\,, \\
  2N \vec{e}^{\,\mathcal{G}}_2 \cdot \left( \left[\boldsymbol{H}_k\, , \, \boldsymbol{\Omega}_6 \right]  \vec{e}^{\,\mathcal{G}}_1 \right) &=& 2N \vec{e}^{\,\mathcal{G}}_1 \cdot \left( \left[\boldsymbol{H}_k\, , \, \boldsymbol{\Omega}_6 \right]  \vec{e}^{\,\mathcal{G}}_2 \right) \nonumber\\
  &=& \left[\frac{\dot{\lambda}_S}{\lambda_D}-\frac{\dot{\mu}_S}{\mu_S}\frac{\lambda_S}{\lambda_D}-N\frac{\mu_S}{\lambda_D}\left(\frac{k}{v^{1/3}}\right)^2\right] \,.
\eea
Using now the fact that $\left(\boldsymbol{\Omega}_6\right)^2 = - \boldsymbol{I}_6$, the above relations lead to
\bea
 N \vec{e}^{\,\mathcal{C}}_1 \cdot \left( \left[\boldsymbol{H}_k\, , \, \boldsymbol{\Omega}_6 \right]  \vec{e}^{\,\mathcal{C}}_1 \right) &=& - \frac{\dot{\lambda}_D}{\lambda_D}\,, \\
  N \vec{e}^{\,\mathcal{C}}_2 \cdot \left( \left[\boldsymbol{H}_k\, , \, \boldsymbol{\Omega}_6 \right]  \vec{e}^{\,\mathcal{C}}_2 \right) &=& - \frac{\dot{\mu}_S}{\mu_S}\,, \\
  2N \vec{e}^{\,\mathcal{C}}_2 \cdot \left( \left[\boldsymbol{H}_k\, , \, \boldsymbol{\Omega}_6 \right]  \vec{e}^{\,\mathcal{C}}_1 \right) &=& 2N \vec{e}^{\,\mathcal{C}}_1 \cdot \left( \left[\boldsymbol{H}_k\, , \, \boldsymbol{\Omega}_6 \right]  \vec{e}^{\,\mathcal{C}}_2 \right)  \nonumber \\
  &=& - \left[\frac{\dot{\lambda}_S}{\lambda_D}-\frac{\dot{\mu}_S}{\mu_S}\frac{\lambda_S}{\lambda_D}-N\frac{\mu_S}{\lambda_D}\left(\frac{k}{v^{1/3}}\right)^2\right]\,.
\eea
One finally notes that, since $\left[\boldsymbol{K}_k\right]_{1,0}=\left[\boldsymbol{K}_k\right]_{0,1}$, one has
\bea
\vec{e}^{\, \mathcal{C}}_2 \cdot \left( \left[\boldsymbol{H}_k\, , \, \boldsymbol{\Omega}_6 \right]  \vec{e}^{\,\mathcal{G}}_1 \right) &=& \vec{e}^{\, \mathcal{C}}_1 \cdot \left( \left[\boldsymbol{H}_k\, , \, \boldsymbol{\Omega}_6 \right]  \vec{e}^{\,\mathcal{G}}_2 \right) .
\eea

\subsection*{Couplings with the gauge degrees of freedom}
The couplings with gauge degrees of freedom were computed in \Eqs{eq:Ka3} and~\eqref{eq:Ka4}.
In particular, the only non-zero components are the terms $\left[\boldsymbol{K}_k\right]_{0,3}$, $\left[\boldsymbol{K}_k\right]_{0,4}$ and $\left[\boldsymbol{K}_k\right]_{1,4}$. These represent couplings between $(Q_1;P_1)$, $(Q_1;P_2)$ and $(Q_2;P_2)$ respectively. Gauge degrees of freedom are therefore only coupled to the constraints. This was to be expected, as it implies that only the constraints source their own evolution which is equivalent to saying that the constraints are preserved on shell. In \App{ssec:dynPhys}, these couplings are gathered in the $2\times 2$ matrix $\boldsymbol{k}_{\mathrm{cg}}$.

\subsection*{Couplings with the constraints}
We now consider the couplings with the constraints, $\left[\boldsymbol{K}_k\right]_{a,0}$ and $\left[\boldsymbol{K}_k\right]_{a,1}$. We have already established how constraints and gauge degrees of freedom are coupled.

Let us start by computing the coupling between the constraints and themselves. For $\left[\boldsymbol{K}_k\right]_{0,0}$, we have
\bea
\left[\boldsymbol{K}_k\right]_{0,0}= -2 N \lambda_D^2 \vec{D}_k \cdot \left( \left[\boldsymbol{H}_k\, , \, \boldsymbol{\Omega}_6 \right]  \boldsymbol{\Omega}_6 \vec{D}_k \right)  .
\eea
Then, using that $\vec{e}_2^{\,\mathcal{C}}=\lambda_S \vec{D}_k+\mu_S \vec{S}_k$, one can rewrite $\left[\boldsymbol{K}_k\right]_{0,1}$ as
\bea
\left[\boldsymbol{K}_k\right]_{0,1} = \frac{\lambda_S}{\lambda_D} \left[\boldsymbol{K}_k\right]_{0,0} - 2 N \mu_S \lambda_D \, \vec{S}_k\cdot \left[\boldsymbol{H}_k\, , \, \boldsymbol{\Omega}_6 \right] \boldsymbol{\Omega}_6 \vec{D}_k \,,
\eea
and $\left[\boldsymbol{K}_k\right]_{1,1}$ as
\bea
\left[\boldsymbol{K}_k\right]_{1,1} = 2 \frac{\lambda_S}{\lambda_D} \left[\boldsymbol{K}_k\right]_{0,1} - \frac{\lambda_S^2}{\lambda_D^2} \left[\boldsymbol{K}_k\right]_{0,0} - 2 N \mu_S^2 \, \vec{S}_k\cdot \left[\boldsymbol{H}_k\, , \, \boldsymbol{\Omega}_6 \right] \boldsymbol{\Omega}_6 \vec{S}_k \,.
\eea
In \App{ssec:dynPhys}, these three couplings are gathered in the matrix denoted $\boldsymbol{k}_{\mathrm{cc}}$.

We then move on with the couplings between the constraints and the physical degrees of freedom. We recall that $\vec{e}^{\,\mathcal{P}}_1$ is given by the normalised projection of $\vec{B}$ onto the physical plane $\mathcal{P}$, \ie
\bea
\label{eq:e1P:def}
\vec{e}^{\,\mathcal{P}}_1 = \frac{1}{\lambda_B}\left(\vec{B} - \B1 \vec{e}_1^{\, \mathcal{C}} -  \B2 \vec{e}_2^{\, \mathcal{C}}\right)
\eea
where $\lambda_B$ is such that $\vec{e}^{\,\mathcal{P}}_1$ is normalised.

 For the first physical degree of freedom, one has
\bea
\left[\boldsymbol{K}_k\right]_{2,0}&=& 2N \lambda_B \vec{e}^{\,\mathcal{G}}_1 \cdot \left( \left[\boldsymbol{H}_k\, , \, \boldsymbol{\Omega}_6 \right]  \vec{B} \right) - \lambda_B  \B1  \left[\boldsymbol{K}_k\right]_{0,0} - \lambda_B  \B2   \left[\boldsymbol{K}_k\right]_{0,1}\,, \label{eq:K02} \\
\left[\boldsymbol{K}_k\right]_{2,1}&=& 2N \lambda_B \vec{e}^{\,\mathcal{G}}_2 \cdot \left( \left[\boldsymbol{H}_k\, , \, \boldsymbol{\Omega}_6 \right]  \vec{B} \right) - \lambda_B  \B1  \left[\boldsymbol{K}_k\right]_{0,1} - \lambda_B  \B2   \left[\boldsymbol{K}_k\right]_{1,1}\,. \label{eq:K12}
\eea
Developing $\vec{e}^{\,\mathcal{G}}_2 \cdot \left( \left[\boldsymbol{H}_k\, , \, \boldsymbol{\Omega}_6 \right]  \vec{B} \right)$ and inserting $\left[\boldsymbol{K}_k\right]_{2,0}$, the expression for $\left[\boldsymbol{K}_k\right]_{2,1}$ boils down to
\bea
\left[\boldsymbol{K}_k\right]_{2,1}&=& \frac{\lambda_S}{\lambda_D} \left(\left[\boldsymbol{K}_k\right]_{2,0} + \lambda_B \B1  \left[\boldsymbol{K}_k\right]_{0,0} + \lambda_B \B2   \left[\boldsymbol{K}_k\right]_{0,1} \right) - \lambda_B  \B1  \left[\boldsymbol{K}_k\right]_{0,1} \nonumber \\
&&- \lambda_B  \B2   \left[\boldsymbol{K}_k\right]_{1,1} - 2N \lambda_B \mu_S \,\vec{S}_k \cdot \left( \left[\boldsymbol{H}_k\, , \, \boldsymbol{\Omega}_6 \right]  \boldsymbol{\Omega}_6\vec{B} \right)\,. 
\eea
For the second physical degree of freedom, one similarly obtains
\bea
\left[\boldsymbol{K}_k\right]_{0,5}&=& - 2N \lambda_B \vec{e}_1^{\, \mathcal{C}} \cdot \left[\boldsymbol{H}_k\, , \, \boldsymbol{\Omega}_6 \right] \vec{B} - 2\lambda_B  \B1  \left[\boldsymbol{K}_k\right]_{0,3} -\lambda_B  \B2   \left[\boldsymbol{K}_k\right]_{0,4} \,,\\
\left[\boldsymbol{K}_k\right]_{1,5}&=& - 2N \lambda_B \vec{e}_2^{\, \mathcal{C}} \cdot \left[\boldsymbol{H}_k\, , \, \boldsymbol{\Omega}_6 \right] \vec{B} - 2\lambda_B  \B2   \left[\boldsymbol{K}_k\right]_{1,4} -\lambda_B  \B1  \left[\boldsymbol{K}_k\right]_{0,4} \, .
\eea
Developing $\vec{e}_2^{\,\mathcal{C}}=\lambda_S \vec{D}_k+\mu_S \vec{S}_k$, the last equation reduces to
\bea
\left[\boldsymbol{K}_k\right]_{1,5} &=& \frac{\lambda_S}{\lambda_D} \left(\left[\boldsymbol{K}_k\right]_{0,5} + 2 \lambda_B  \B1  \left[\boldsymbol{K}_k\right]_{0,3} + \lambda_B  \B2   \left[\boldsymbol{K}_k\right]_{0,4} \right) - \lambda_B  \B1  \left[\boldsymbol{K}_k\right]_{0,4}\nonumber  \\
&& - 2 \lambda_B  \B2   \left[\boldsymbol{K}_k\right]_{1,4} - 2N \lambda_B \mu_S \,\vec{S}_k\cdot \left[\boldsymbol{H}_k\, , \, \boldsymbol{\Omega}_6 \right] \vec{B}  \,.
\eea
In \App{ssec:dynPhys}, these four terms correspond to the entries of $\boldsymbol{k}_{\mathrm{cp}}$. Alternatively, one could compute them by rewriting $\left[\boldsymbol{K}_k\right]_{a,2}$ and $\left[\boldsymbol{K}_k\right]_{a,5}$ as:
\bea
\left[\boldsymbol{K}_k\right]_{a,2}&=&  \lambda_B \vec{e}_a \cdot \left(2N \boldsymbol{H}_k \vec{B}+ \boldsymbol{\Omega}_6 \dot{\vec{B}} \right) - \frac{\dot{\lambda}_E}{\lambda_B} \underbrace{\vec{e}_a \cdot \vec{e}^{\, \mathcal{P}}_2}_{=\delta_{a,5}} + \lambda_B \left( \dot{B}^{\, \mathcal{C}}_2 + B^{\, \mathcal{C}}_2 \frac{\dot{\mu}_S}{\mu_S} \right) \underbrace{\vec{e}_a \cdot \vec{e}^{\, \mathcal{G}}_2}_{=\delta_{a,4}}\nonumber \\
&& + \lambda_B \left\{\dot{B}^{\, \mathcal{C}}_1 + B^{\, \mathcal{C}}_1 \frac{\dot{\lambda}_D}{\lambda_D} + B^{\, \mathcal{C}}_2 \left[\frac{\dot{\lambda}_S}{\lambda_D}-\frac{\dot{\mu}_S}{\mu_S}\frac{\lambda_S}{\lambda_D}-N\frac{\mu_S}{\lambda_D}\left(\frac{k}{v^{1/3}}\right)^2\right] \right\} \underbrace{\vec{e}_a \cdot \vec{e}^{\, \mathcal{G}}_1}_{=\delta_{a,3}}  \nonumber \\
&&  - 2N \lambda_B B^{\, \mathcal{C}}_1 \vec{e}_a \cdot \left( \left[\boldsymbol{H}_k\, , \, \boldsymbol{\Omega}_6 \right]  \vec{e}^{\,\mathcal{G}}_1 \right)  - 2N \lambda_B B^{\, \mathcal{C}}_2 \vec{e}_a \cdot \left( \left[\boldsymbol{H}_k\, , \, \boldsymbol{\Omega}_6 \right]  \vec{e}^{\,\mathcal{G}}_2 \right) \,, 
\eea
and
\bea
\left[\boldsymbol{K}_k\right]_{a,5}&=&  - \lambda_B \vec{e}_a \cdot \left(2N \boldsymbol{H}_k \boldsymbol{\Omega}_6 \vec{B} - \dot{\vec{B}} \right) + \frac{\dot{\lambda}_E}{\lambda_B} \underbrace{\vec{e}_a \cdot \vec{e}^{\, \mathcal{P}}_1}_{=\delta_{a,2}} - \lambda_B \left( \dot{B}^{\, \mathcal{C}}_2 + B^{\, \mathcal{C}}_2 \frac{\dot{\mu}_S}{\mu_S} \right) \underbrace{\vec{e}_a \cdot \vec{e}^{\, \mathcal{C}}_2}_{=\delta_{a,1}} \nonumber \\
&& - \lambda_B \left\{\dot{B}^{\, \mathcal{C}}_1 + B^{\, \mathcal{C}}_1 \frac{\dot{\lambda}_D}{\lambda_D} + B^{\, \mathcal{C}}_2 \left[\frac{\dot{\lambda}_S}{\lambda_D}-\frac{\dot{\mu}_S}{\mu_S}\frac{\lambda_S}{\lambda_D}-N\frac{\mu_S}{\lambda_D}\left(\frac{k}{v^{1/3}}\right)^2\right] \right\} \underbrace{\vec{e}_a \cdot \vec{e}^{\, \mathcal{C}}_1}_{=\delta_{a,0}} \,. \quad\quad  
\eea

\subsection*{Couplings with the physical degrees of freedom}

We are left with the couplings between physical degrees of freedom, corresponding to the entries of $\boldsymbol{k}_{\mathrm{pp}}$ in \App{ssec:dynPhys}. Those components were already computed in \Refc{Boldrin:2022vcp}, with which we check that the results derived below are consistent.

For the self-coupling of the physical configuration degree of freedom, inserting \Eq{eq:e1P:def} into \Eq{eq:Kab-app}, one obtains
\bea
\left[\boldsymbol{K}_k\right]_{2,2} &=& 2N \lambda_B^2 \vec{B}\cdot \left(\boldsymbol{H}_k \vec{B} \right)  - 4 N \lambda_B^2 \left(\B1  \vec{e}^{\,\mathcal{G}}_1 + \B2   \vec{e}^{\,\mathcal{G}}_2 \right) \cdot \left( \left[\boldsymbol{H}_k\, , \, \boldsymbol{\Omega}_6 \right]  \vec{B} \right)  \nonumber  \\
&& + \lambda_B^2  \left( \B1 \right)^2 \left[\boldsymbol{K}_k\right]_{0,0} + \lambda_B^2  \left( \B2  \right)^2 \left[\boldsymbol{K}_k\right]_{1,1} + 2\lambda_B^2   \B1   \B2   \left[\boldsymbol{K}_k\right]_{0,1}\,.
\eea
Let us now observe that the term $\vec{e}^{\,\mathcal{G}}_\mu \cdot ( \left[\boldsymbol{H}_k\, , \, \boldsymbol{\Omega}_6 \right]  \vec{B})$ is also present in $\left[\boldsymbol{K}_k\right]_{0,2}$ and in $\left[\boldsymbol{K}_k\right]_{1,2}$, see \Eqs{eq:K02} and~\eqref{eq:K12} respectively. The above formula can therefore be recast into
\bea
\left[\boldsymbol{K}_k\right]_{2,2} &=& 2N \lambda_B^2 \vec{B}\cdot \left(\boldsymbol{H}_k \vec{B} \right) - 2 \lambda_B  \B1  \left[\boldsymbol{K}_k\right]_{0,2} - 2 \lambda_B  \B1  \left[\boldsymbol{K}_k\right]_{1,2} \nonumber \\
&& - \lambda_B^2  \left( \B1 \right)^2 \left[\boldsymbol{K}_k\right]_{0,0} - \lambda_B^2  \left( \B2  \right)^2 \left[\boldsymbol{K}_k\right]_{1,1} - 2\lambda_B^2   \B1   \B2   \left[\boldsymbol{K}_k\right]_{0,1}  \,.
\eea

For the coupling between the physical configuration and its conjugated momentum, one finds
\bea
\left[\boldsymbol{K}_k\right]_{2,5} &=& - N\lambda_B^2 \vec{B}\cdot \left[\boldsymbol{H}_k\, , \, \boldsymbol{\Omega}_6 \right] \vec{B} + 2N \lambda_B^2 \left(\B1  \vec{e}^{\,\mathcal{C}}_1 + \B2   \vec{e}^{\,\mathcal{C}}_2 \right) \cdot \left( \left[\boldsymbol{H}_k\, , \, \boldsymbol{\Omega}_6 \right]  \vec{B} \right)\nonumber  \\
&& + 2 \lambda_B^2 \left(\B1  \dot{B}_1^{\,\mathcal{C}} + \B2   \dot{B}_2^{\,\mathcal{C}} \right) + \lambda_B^2 \left(\B1 \right)^2 \left[\boldsymbol{K}_k\right]_{0,3} + \lambda_B^2 \left(\B2  \right)^2 \left[\boldsymbol{K}_k\right]_{1,4} \nonumber \\
&& + \lambda_B^2 \B1  \B2   \left[\boldsymbol{K}_k\right]_{0,4}  \,,
\eea
and using the expressions obtained above for $\left[\boldsymbol{K}_k\right]_{0,5}$ and $\left[\boldsymbol{K}_k\right]_{1,5}$, this can be written as
\bea
\left[\boldsymbol{K}_k\right]_{2,5} &=& - N\lambda_B^2 \vec{B}\cdot \left[\boldsymbol{H}_k\, , \, \boldsymbol{\Omega}_6 \right] \vec{B} + 2 \lambda_B^2 \left(\B1  \dot{B}_1^{\,\mathcal{C}} + \B2   \dot{B}_2^{\,\mathcal{C}} \right) \nonumber \\
&& - \lambda_B^2 \left(\B1 \right)^2 \left[\boldsymbol{K}_k\right]_{0,3} - \lambda_B^2 \left(\B2  \right)^2 \left[\boldsymbol{K}_k\right]_{1,4} - \lambda_B^2 \B1  \B2   \left[\boldsymbol{K}_k\right]_{0,4} \nonumber \\
&& - \lambda_B \B1  \left[\boldsymbol{K}_k\right]_{0,5} - \lambda_B \B2   \left[\boldsymbol{K}_k\right]_{1,5}\, .
\eea

Finally, for the self-coupling of the physical momentum, one has
\bea
\left[\boldsymbol{K}_k\right]_{5,5} &=& 2N \lambda_B^2 \left(\boldsymbol{\Omega}_6 \vec{B} \right) \cdot \left( 
 \boldsymbol{H}_k \, \boldsymbol{\Omega}_6 \vec{B} \right) \, ,
\eea
and a direct calculation shows that this component simply reduces to $\lambda_B^2$.

\subsection*{Components of $\left[\boldsymbol{K}_k\right]_{a,b}$}

Combining and summarising the above results, one obtains
\bea
\left[\boldsymbol{K}_k\right]_{0,0} &=& \frac{\lambda^2_D N}{12 \lambda^6 \Mp^6 v^{
 11/3}} \left[4 \lambda^8 \pi_\phi^4 + 
   2 \lambda^4 v^{10/3} (48 \Mp^{12} v^2 - \lambda^4 \Mp^2 k^2 V + 20 \lambda^2 \Mp^6 v^{4/3} V + 
      2 \lambda^4 v^{2/3} V^2) \right. \nonumber \\
      && + \pi_\phi^2 (- \lambda^8 k^2 \Mp^2 v^{4/3} + 8 \lambda^6 \Mp^6 v^{8/3} + 12 k^2 \Mp^{10} v^4 + 10 \lambda^8 v^2 V + 12 \Mp^{10} v^{14/3} V_{,\phi,\phi}) \nonumber \\
      && \left.+ 
   6 \lambda^4 \Mp^4 \pi_\phi v^{13/3} V_\phi \theta \right] \,, \\
\left[\boldsymbol{K}_k\right]_{0,1} &=& \frac{\lambda_S}{\lambda_D} \left[\boldsymbol{K}_k\right]_{0,0} - 2 N \mu_S \lambda_D \, \left[ \frac{1}{4\lambda^2\Mp^2} \frac{\pi_\phi^3}{v^{4/3}} V_{,\phi} - \frac{1}{4\lambda^2}(2\lambda^2-k^2) \pi_\phi V_{,\phi} \right. \nonumber\\
&& - \frac{\Mp^4}{2\lambda^6} k^2 v^{4/3} \pi_\phi V_{,\phi}  - \frac{1}{4\lambda^2 \Mp^2} v^{2/3} \pi_\phi V V_{,\phi}  - \frac{\Mp^4}{2\lambda^6} v^2 \pi_\phi V_{,\phi} V_{,\phi,\phi} + \frac{\lambda^2}{18 \Mp^8} \frac{\pi_\phi^4 \theta}{v^{11/3}}  \nonumber\\
&& + \frac{\lambda^2}{8 \Mp^6} (\lambda^2+k^2) \frac{\pi_\phi^2 \theta}{v^{7/3}} - \frac{\lambda^2}{36 \Mp^8} \frac{\pi_\phi^2 \theta}{v^{5/3}} V + \frac{1}{4\Mp^2} \frac{\pi_\phi^2 \theta }{v} - \frac{\lambda^2}{16 \Mp^4} k^4 \frac{\theta}{v}  + \frac{\lambda^2}{8\Mp^6} k^2 \frac{\theta V}{v^{1/3}} \nonumber \\
&& \left. + \frac{1}{2} k^2 v^{1/3} \theta - \frac{\lambda^2}{36 \Mp^8} v^{1/3} \theta V^2 - \frac{1}{2 \Mp^2} v \theta V + \frac{2 \Mp^4}{\lambda^2} v^{5/3}\theta - \frac{1}{8 \lambda^2 \Mp^2} v^{5/3} \theta V_{,\phi}^2 \right]  \,,\\
\left[\boldsymbol{K}_k\right]_{0,2} &=& - \lambda_B  \B1  \left[\boldsymbol{K}_k\right]_{0,0} - \lambda_B  \B2   \left[\boldsymbol{K}_k\right]_{0,1} + \sqrt{N} \lambda_B \lambda_D \left( \frac{1}{2\lambda^2} \frac{\pi_\phi^2 V_{,\phi}}{ v^{5/6} \theta} + \frac{v^{7/6} \theta V_{,\phi}}{4 \lambda^2 \Mp^2}  + \frac{\pi_\phi}{ \sqrt{v}} \right. \nonumber \\
&& \left.+ \frac{\lambda^2}{3 \Mp^6} \frac{\pi_\phi^3}{v^{19/6}} - \frac{\lambda^2 k^2}{8 \Mp^4} \frac{\pi_\phi}{v^{11/6}} + \frac{k^2 \Mp^4}{\lambda^6} \pi_\phi v^{5/6} + \frac{\lambda^2}{6 \Mp^6} \frac{\pi_\phi V}{v^{7/6}} + \frac{\Mp^4}{\lambda^6} \pi_\phi v^{3/2} V_{,\phi,\phi} \right)  \,,\\
\left[\boldsymbol{K}_k\right]_{0,3} &=& \frac{\dot{\lambda}_D}{\lambda_D} \,,\\
\left[\boldsymbol{K}_k\right]_{0,4} &=& \frac{\dot{\lambda}_S}{\lambda_D}-\frac{\dot{\mu}_S}{\mu_S}\frac{\lambda_S}{\lambda_D}-N\frac{\mu_S}{\lambda_D}\left(\frac{k}{v^{1/3}}\right)^2 \,,\\
\left[\boldsymbol{K}_k\right]_{0,5} &=& - 2\lambda_B  \B1  \left[\boldsymbol{K}_k\right]_{0,3} -\lambda_B  \B2   \left[\boldsymbol{K}_k\right]_{0,4} + \sqrt{N} \lambda_B \lambda_D \left( - \frac{\lambda}{2 \Mp^2} \frac{\pi_\phi^3}{v^{17/6} \theta} \right.  \nonumber\\
&& \left.+ \frac{4 \Mp^4}{\lambda} \frac{\pi_\phi}{v^{1/6} \theta } + \frac{\lambda}{4\Mp^4} \frac{\pi_\phi \theta }{v^{5/6}} + \frac{\Mp^2}{\lambda^3} v^{3/2} V_{,\phi} \right)\,,\\
\left[\boldsymbol{K}_k\right]_{1,1} &=& 2 \frac{\lambda_S}{\lambda_D} \left[\boldsymbol{K}_k\right]_{0,1} - \frac{\lambda_S^2}{\lambda_D^2} \left[\boldsymbol{K}_k\right]_{0,0} + N \mu_S^2 \left[ \frac{2 \lambda^2}{9 \Mp^8}\frac{\pi_\phi^6}{v^{17/3}}  + \frac{\lambda^2}{18 \Mp^6}\frac{(18\lambda^2 + 7k^2) \pi_\phi^4}{v^{13/3}} \right.\nonumber \\
&& - \frac{2}{\Mp^2}\frac{\pi_\phi^4}{v^3} - \frac{\lambda^2}{3\Mp^8}\frac{\pi_\phi^4 V}{v^{11/3}} - \frac{\lambda^2 k^6}{8\Mp^2 v^{5/3}} + \frac{\lambda^2 k^4}{3\Mp^4}\frac{V}{v}  - \frac{5 \lambda^2 k^2}{18\Mp^6}\frac{V^2}{v^{1/3}} - \frac{10 k^2}{3}v^{1/3} V \nonumber\\
&& + \frac{\lambda^2}{9 \Mp^8}v^{1/3} V^3 + \frac{k^2\Mp^4}{\lambda^6} v^{7/3} V_{,\phi}^2 - \frac{8 \Mp^4}{\lambda^2} v^{5/3} V  + \frac{1}{\lambda^2 \Mp^2} v^{5/3} V V_{,\phi}^2 \nonumber\\
&&+ \frac{2}{\Mp^2} v V^2 - \frac{1}{\lambda^2}(\lambda^2+k^2) v V_{,\phi}^2 + \frac{\Mp^4}{\lambda^6} v^{3} V_{,\phi}^2 V_{,\phi,\phi}  + \frac{\lambda^2}{6\Mp^4}\frac{(6\lambda^4 + 6\lambda^2 k^2 + k^4) \pi_\phi^2}{v^3}  \nonumber\\
&&- \frac{2 k^2}{3}\frac{\pi_\phi^2}{v^{5/3}} - \frac{\lambda^2}{9 \Mp^6}\frac{(9\lambda^2 + k^2) \pi_\phi^2 V}{v^{7/3}} - \frac{4 \Mp^4}{\lambda^2} \frac{\pi_\phi^2}{v^{1/3}}  - \frac{1}{\lambda^2\Mp^2}\frac{\pi_\phi^2 V_{,\phi}^2}{v^{1/3}} \nonumber\\
&&\left. - \frac{6}{\Mp^2}\frac{\pi_\phi^2 V}{v}  - \frac{\pi_\phi^2 V_{,\phi,\phi}}{v} + \frac{6}{\Mp^2} \pi_\phi \theta V_{,\phi}  \right] \,,\\
\left[\boldsymbol{K}_k\right]_{1,2} &=& \frac{\lambda_S}{\lambda_D} \left(\left[\boldsymbol{K}_k\right]_{0,2} + \lambda_B  \B1  \left[\boldsymbol{K}_k\right]_{0,0} + \lambda_B  \B2   \left[\boldsymbol{K}_k\right]_{0,1} \right) - \lambda_B  \B1  \left[\boldsymbol{K}_k\right]_{0,1} \nonumber \\
&& - \lambda_B  \B2   \left[\boldsymbol{K}_k\right]_{1,1}  + \sqrt{N} \lambda_B \mu_S \left[ - \frac{2\lambda^2}{9 \Mp^6} \frac{\pi_\phi^5}{v^{31/6} \theta} - \frac{\lambda^2}{4 \Mp^4} \frac{(2\lambda^2+k^2) \pi_\phi^3}{v^{23/6} \theta} \right. \nonumber \\
&& - \frac{2\pi_\phi^3}{3 v^{5/2} \theta} + \frac{\lambda^2}{9 \Mp^6} \frac{\pi_\phi^3 V}{v^{19/6} \theta} - \frac{1}{2 \lambda^2 \Mp^2} \frac{\pi_\phi^2 V_{,\phi}}{v^{5/6}} - \frac{1}{2\lambda^2} \left(k^2+2\lambda^2\right) \sqrt{v}V_{,\phi} \nonumber \\
&&  + \frac{k^2 \Mp^4}{\lambda^6} v^{11/6} V_{,\phi} + \frac{1}{2 \Mp^2\lambda^2} v^{7/6} V V_{,\phi} + \frac{\Mp^4}{\lambda^6} v^{5/2} V_{,\phi} V_{,\phi,\phi} +  \frac{ \lambda^2 k^4}{8 \Mp^2} \frac{\pi_\phi}{v^{5/2} \theta} \nonumber \\
&&\left. - \frac{\lambda^2 k^2}{4 \Mp^4} \frac{\pi_\phi V}{v^{11/6} \theta} +\frac{\lambda^2}{9 \Mp^6} \frac{\pi_\phi V^2}{v^{7/6} \theta}  + \frac{1}{2\lambda^2} \frac{\pi_\phi v^{1/6} V_{,\phi}^2}{\theta} + \frac{2\pi_\phi V}{3 \sqrt{v} \theta} + \frac{5}{2\Mp^2} \frac{\pi_\phi \theta}{\sqrt{v}} \right] \\
\left[\boldsymbol{K}_k\right]_{1,3} &=& 0\,,\\
\left[\boldsymbol{K}_k\right]_{1,4} &=& \frac{\dot{\mu}_S}{\mu_S}\,,\\
\left[\boldsymbol{K}_k\right]_{1,5} &=& \frac{\lambda_S}{\lambda_D} \left(\left[\boldsymbol{K}_k\right]_{0,5} + 2 \lambda_B  \B1  \left[\boldsymbol{K}_k\right]_{0,3} + \lambda_B  \B2   \left[\boldsymbol{K}_k\right]_{0,4} \right) - \lambda_B  \B1  \left[\boldsymbol{K}_k\right]_{0,4} \nonumber  \\
&& - 2 \lambda_B  \B2   \left[\boldsymbol{K}_k\right]_{1,4}  + \sqrt{N} \lambda_B \mu_S \left[ - \frac{\lambda}{ \Mp^2} \frac{ \pi_\phi^2 V_\phi}{v^{11/6} \theta}   + \frac{3\lambda}{2 \Mp^4} \frac{\pi_\phi^3}{v^{17/6}} + \frac{\lambda}{4 \Mp^2} \frac{(4\lambda^2+k^2) \pi_\phi}{v^{3/2}} \right. \nonumber \\
&& \left. + \frac{\Mp^2}{\lambda^3} \frac{(2\lambda^2-k^2) \pi_\phi}{v^{1/6}} - \frac{\lambda}{2\Mp^4} \frac{\pi_\phi V}{v^{5/6}}  - \frac{\Mp^2}{\lambda^3} \sqrt{v} \pi_\phi V_{,\phi,\phi} + \frac{3}{2\lambda^3} v^{3/2} V_{,\phi} \theta \right]\,,\\
\left[\boldsymbol{K}_k\right]_{2,2} &=& \lambda_B^2 \left[ \frac{\lambda^2}{3\Mp^4} \frac{\pi_\phi^4}{v^{14/3} \theta^2} - \frac{\lambda^2 k^2}{8 \Mp^2} \frac{\pi_\phi^2}{v^{10/3}\theta^2} + \frac{\lambda^2}{6 \Mp^4} \frac{\pi_\phi^2 V}{v^{8/3}\theta^2} + \frac{1}{\lambda^2} \frac{\pi_\phi V_{,\phi}}{v^{1/3} \theta} \right. \nonumber  \\
&& \left. + \frac{\Mp^4}{\lambda^6} v^{4/3} \left(k^2 + v^{2/3} V_{,\phi,\phi}\right) \right] - 2 \lambda_B  \B1  \left[\boldsymbol{K}_k\right]_{0,2} - 2 \lambda_B  \B1  \left[\boldsymbol{K}_k\right]_{1,2} \nonumber \\
&&  - \lambda_B^2  \left( \B1 \right)^2 \left[\boldsymbol{K}_k\right]_{0,0} - \lambda_B^2  \left( \B2  \right)^2 \left[\boldsymbol{K}_k\right]_{1,1} - 2\lambda_B^2   \B1   \B2   \left[\boldsymbol{K}_k\right]_{0,1} 
\,,\\
\left[\boldsymbol{K}_k\right]_{2,3} &=& 0\,,\\
\left[\boldsymbol{K}_k\right]_{2,4} &=& 0\,,\\
\left[\boldsymbol{K}_k\right]_{2,5} &=& - \lambda_B^2 \frac{\lambda}{2\Mp^2} \frac{\pi_\phi^2}{\theta v^{7/3}} + 2 \lambda_B^2 \left(\B1  \dot{B}_1^{\,\mathcal{C}} + \B2   \dot{B}_2^{\,\mathcal{C}} \right)  - \lambda_B^2 \left(\B1 \right)^2 \left[\boldsymbol{K}_k\right]_{0,3} \nonumber \\
&& - \lambda_B^2 \left(\B2  \right)^2 \left[\boldsymbol{K}_k\right]_{1,4} - \lambda_B^2 \B1  \B2   \left[\boldsymbol{K}_k\right]_{0,4} - \lambda_B \B1  \left[\boldsymbol{K}_k\right]_{0,5} - \lambda_B \B2   \left[\boldsymbol{K}_k\right]_{1,5}  \,,\\
\left[\boldsymbol{K}_k\right]_{3,3} &=& 0\,,\\
\left[\boldsymbol{K}_k\right]_{3,4} &=& 0\,,\\
\left[\boldsymbol{K}_k\right]_{3,5} &=& 0\,,\\
\left[\boldsymbol{K}_k\right]_{4,4} &=& 0\,,\\
\left[\boldsymbol{K}_k\right]_{4,5} &=& 0\,,\\
\left[\boldsymbol{K}_k\right]_{5,5} &=& \lambda_B^2 \,.
\eea

\section{Dynamical basis}
\label{ssec:dynPhys}

As explained in \Sec{sec:PSgen}, the use of the \Pbasis allows one to separate the physical sector from the unphysical one at the {\it kinematical level}. However, at the {\it dynamical level}, the Hamiltonian exhibits some couplings between the two sectors. Though a clear separation of the physical sector from the unphysical one is effectively recovered providing that the dynamics is solved on the surface of constraints [see \Eqs{eq:eomQ} and \eqref{eq:eomP}], the gauge degrees of freedom are not entirely decoupled from the physical degrees of freedom at the dynamical level [see \eg \Eqs{eq:PDsol} and~\eqref{eq:PSsol}].  In practice this is sufficient since one is mainly interested in the physical degrees of freedom. Yet, formal studies of first-class constrained systems show that such a separation can be performed directly at the level of the Hamiltonian \cite{doi:10.1063/1.529065}. This is the topic of this appendix. 

Consider $n$ canonical pairs, $(q_\mu,p^\mu)$ with the constrained Hamiltonian 
\bea
	H(q_\mu,p^\mu)=h(q_\mu,p^\mu)+\lambda^\sigma C_\sigma(q_\mu,p^\mu)\, ,
\eea
where  $C_\sigma$ stands for $m$ constraints. We assume that constraints are first class and preserved through evolution. It is thus possible to perform a canonical transformation such that the $m$ new configuration variables are $Q_a=C_a$. The $m$ associated new momenta are denoted $P^a$ and stand for the gauge degrees of freedom in the context of cosmological perturbations. The remaining $(n-m)$ canonical pairs are denoted $(Q_i,P^i)$ and correspond to the physical degrees of freedom. Moreover, the canonical transformation can be chosen such that the new Hamiltonian reads~\cite{doi:10.1063/1.529065} 
\bea
	K(Q_\mu,P^\mu)=k(Q_i,P^i)+\Lambda^aQ_a \label{eq:hamref7}
\eea
where $k(Q_i,P^i)$ depends on the physical degrees of freedom only, and where the $\Lambda_a$'s are Lagrange multipliers whose specific choice fixes the specific values of the gauge degrees of freedom $P^a$ (more precisely it fixes their time dependence). In this setup, physical and unphysical degrees of freedom fully decouple, since the Hamiltonian is separable. 

In this appendix, we explicitly construct the canonical transformation allowing for the Hamiltonian of cosmological perturbations to be expressed in a form identical to the one given in \Eq{eq:hamref7}, starting from the Hamiltonian expressed in the \Pbasis. 

\subsection{Reordering the phase space}
\label{sssec:Reorder}

Starting from \Eq{eq:deltaQ}, we first re-order the phase space as $\Gamma=\Gamma_{\mathrm{unphys}}\otimes\Gamma_{\mathrm{phys}}$, where $\Gamma_{\mathrm{unphys}}=\left\{(Q_1,Q_2;P_1,P_2)\right\}$ contains the unphysical degrees of freedom and $\Gamma_{\mathrm{phys}}=\{(Z_1;Z_2)\}$ contains the physical ones. A vector in the full phase space is now 
\bea
    \vec{z}_{\vec{k}}=\left(Q_1(\vec{k}),\,Q_2(\vec{k}),\,P_1(\vec{k}),\,P_2(\vec{k});\,Z_1(\vec{k}),\,Z_2(\vec{k})\right)^\mathrm{T},
\eea
and the symplectic form is given by the direct sum $\boldsymbol{\Omega}_4\oplus\boldsymbol{\Omega}_2$, that is:
\bea
	\boldsymbol{\Omega}_4\oplus\boldsymbol{\Omega}_2=\left(\begin{array}{ccc}
		\boldsymbol{0} & \boldsymbol{I}_2 & \boldsymbol{0} \\
		-\boldsymbol{I}_2 & \boldsymbol{0} & \boldsymbol{0} \\
		\boldsymbol{0} & \boldsymbol{0} & \boldsymbol{\Omega}_2
	\end{array}\right).
\eea
The Hamiltonian~\eqref{eq:hamphys} then reads
\bea
	{K}=\ds\int\dd^3\vec{k}\left\{\left[\Lambda_1^\star (\vec{k})Q_1(\vec{k})+\Lambda_2^\star(\vec{k}) Q_2(\vec{k}) +\mathrm{c.c.}\right]+\vec{z}^{\,\dag}_{\vec{k}}\left(\begin{array}{ccc}
		\boldsymbol{k}_{\mathrm{cc}} & \boldsymbol{k}_{\mathrm{cg}} & \boldsymbol{k}_{\mathrm{cp}} \\
		\boldsymbol{k}_{\mathrm{cg}}^\mathrm{T} & \boldsymbol{0} & \boldsymbol{0} \\
		\boldsymbol{k}_{\mathrm{cp}}^\mathrm{T} & \boldsymbol{0} & \boldsymbol{k}_{\mathrm{pp}}
	\end{array}\right)\vec{z}_{\vec{k}}\right\}, \label{eq:KamilReorder}
\eea
where the nonvanishing blocks $\boldsymbol{k}_{\mathrm{cc}}$, $\boldsymbol{k}_{\mathrm{cg}}$, $\boldsymbol{k}_{\mathrm{cp}}$ and $\boldsymbol{k}_{\mathrm{pp}}$ can be found in \App{app:HamKinBasis} (note that all the vanishing blocks, $\boldsymbol{k}_{\mathrm{gg}}=\boldsymbol{k}_{\mathrm{gp}}=0$, come from  the fact that gauge degrees of freedom are solely coupled to the constraints). In the following, lower-case latin letters in boldface format stand for $2\times2$ matrices.

In that ordering, a Hamiltonian of the form given by \Eq{eq:hamref7} reads
\bea
	\widetilde{K}=\ds\int\dd^3k\left\{\left[\widetilde{\Lambda}_1^\star(\vec{k}) \widetilde{Q}_1(\vec{k})+\widetilde{\Lambda}_2^\star (\vec{k})\widetilde{Q}_2(\vec{k}) +\mathrm{c.c.}\right]+\widetilde{\vec{z}}^{\,\dag}_{\vec{k}}\left(\begin{array}{ccc}
		\boldsymbol{0} & \boldsymbol{0} & \boldsymbol{0} \\
		\boldsymbol{0} & \boldsymbol{0} & \boldsymbol{0} \\
		\boldsymbol{0} & \boldsymbol{0} & \widetilde{\boldsymbol{k}}_{\mathrm{pp}}
	\end{array}\right) \widetilde{\vec{z}}_{\vec{k}} \right\}, \label{eq:dynsepham}
\eea
where $\widetilde{\vec{z}}_{\vec{k}}$ stands for the new set of canonical variables and where the $\widetilde{\Lambda}_\mu$'s are new Lagrange multipliers given by combinations of the old Lagrange multipliers only.

\subsection{Canonical transformation}

One seeks a linear canonical transformation, $\widetilde{\vec{z}}_{\vec{k}}=\boldsymbol{C}\vec{z}_{\vec{k}}$, such that the new Hamiltonian has the form given in \Eq{eq:dynsepham}. Its shape is constrained using prescriptions at the kinematical level, and prescriptions at the dynamical level. 

\subsubsection*{Kinematical prescription} The matrix $\boldsymbol{C}$ generating the canonical transformation has to be symplectic. It is further constrained to ensure that the linear part of the new Hamiltonian is only composed of the constraints and that the new physical degrees of freedom do not receive any contribution from the gauge degrees of freedom. This requires $\boldsymbol{C}$ to be of the form
\bea
\label{eq:C:def}
	\boldsymbol{C}=\left(\begin{array}{ccc}
		\boldsymbol{m}_{\mathrm{c}} & \boldsymbol{0} & \boldsymbol{0} \\
		\boldsymbol{n} & \boldsymbol{m}_{\mathrm{g}} & \boldsymbol{b} \\
		\boldsymbol{a} & \boldsymbol{0} & \boldsymbol{m}_{\mathrm{p}}
	\end{array}\right),
\eea
where its different blocks have to satisfy
\bea
	\boldsymbol{m}_{\mathrm{p}}^\mathrm{T}\boldsymbol{\Omega}_2\boldsymbol{m}_{\mathrm{p}}=\boldsymbol{\Omega}_2, \label{eq:mpSymplectic} \\
	\boldsymbol{m}_{\mathrm{g}}=\boldsymbol{m}_{\mathrm{c}}^{-1\,\mathrm{T}}, \label{eq:mg.mc} \\
	\boldsymbol{m}_{\mathrm{c}}^\mathrm{T}\boldsymbol{b}+\boldsymbol{a}^\mathrm{T}\boldsymbol{\Omega}_2\boldsymbol{m}_{\mathrm{p}}=\boldsymbol{0}, \\
	\boldsymbol{m}_{\mathrm{c}}^\mathrm{T}\boldsymbol{n}-\boldsymbol{n}^\mathrm{T}\boldsymbol{m}_{\mathrm{c}}+\boldsymbol{a}^\mathrm{T}\boldsymbol{\Omega}_2\boldsymbol{a}=\boldsymbol{0},
\eea
for $\boldsymbol{C}$ to be symplectic. We stress that the first condition simply states that $\boldsymbol{m}_{\mathrm{p}}$ belongs to the symplectic group $\mathrm{Sp}(2,\mathbb{R})$. 

The inverse of $\boldsymbol{C}$ is obtained using blockwise inversion.\footnote{In practice, $\boldsymbol{C}$ is first partitionned as
\bea
	\boldsymbol{C}=\left(\begin{array}{cc}
		\boldsymbol{A}_{4,4} & \boldsymbol{B}_{4,2} \\
		\boldsymbol{C}_{2,4} & \boldsymbol{D}_{2,2}
	\end{array}\right)
\eea
where the subscripts give the dimension of each block. Block inversion requires to compute $\boldsymbol{D}_{2,2}^{-1}=\boldsymbol{m}_{\mathrm{p}}^{-1}$, and $\boldsymbol{A}^{-1}_{4,4}$, which can itself be inverted blockwise by partitioning it into four square blocks of dimension $2\times2$.} It boils down to
\bea
	\boldsymbol{C}^{-1}=\left(\begin{array}{ccc}
		\boldsymbol{m}_{\mathrm{c}}^{-1} & \boldsymbol{0} & \boldsymbol{0} \\
		\boldsymbol{\alpha} & \boldsymbol{m}_{\mathrm{c}}^\mathrm{T} & \boldsymbol{\beta} \\
		\boldsymbol{\gamma} & \boldsymbol{0} & \boldsymbol{m}^{-1}_{\mathrm{p}}
	\end{array}\right), \label{eq:invCmat}
\eea
where
\bea
	\boldsymbol{\alpha}&=&-\boldsymbol{m}^\mathrm{T}_{\mathrm{c}}\left(\boldsymbol{n}-\boldsymbol{b}\boldsymbol{m}^{-1}_{\mathrm{p}}\boldsymbol{a}\right)\boldsymbol{m}^{-1}_{\mathrm{c}}\, , \\
	\boldsymbol{\beta}&=&-\boldsymbol{m}^\mathrm{T}_{\mathrm{c}}\boldsymbol{b}\boldsymbol{m}^{-1}_{\mathrm{p}}\, , \\
	\boldsymbol{\gamma}&=&-\boldsymbol{m}^{-1}_{\mathrm{p}}\boldsymbol{a}\boldsymbol{m}_{\mathrm{c}}^{-1}\, , 
\eea
where we have inserted the expression of $\boldsymbol{\gamma}$ in $\boldsymbol{\alpha}$ and we have used \Eq{eq:mg.mc}. Using the remaining symplectic constraints we easily obtain that $\boldsymbol{\beta}=\boldsymbol{a}^\mathrm{T}\boldsymbol{\Omega}_2$ and $\boldsymbol{\alpha}=-\boldsymbol{n}^\mathrm{T}$.

\subsubsection*{Dynamical prescription}
Starting from \Eq{eq:invCmat}, the new Hamiltonian reads
\bea
	\widetilde{K}&=& \ds\int\dd^3\vec{k}\left\{\left[\widetilde{\Lambda}_1^\star(\vec{k}) \widetilde{Q}_1(\vec{k})+\widetilde{\Lambda}_2^\star(\vec{k}) \widetilde{Q}_2(\vec{k})+\mathrm{c.c.}\right]+\widetilde{\vec{z}}^{\,\dagger}_{\vec{k}}\left(\begin{array}{ccc}
		\widetilde{\boldsymbol{k}}_{\mathrm{cc}} & \widetilde{\boldsymbol{k}}_{\mathrm{cg}} & \widetilde{\boldsymbol{k}}_{\mathrm{cp}} \\
		\widetilde{\boldsymbol{k}}_{\mathrm{cg}}^\mathrm{T} & \boldsymbol{0} & \boldsymbol{0} \\
		\widetilde{\boldsymbol{k}}_{\mathrm{cp}}^\mathrm{T} & \boldsymbol{0} & \widetilde{\boldsymbol{k}}_{\mathrm{pp}}
	\end{array}\right)\widetilde{\vec{z}}_{\vec{k}}\right\},
\eea
where the new Lagrange multipliers and the new constraints are
\bea
	\left(\begin{array}{c}
		\widetilde{\Lambda}_D \\
		\widetilde{\Lambda}_S
	\end{array}\right)=\boldsymbol{m}^{-1\,\mathrm{T}}_{\mathrm{c}}\left(\begin{array}{c}
		{\Lambda}_D \\
		{\Lambda}_S
	\end{array}\right) \,\,\,\,\, \mathrm{and} \,\,\,\,\, \left(\begin{array}{c}
		\widetilde{Q}_1 \\
		\widetilde{Q}_2
	\end{array}\right)=\boldsymbol{m}_{\mathrm{c}}\left(\begin{array}{c}
		{Q}_1 \\
		{Q}_2
	\end{array}\right) ,
\eea
and where the quadratic part is given by
\bea
\left(\begin{array}{ccc}
		\widetilde{\boldsymbol{k}}_{\mathrm{cc}} & \widetilde{\boldsymbol{k}}_{\mathrm{cg}} & \widetilde{\boldsymbol{k}}_{\mathrm{cp}} \\
		\widetilde{\boldsymbol{k}}_{\mathrm{cg}}^\mathrm{T} & \boldsymbol{0} & \boldsymbol{0} \\
		\widetilde{\boldsymbol{k}}_{\mathrm{cp}}^\mathrm{T} & \boldsymbol{0} & \widetilde{\boldsymbol{k}}_{\mathrm{pp}}
	\end{array}\right)
= \boldsymbol{C}^{-1\,\mathrm{T}}\left(\begin{array}{ccc}
		\boldsymbol{k}_{\mathrm{cc}} & \boldsymbol{k}_{\mathrm{cg}} & \boldsymbol{k}_{\mathrm{cp}} \\
		\boldsymbol{k}_{\mathrm{cg}}^\mathrm{T} & \boldsymbol{0} & \boldsymbol{0} \\
		\boldsymbol{k}_{\mathrm{cp}}^\mathrm{T} & \boldsymbol{0} & \boldsymbol{k}_{\mathrm{pp}}
	\end{array}\right) \boldsymbol{C}^{-1} + \boldsymbol{C}^{-1\,\mathrm{T}} \left(\boldsymbol{\Omega}_4\oplus\boldsymbol{\Omega}_2\right) \frac{\dd \boldsymbol{C}^{-1}}{\dd \tau} \, .
\eea
The structure of $\widetilde{K}$ is similar to the one of $K$: its linear part is composed of the constraints only, and its quadratic part has the same vanishing blocks as $K$.  However, one can now tune the matrix $\boldsymbol{C}$ such that $\widetilde{K}$ has the form given in \Eq{eq:dynsepham}, which amounts to canceling the blocks $\widetilde{\boldsymbol{k}}_{\mathrm{cc}}$, $\widetilde{\boldsymbol{k}}_{\mathrm{cg}}$, and $\widetilde{\boldsymbol{k}}_{\mathrm{cp}}$ in the Hamiltonian kernel while the block $\widetilde{\boldsymbol{k}}_{\mathrm{pp}}$ remains unconstrained.

We start by providing the expression of the non-vanishing block,
\bea
	\widetilde{\boldsymbol{k}}_{\mathrm{pp}}=\boldsymbol{m}_{\mathrm{p}}^{-1\,\mathrm{T}}\boldsymbol{k}_{\mathrm{pp}}\boldsymbol{m}_{\mathrm{p}}^{-1}+\boldsymbol{m}_{\mathrm{p}}^{-1\,\mathrm{T}}\boldsymbol{\Omega}_2\frac{\dd\boldsymbol{m}_{\mathrm{p}}^{-1}}{\dd \tau}. \label{eq:tildekphys}
\eea
Since $\boldsymbol{m}_{\mathrm{p}}\in\mathrm{Sp}(2,\mathbb{R})$ and $\boldsymbol{k}_{\mathrm{pp}}$ is symmetric, this is just the usual canonical transformation, internal to the 2-dimensional phase space of physical degrees of freedom. The new Hamiltonian kernel of the physical degrees of freedom, $\widetilde{\boldsymbol{k}}_{\mathrm{pp}}$, is symmetric, and can otherwise assume any form, depending on the choice of $\boldsymbol{m}_{\mathrm{p}}$.

Second, we consider the couplings between constraint and gauge degrees of freedom, which are given by
\bea
	\widetilde{\boldsymbol{k}}_{\mathrm{cg}}=\boldsymbol{m}^{-1\,\mathrm{T}}_{\mathrm{c}}\left(\frac{\dd \boldsymbol{m}_{\mathrm{c}}^\mathrm{T}}{\dd \tau}+\boldsymbol{k}_{\mathrm{cg}}\boldsymbol{m}_{\mathrm{c}}^\mathrm{T}\right).
\eea
This block is set to zero by choosing  $\boldsymbol{m}_{\mathrm{c}}^\mathrm{T}$ to be a non-singular solution of 
\bea
	\frac{\dd \boldsymbol{m}_{\mathrm{c}}^\mathrm{T}}{\dd \tau}+\boldsymbol{k}_{\mathrm{cg}}\boldsymbol{m}_{\mathrm{c}}^\mathrm{T}=\boldsymbol{0}. \label{eq:kcg=0}
\eea
In the following, we will denote by $\boldsymbol{m}^{\mathrm{T}}_\mathrm{sol}$ the matrices that are solutions of \Eq{eq:kcg=0}. They read
\bea
	\boldsymbol{m}_\mathrm{sol}^\mathrm{T}(\tau)=\left(\begin{array}{ccc}
		\ds\frac{1}{\lambda_D(\tau)}\left[\alpha_1-\beta_1H_k(\tau)\right] &\quad\quad\quad & \ds\frac{1}{\lambda_D(\tau)}\left[\alpha_2-\beta_2H_k(\tau)\right] \\
		\ds\frac{\beta_1}{\mu_S(\tau)} &\quad\quad\quad &\ds\frac{\beta_2}{\mu_S(\tau)} 
	\end{array}\right), \label{eq:solmc}
\eea
where 
\bea
	H_k(\tau)=\ds\int^\tau_{\tau_\uin}\dd \tau'\left[\frac{\dot{\lambda}_S}{\mu_S}-\frac{\dot{\mu}_S}{\mu_S^2}\lambda_S-N\left(\frac{k}{v^{1/3}}\right)^2\right] ,
\eea	
with $\alpha_1$, $\alpha_2$, $\beta_1$ and $\beta_2$ being four integration constants satisfying $\alpha_1\beta_2-\alpha_2\beta_1\neq0$ for the matrix to be invertible.\footnote{A simple case is $\alpha_1=\beta_2=1$ and $\alpha_2=\beta_1=0$, which gives an upper triangular matrix reading
\bea
	\boldsymbol{m}_{\mathrm{c}}^\mathrm{T}=\left(\begin{array}{cc}
		\ds\frac{1}{\lambda_D} & \ds-\frac{H_k(t)}{\lambda_D} \\
		\ds0 & \ds\frac{1}{\mu_S} 
	\end{array}\right).
\eea}  
We note that, in full generality, $H_k$ is defined up to an integration constant that can be absorbed in the constants $\alpha_\mu$. This shows that the coupling between constraint and gauge degrees of freedom can systematically be canceled out and this only requires the knowledge of $\boldsymbol{k}_{\mathrm{cg}}$.

Third, we turn our attention to the block leading to quadratic contributions of the constraints, \ie $\widetilde{\boldsymbol{k}}_{\mathrm{cc}}$. This block reads
\bea
	\widetilde{\boldsymbol{k}}_{\mathrm{cc}}=-\boldsymbol{m}^{-1\,\mathrm{T}}_{\mathrm{c}}\boldsymbol{k}_{\mathrm{cg}}\boldsymbol{n}^\mathrm{T}-\boldsymbol{n}\boldsymbol{k}_{\mathrm{cg}}^\mathrm{T}\boldsymbol{m}^{-1}_{\mathrm{c}}-\boldsymbol{m}^{-1\,\mathrm{T}}_{\mathrm{c}}\dot{\boldsymbol{n}}^\mathrm{T}+\boldsymbol{n}\dot{(\boldsymbol{m}^{-1}_{\mathrm{c}})}+\boldsymbol{r}, \label{eq:exptildekcc}
\eea
where 
\bea
	\boldsymbol{r}&=&\boldsymbol{m}^{-1\,\mathrm{T}}_{\mathrm{c}}\left(\boldsymbol{k}_{\mathrm{cc}}-\boldsymbol{k}_{\mathrm{cp}}\boldsymbol{m}^{-1}_{\mathrm{p}}\boldsymbol{a}-\boldsymbol{a}^\mathrm{T}\boldsymbol{m}^{-1\,\mathrm{T}}_{\mathrm{p}}\boldsymbol{k}_{\mathrm{cp}}^\mathrm{T}+\boldsymbol{a}^\mathrm{T}\boldsymbol{m}^{-1\,\mathrm{T}}_{\mathrm{p}}\boldsymbol{k}_{\mathrm{pp}}\boldsymbol{m}^{-1}_{\mathrm{p}}\boldsymbol{a}\right)\boldsymbol{m}^{-1}_{\mathrm{c}} \nonumber \\
	&&+\boldsymbol{m}^{-1\,\mathrm{T}}_{\mathrm{c}}\boldsymbol{a}^\mathrm{T}\boldsymbol{m}^{-1\,\mathrm{T}}_{\mathrm{p}}\boldsymbol{\omega}\frac{\dd}{\dd t}{(\boldsymbol{m}^{-1}_{\mathrm{p}}\boldsymbol{a}\boldsymbol{m}^{-1}_{\mathrm{c}})}\, .
	\label{eq:r:def}
\eea
We now suppose that $\boldsymbol{m}_{\mathrm{c}}$ is chosen to cancel $\widetilde{\boldsymbol{k}}_{\mathrm{cg}}$. By further using $\dot{(\boldsymbol{m}^{-1}_{\mathrm{c}})}=-\boldsymbol{m}^{-1}_{\mathrm{c}}\dot{(\boldsymbol{m}_{\mathrm{c}})}\boldsymbol{m}^{-1}_{\mathrm{c}}$ and making use of \Eq{eq:kcg=0}, one easily obtains that $\widetilde{\boldsymbol{k}}_{\mathrm{cc}}$ is set equal to zero providing that we choose $\boldsymbol{n}$ such that
\bea
	\dot{\boldsymbol{n}}^\mathrm{T}+\boldsymbol{k}_{\mathrm{cg}}\boldsymbol{n}^\mathrm{T}&=&\boldsymbol{m}^\mathrm{T}_\mathrm{sol}\boldsymbol{r}_\mathrm{sol},
\eea
where $\boldsymbol{r}_\mathrm{sol}$ stands for the expression~\eqref{eq:r:def} where $\boldsymbol{m}_{\mathrm{c}}$ has been replaced by $\boldsymbol{m}_\mathrm{sol}$. The general solution is easily built from the solutions of \Eq{eq:kcg=0}. It is given by
\bea
	\boldsymbol{n}^{\mathrm{T}}_{\mathrm{sol}}(\tau)=\boldsymbol{m}^\mathrm{T}_\mathrm{sol}(\tau)\,\left[\ds\int^\tau\dd\tau'\boldsymbol{r}_\mathrm{sol}(\tau')\right],
\eea
where $\boldsymbol{m}^\mathrm{T}_\mathrm{sol}$ is given by \Eq{eq:solmc}. Because of the source matrix $\boldsymbol{r}_\mathrm{sol}$, the matrix $\boldsymbol{n}_{\mathrm{sol}}$ depends on $\boldsymbol{a}$ and $\boldsymbol{m}_{\mathrm{p}}\in\mathrm{Sp}(2,\mathbb{R})$ which have not been specified yet. However the above solution exists irrespectively of the choice of these two matrices. This shows that canceling the terms quadratic in the constraints is systematically feasible provided that the coupling between constraints and gauge degrees of freedom is  canceled at the same time.

Fourth, we consider the couplings between the constraints and the physical degrees of freedom, which after a bit of algebra are given by
\bea
	\widetilde{\boldsymbol{k}}_{\mathrm{cp}}&=&\boldsymbol{m}^{-1\,\mathrm{T}}_{\mathrm{c}}\left[\dot{\boldsymbol{a}}^\mathrm{T}\boldsymbol{\Omega}_2-\boldsymbol{a}^\mathrm{T}\boldsymbol{m}_{\mathrm{p}}^{-1\,\mathrm{T}}\boldsymbol{\Omega}_2\dot{(\boldsymbol{m}^{-1}_{\mathrm{p}})}+\boldsymbol{k}_{\mathrm{cg}}\boldsymbol{a}^\mathrm{T}\boldsymbol{\Omega}_2-\boldsymbol{a}^\mathrm{T}\boldsymbol{m}^{-1\,\mathrm{T}}_{\mathrm{p}}\boldsymbol{k}_{\mathrm{pp}}\boldsymbol{m}^{-1}_{\mathrm{p}}+\boldsymbol{k}_{\mathrm{cp}}\boldsymbol{m}^{-1}_{\mathrm{p}}\right]. \nonumber \\
\eea
This block can be set to zero if the matrix $\boldsymbol{a}$ is solution of the following equation
\bea
	\dot{\boldsymbol{a}}^\mathrm{T}+\boldsymbol{k}_{\mathrm{cg}}\boldsymbol{a}^\mathrm{T}+\boldsymbol{a}^\mathrm{T}\left[\boldsymbol{m}_{\mathrm{p}}^{-1\,\mathrm{T}}\boldsymbol{\Omega}_2\dot{(\boldsymbol{m}^{-1}_{\mathrm{p}})}+\boldsymbol{m}^{-1\,\mathrm{T}}_{\mathrm{p}}\boldsymbol{k}_{\mathrm{pp}}\boldsymbol{m}^{-1}_{\mathrm{p}}\right]\boldsymbol{\Omega}_2=\boldsymbol{k}_{\mathrm{cp}}\boldsymbol{m}^{-1}_{\mathrm{p}}\boldsymbol{\Omega}_2. \label{eq:kcp=0}
\eea
To solve the above, we write $\boldsymbol{a}^\mathrm{T}=\boldsymbol{m}_\mathrm{sol}^\mathrm{T}\boldsymbol{\mathfrak{a}}^\mathrm{T}$. Since $\boldsymbol{m}_\mathrm{sol}^\mathrm{T}$ is solution of \Eq{eq:kcg=0}, it yields the following differential equation for $\boldsymbol{\mathfrak{a}}$,
\bea
	\dot{\boldsymbol{\mathfrak{a}}}^\mathrm{T}+\boldsymbol{\mathfrak{a}}^\mathrm{T}\widetilde{\boldsymbol{k}}_{\mathrm{pp}}\boldsymbol{\Omega}_2=\boldsymbol{m}^{-1\,\mathrm{T}}_\mathrm{sol}\boldsymbol{k}_{\mathrm{cp}}\boldsymbol{m}^{-1}_{\mathrm{p}}\boldsymbol{\Omega}_2, \label{eq:diffafrak}
\eea
where we further make use of \Eq{eq:tildekphys}. The homogeneous equation, $\dot{\boldsymbol{\mathfrak{a}}}^\mathrm{T}+\boldsymbol{\mathfrak{a}}^\mathrm{T}\widetilde{\boldsymbol{k}}_{\mathrm{pp}}\boldsymbol{\Omega}_2=0$, describes standard Hamiltonian dynamics since $\widetilde{\boldsymbol{k}}_{\mathrm{pp}}$ is symmetric. Hence it necessarily has solutions given by matrices of the symplectic group $\mathrm{Sp}(2,\mathbb{R})$ (see \eg \Refc{Grain:2019vnq}). On denoting $\boldsymbol{\mathfrak{a}}_\mathrm{hom}(\tau)$ a solution of the homogeneous equation, the solutions of the inhomogeneous ones are given by
\bea
	\boldsymbol{\mathfrak{a}}^\mathrm{T}_{\mathrm{sol}}(\tau)=\left[\ds\int^\tau\dd\tau'\boldsymbol{m}^{-1\,\mathrm{T}}_\mathrm{sol}\boldsymbol{k}_{\mathrm{cp}}\boldsymbol{m}^{-1}_{\mathrm{p}}\boldsymbol{\Omega}_2\boldsymbol{\mathfrak{a}}^{-1\,\mathrm{T}}_\mathrm{hom}(\tau')\right] \boldsymbol{\mathfrak{a}}^\mathrm{T}_\mathrm{hom}(\tau).
\eea
This finalises the proof that there exists a matrix $\boldsymbol{a}$ ensuring the couplings between the constraints and the physical degrees of freedom to be set to zero. The cancellation of $\widetilde{\boldsymbol{k}}_{\mathrm{cp}}$ is thus always possible. The final solution depends on $\boldsymbol{m}_{\mathrm{p}}\in\mathrm{Sp}(2,\mathbb{R})$ but the solution always exists irrespectively of that choice.  We note that despite of having used $\boldsymbol{m}_\mathrm{sol}$ to find a solution to $\widetilde{\boldsymbol{k}}_{\mathrm{cp}}=0$, it is not required to cancel the couplings between constraints and gauge degrees of freedom since the differential equation for $\boldsymbol{a}$ does not depend on $\boldsymbol{m}_{\mathrm{c}}$. This is unlike canceling $\widetilde{\boldsymbol{k}}_{\mathrm{cc}}$ for which we had to set $\boldsymbol{m}_{\mathrm{c}}$ to $\boldsymbol{m}_\mathrm{sol}$ in order to make the coupling between constraint and gauge degrees of freedom vanish. This means that canceling the block $\widetilde{\boldsymbol{k}}_{\mathrm{cp}}$ can be done independently of canceling the blocks $\widetilde{\boldsymbol{k}}_{\mathrm{cc}}$ and $\widetilde{\boldsymbol{k}}_{\mathrm{cg}}$. However, it is worth mentioning that to get the expression for $\boldsymbol{a}$ that cancels the couplings between the physical degrees of freedom and the constraints, it is needed to solve the homogeneous part of \Eq{eq:diffafrak}. Though we have argued that such an equation has solutions, we have not shown any concrete example of them (see the discussion below in \Sec{sssec:eomdynphysbas}).

We finally stress that the process of making $\widetilde{\boldsymbol{k}}_{\mathrm{cg}}$, $\widetilde{\boldsymbol{k}}_{\mathrm{cc}}$, and $\widetilde{\boldsymbol{k}}_{\mathrm{cp}}$ equal to zero can be done irrespectively of the choice of the matrix $\boldsymbol{m}_{\mathrm{p}}$ which only generates a canonical transformation in the 2-dimensional sub-phase space of the physical degrees of freedom.

\subsection{Kinematics} 

The canonical transformation generated by $\boldsymbol{C}$ allows one to define a new set of basis vectors on which solutions can be decomposed, and that we will call the {\textit{\Dbasis}}. On denoting its vectors by $\left\{\vec{v}_a\right\}_{a\in\{0,1,2,3,4,5\}}$, the \Dbasis is related to the \Pbasis $\{\vec{e}_a\}$ by
\bea
	\vec{v}_a=\ds\sum_{b=0}^5\left[\boldsymbol{C}^{-1\,\mathrm{T}}\right]_{ab}\,\vec{e}_b.
\eea
Although the \Pbasis is orthonormal, the matrix $\boldsymbol{C}$ is not an orthogonal matrix. This implies that the \Dbasis is not an orthonormal basis. As a consequence, the different degrees of freedom in the \Dbasis cannot be extracted out of $ \dz $ by projecting it onto the $\vec{v}_a$'s. Instead, its components are given by 
\bea
	\left[\widetilde{\vec{z}}_{\vec{k}}\right]_a=\ds\sum_{b=0}^5\left[\boldsymbol{C}\right]_{ab}\left(\vec{e}_b\cdot \dz \right).
\eea
On denoting 
\bea
	\vec{w}_a=\ds\sum_{b=0}^5[\boldsymbol{C}]_{ab}\,\vec{e}_b\,,
\eea
any solution is decomposed on the \Dbasis as
\bea
	 \dz =\ds\sum_{a}\left(\vec{w}_a\cdot \dz \right)\,\vec{v}_a. \label{eq:soldecdyn}
\eea

Let us briefly discuss the new variables that are obtained by projecting $ \dz $ onto the $\vec{w}_a$'s. Using \Eqs{eq:C:def} and~\eqref{eq:solmc}, one can readily show that the new constraints are given by $\widetilde{Q}_\mu=\vec{w}^{\,\mathcal{C}}_\mu\cdot \dz $ where the two vectors $\vec{w}^{\,\mathcal{C}}_\mu$ are
\bea
	\vec{w}^{\,\mathcal{C}}_\mu=\beta_\mu\vec{S}_k+\left[\alpha_\mu+\beta_\mu\left(\frac{\lambda_S}{\mu_S}-H_k\right)\right]\vec{D}_k.
\eea
These two vectors belong to the plane of constraints. However, since $\alpha_\mu$ and $\beta_\mu$ must satisfy $\alpha_1\beta_2-\alpha_2\beta_1\neq1$, it is not possible to find a choice of these four parameters such that \eg $\vec{w}^{\,\mathcal{C}}_1=\vec{D}_k$ and $\vec{w}^{\,\mathcal{C}}_2=\vec{S}_k$. The new Lagrange multipliers associated to these constraints are
\bea
	\left(\alpha_1\beta_2-\alpha_2\beta_1\right)\,\widetilde{\Lambda}_\mu=\frac{\beta_\mu}{\lambda_D}\,k\delta N_1-\frac{1}{\lambda_D}\left[\alpha_\mu+\beta_\mu\left(H_k+\frac{\lambda_S}{\mu_S}\right)\right]\,\delta N.
\eea

For the new gauge degrees of freedom, the vectors are
\bea
	\vec{w}^{\,\mathcal{G}}_\mu=\ds\sum_{\mu'=1}^2\left\{\left[\boldsymbol{m}^{-1\,\mathrm{T}}_\mathrm{sol}\right]_{\mu\mu'}\,\vec{e}^{\,\mathcal{G}}_{\mu'}+\left[\boldsymbol{n}_\mathrm{sol}\right]_{\mu\mu'}\,\vec{e}^{\,\mathcal{C}}_{\mu'}-\left[\boldsymbol{m}^{-1\,\mathrm{T}}_\mathrm{sol}\boldsymbol{a}^\mathrm{T}_\mathrm{sol}\boldsymbol{\Omega}_2\boldsymbol{m}_{\mathrm{p}}\right]_{\mu\mu'}\,\vec{e}^{\,\mathcal{P}}_{\mu'}\right\}.
\eea
The first term guarantees that the new gauge degrees of freedom are canonically conjugated to the new constraints. The second and third terms guarantee the constraints and the physical degrees of freedom, respectively, to have vanishing contribution in the equations of motion of the gauge degrees of freedom.

Finally, the physical degrees of freedom are obtained by projecting onto the following vectors
\bea
	\vec{w}^{\,\mathcal{P}}_\mu=\ds\sum_{\mu'=1}^2\left\{\left[\boldsymbol{m}_{\mathrm{p}}\right]_{\mu\mu'}\,\vec{e}^{\,\mathcal{P}}_{\mu'}+\left[\boldsymbol{a}_\mathrm{sol}\right]_{\mu\mu'}\,\vec{e}^{\,\mathcal{C}}_{\mu'}\right\}.
\eea
The first term corresponds to a canonical transformation internal to the 2-dimensional phase space of physical degrees of freedom, while the second term ensures that constraints no longer contribute to the dynamics of the physical degrees of freedom.

\subsection{Equations of motion}
\label{sssec:eomdynphysbas}

The procedure detailed above leads to a Hamiltonian of the desired form~\eqref{eq:dynsepham}, from which the equations of motion for the new set of variables are easily deduced. They read
\bea
	\dot{\widetilde{Q}}_\mu&=&0, \label{eq:tildeCeom} \\
	\dot{\widetilde{P}}_\mu&=&-\widetilde{\Lambda}_\mu, \label{eq:tildeGeom} \\
	\left(\begin{array}{c}
		\dot{\widetilde{Z}}_1 \\
		\dot{\widetilde{Z}}_2
	\end{array}\right)&=&\boldsymbol{\Omega}_2\widetilde{\boldsymbol{k}}_{\mathrm{pp}}\left(\begin{array}{c}
		\widetilde{Z}_1 \\
		\widetilde{Z}_2
	\end{array}\right), \label{eq:tildePeom}
\eea
where $\mu$ runs over 1, 2. The dynamical equations for the gauge degrees of freedom and the physical degrees of freedom now hold whether one solves the entire system on the surface of constraints or not. We further stress that the gauge degrees of freedom are no longer sourced by the physical ones. Their time-derivatives are only determined by the Lagrange multipliers, showing that these degrees of freedom are totally arbitrary and carry no physical information.

The unphysical sector is trivially solved and gives 
\bea
	\widetilde{Q}_\mu(\tau)=\mathrm{constant},
\eea
which is zero if initial conditions are selected to lie on the surface of constraints, and 
\bea
	\widetilde{P}_\mu(\tau)=\widetilde{P}_\mu(\tau_\uin)+\ds\int^\tau_{\tau_\uin}\dd\tau'\widetilde{\Lambda}_\mu(\tau')\, ,
\eea
 where $\widetilde{P}_\mu(\tau_\uin)$ are arbitrary initial conditions for the gauge parameters. However, generic solutions cannot be found for the physical sector since this is where all the nontrivial dynamics enters. 
 
 We finally mention that if two linearly independent solutions of the dynamical equations of the physical degrees of freedom are found, one can built the fundamental matrix, $\boldsymbol{u}_\mathrm{phys}$, which is solution of 
 \bea
 	\dot{\boldsymbol{u}}_\mathrm{phys}-\boldsymbol{\Omega}_2\widetilde{\boldsymbol{k}}_{\mathrm{pp}}\boldsymbol{u}_\mathrm{phys}=\boldsymbol{0}.
\eea
This is the same equation as the transpose of the homogeneous part of \Eq{eq:diffafrak}, whose solutions are needed to cancel the couplings between the physical degrees of freedom and the constraints. As a consequence, finding the part of the canonical transformation that cancels $\widetilde{\boldsymbol{k}}_{\mathrm{cp}}$ is equivalent to (and as difficult as) solving for the physical degrees of freedom $(\widetilde{Z}_1,\widetilde{Z}_2)$.\footnote{One might be tempted to look for a specific choice of $\boldsymbol{m}_{\mathrm{p}}$that simplifies the resolution of \Eq{eq:diffafrak}. Up to our investigations, there is no such general choice for the matrix $\boldsymbol{m}_{\mathrm{p}}$. A first example consists in setting $\boldsymbol{m}_{\mathrm{p}}=\boldsymbol{I}_2$. The equation of motion for $\boldsymbol{\mathfrak{a}}$ is then given by \Eq{eq:diffafrak} where $\widetilde{\boldsymbol{k}}_{\mathrm{pp}}$ is replaced by $\boldsymbol{k}_{\mathrm{pp}}$. Hence solving for $\boldsymbol{\mathfrak{a}}$ is as complicated as solving for the dynamics of $(Z_1,Z_2)$. A second example consists in choosing $\boldsymbol{m}_{\mathrm{p}}$ such that
\bea
	\dot{(\boldsymbol{m}^{-1}_{\mathrm{p}})}-\boldsymbol{\Omega}_2\boldsymbol{k}_{\mathrm{pp}}\boldsymbol{m}^{-1}_{\mathrm{p}}=\boldsymbol{0}.
\eea
We denote solutions of the above by $\boldsymbol{m}_{\mathrm{p},\mathrm{sol}}^{-1}$. The differential equation driving $\boldsymbol{a}$, \Eq{eq:kcp=0}, then simplifies to
\bea
	\dot{\boldsymbol{a}}^\mathrm{T}+\boldsymbol{k}_{\mathrm{cg}}\boldsymbol{a}^\mathrm{T}=\boldsymbol{k}_{\mathrm{cp}}\boldsymbol{m}^{-1}_{\mathrm{p},\mathrm{sol}}\boldsymbol{\Omega}_2,
\eea
which is solved as
\bea
	\boldsymbol{a}^\mathrm{T}(\tau)=\boldsymbol{m}^\mathrm{T}_{\mathrm{sol}}(\tau)\,\left[\ds\int^\tau\dd\tau'\boldsymbol{m}^{-1\,\mathrm{T}}_{\mathrm{sol}}\boldsymbol{k}_{\mathrm{cp}}\boldsymbol{m}^{-1}_{\mathrm{p},\mathrm{sol}}\boldsymbol{\Omega}_2\right].
\eea
However, finding $\boldsymbol{m}_{\mathrm{p},\mathrm{sol}}$ is as complicated as finding $\boldsymbol{\mathfrak{a}}_\mathrm{hom}$ when setting  $\boldsymbol{m}_{\mathrm{p}}=\boldsymbol{I}_2$.}

\subsection{Discussion}

Let us further compare the \Dbasis and the \Pbasis.  Any solution can be decomposed on the \Dbasis using \Eq{eq:soldecdyn}, in which two sets of vectors need to be introduced, $\vec{v}_a$ and $\vec{w}_a$. This has to be contrasted with the decomposition on the \Pbasis,
\bea
	 \dz =\ds\sum_{a}\left(\vec{e}_a\cdot \dz \right)\,\vec{e}_a,
\eea
which relies on one set of vectors only, thanks to the orthonormality of the \Pbasis. In that respect, the use of the \Pbasis is easier.

Obviously, the real advantage of the \Dbasis is at the level of the equations of motion. However, this supposes that one is able to find the matrix $\boldsymbol{a}_\mathrm{sol}$, which boils down to a dynamical problem equivalent to solving for the physical degrees of freedom. In practice then, the use of the \Dbasis does not really simplify the problem. 

Nevertheless, it is still practically possible to set $\boldsymbol{a}$ equal to the null matrix, making $\widetilde{\boldsymbol{k}}_{\mathrm{cp}}\neq\boldsymbol{0}$  but still ensuring that $\widetilde{\boldsymbol{k}}_{\mathrm{cc}}$ and $\widetilde{\boldsymbol{k}}_{\mathrm{cg}}$ are vanishing. Fixing $\boldsymbol{a}=\boldsymbol{0}$  leads to $\boldsymbol{b}=\boldsymbol{0}$. In this case the matrix $\boldsymbol{C}$ is the direct sum of two matrices, \ie $\boldsymbol{C}=\boldsymbol{M}_\mathrm{unphys}\oplus\boldsymbol{m}_{\mathrm{p}}$ where $\boldsymbol{m}_{\mathrm{p}}\in\mathrm{Sp}(2,\mathbb{R})$ generates a canonical transformation in $\mathcal{P}$, and $\boldsymbol{M}_\mathrm{unphys}\in\mathrm{Sp}(4,\mathbb{R})$ generates a canonical transformation in $\mathcal{C}\otimes\mathcal{G}$.  This leads to the following equations of motion
\bea
	\dot{\widetilde{Q}}_\mu&=&0, \label{eq:tildeCappeom}\\
	\dot{\widetilde{P}}_\mu&=&-\widetilde{\Lambda}_\mu-\left[\widetilde{\boldsymbol{k}}_{\mathrm{cp}}\right]_{\mu,1}\widetilde{Z}_1-\left[\widetilde{\boldsymbol{k}}_{\mathrm{cp}}\right]_{\mu,2}\widetilde{Z}_2, \label{eq:tildeGappeom}\\
	\left(\begin{array}{c}
		\dot{\widetilde{Z}}_1 \\
		\dot{\widetilde{Z}}_2
	\end{array}\right)&\approx&\boldsymbol{\Omega}_2\widetilde{\boldsymbol{k}}_{\mathrm{pp}}\left(\begin{array}{c}
		\widetilde{Z}_1 \\
		\widetilde{Z}_2
	\end{array}\right). \label{eq:tildePappeom}
\eea
First, this has to be compared with the equations of motion in the \Dbasis, \Eqs{eq:tildeCeom}, \eqref{eq:tildeGeom} and~\eqref{eq:tildePeom}. The dynamics of the constraints remains unchanged but the gauge degrees of freedom receive contributions from the physical ones. The dynamics of physical degrees of freedom remains unaffected at the condition that it is solved on the surface of constraints (which, we recall, is the meaning of the symbol ``$\approx$''). Second, the above has to be compared with the equations of motion~\eqref{eq:EOMphysgen} in the \Pbasis. Using the above compact notation, we remind that they are given by
\bea
	\dot{Q}_\mu&=&\ds\sum_{\mu'=1}^2\left[\boldsymbol{k}_{\mathrm{cg}}\right]_{\mu\mu'}\,Q_{\mu'}, \label{eq:Ceom} \\
	\dot{P}_\mu&\approx&-{\Lambda}_\mu-\left[{\boldsymbol{k}}_{\mathrm{cp}}\right]_{\mu,1}{Z}_1-\left[{\boldsymbol{k}}_{\mathrm{cp}}\right]_{\mu,2}{Z}_2, \label{eq:Geom} \\
	\left(\begin{array}{c}
		\dot{{Z}}_1 \\
		\dot{{Z}}_2
	\end{array}\right)&\approx&\boldsymbol{\Omega}_2{\boldsymbol{k}}_{\mathrm{pp}}\left(\begin{array}{c}
		{Z}_1 \\
		{Z}_2
	\end{array}\right). \label{eq:Peom}
\eea
The dynamics of the constraints is now affected but it remains trivially solved. The gauge degrees of freedom have a similar dynamics provided that they are now solved on the surface of constraints. Finally, the physical degrees of freedom have an identical dynamics up to a canonical transformation internal to the 2-dimensional phase space of physical degrees of freedom. Indeed, it is possible to cast the equation of motion~\eqref{eq:Peom} in the form of \Eq{eq:tildePappeom} without changing the structure of \Eqs{eq:Ceom} and~\eqref{eq:Geom}, by introducing the canonical transformation $\vec{z}\to(\boldsymbol{I}_4\oplus\boldsymbol{m}_{\mathrm{p}})\vec{z}$, which leaves the space of constraints and of gauge degrees of freedom unchanged while the phase space of physical degrees of freedom is transformed along the internal canonical transformation generated by $\boldsymbol{m}_{\mathrm{p}}\in\mathrm{Sp}(2,\mathbb{R})$. 

\section{Lagrange multipliers in the \Pbasis}
\label{app:LagMultKin}

Once the gauge is fixed, the Lagrange multipliers can be expressed as functions of the physical degrees of freedom only. To this end, we decompose the gauge conditions, $\vec{G}_i\cdot \dz =0$, and \Eq{eq:LMfix}, on the \Pbasis. To lighten the calculation, we will directly work on the surface of constraints. Hence the two relevant planes are the plane of gauge degrees of freedom and the plane of physical degrees of freedom.

Let us first introduce the matrix made from the components of the gauge vectors in the plane of physical degrees of freedom, \ie
\bea
    \boldsymbol{G}_{\mathrm{p}}=\left(\begin{array}{cc}
        \vec{G}_1\cdot\vec{e}^{\,\mathcal{P}}_1 & \vec{G}_1\cdot\vec{e}^{\,\mathcal{P}}_2 \\
        \vec{G}_2\cdot\vec{e}^{\,\mathcal{P}}_1 & \vec{G}_2\cdot\vec{e}^{\,\mathcal{P}}_2
    \end{array}\right) .
\eea
The two gauge conditions can be decomposed on the \Pbasis and further expressed as an equality between vectors:
\bea
    \boldsymbol{G}\left(\begin{array}{c}
        P_1 \\
        P_2
    \end{array}\right)+\boldsymbol{G}_{\mathrm{p}}\left(\begin{array}{c}
        Z_1 \\
        Z_2
    \end{array}\right)\approx\boldsymbol{0},
\eea
where $\boldsymbol{G}$ is defined in \Eq{eq:gaugephyssystem} and is built from the components of the gauge vectors in the plane of gauge degrees of freedom. Since non-pathological gauges are considered here, the matrix $\boldsymbol{G}$ is invertible. This yields
\bea
    \left(\begin{array}{c}
        P_1 \\
        P_2
    \end{array}\right)\approx-\boldsymbol{G}^{-1}\boldsymbol{G}_{\mathrm{p}}\left(\begin{array}{c}
        Z_1 \\
        Z_2
    \end{array}\right),
\eea
which can be used to replace the gauge degrees of freedom by the physical degrees of freedom in \Eq{eq:LMfix}.

Decomposing now the right-hand side of \Eq{eq:LMfix} on the \Pbasis leads to
\bea
    \left(\begin{array}{c}
        \boldsymbol{\nabla}_\tau\vec{G}_1\cdot \dz  \\
        \boldsymbol{\nabla}_\tau\vec{G}_2\cdot \dz  
    \end{array}\right)\approx\left(\dot{\boldsymbol{G}}+\boldsymbol{GK}_1+\boldsymbol{G}_{\mathrm{p}}\boldsymbol{K}_2\right)\left(\begin{array}{c}
        P_1 \\
        P_2
    \end{array}\right)+\left(\dot{\boldsymbol{G}}_{\mathrm{p}}+\boldsymbol{GK}_3+\boldsymbol{G}_{\mathrm{p}}\boldsymbol{K}_4\right)\left(\begin{array}{c}
        Z_1 \\
        Z_2
    \end{array}\right),
\eea
where the matrices $\boldsymbol{K}_i$ read
\bea
    \left[\boldsymbol{K}_1\right]_{\mu\nu}&=&\vec{e}^{\,\mathcal{G}}_\nu\cdot\left(\boldsymbol{\nabla}_\tau\vec{e}^{\,\mathcal{G}}_\mu\right), \\
    \left[\boldsymbol{K}_2\right]_{\mu\nu}&=&\vec{e}^{\,\mathcal{G}}_\nu\cdot\left(\boldsymbol{\nabla}_\tau\vec{e}^{\,\mathcal{P}}_\mu\right), \\
    \left[\boldsymbol{K}_3\right]_{\mu\nu}&=&\vec{e}^{\,\mathcal{P}}_\nu\cdot\left(\boldsymbol{\nabla}_\tau\vec{e}^{\,\mathcal{G}}_\mu\right), \\
    \left[\boldsymbol{K}_4\right]_{\mu\nu}&=&\vec{e}^{\,\mathcal{P}}_\nu\cdot\left(\boldsymbol{\nabla}_\tau\vec{e}^{\,\mathcal{P}}_\mu\right).
\eea    
These matrices are easily related to the elements of the Hamiltonian in the \Pbasis and we will use the notations introduced in \Sec{ssec:dynPhys}. First, $\vec{e}^{\,\mathcal{G}}_\mu=-\boldsymbol{\Omega}_6\vec{e}^{\,\mathcal{C}}_\mu$, hence $\boldsymbol{K}_1$ is related to the couplings between the constraints and the gauge degrees of freedom, \ie  $\boldsymbol{K}_1=-\boldsymbol{k}_{\mathrm{cg}}$. Second, $\vec{e}^{\,\mathcal{P}}_2=-\boldsymbol{\Omega}_6\vec{e}^{\,\mathcal{P}}_1$ (which also leads to $\vec{e}^{\,\mathcal{P}}_1=\boldsymbol{\Omega}_6\vec{e}^{\,\mathcal{P}}_2$), hence $\boldsymbol{K}_2$ is given by the couplings between the gauge degrees of freedom and the physical degrees of freedom, which have been shown to vanish, \ie $\boldsymbol{K}_2=\boldsymbol{0}$. Third, the matrix $\boldsymbol{K}_3$ is shown to be given by the couplings between the constraints and the physical degrees of freedom, \ie $\boldsymbol{K}_3=-\boldsymbol{k}_{\mathrm{cg}}$. Fourth, we obtain that the last matrix is related to the couplings internal to the physical degrees of freedom, \ie $\boldsymbol{K}_4=\boldsymbol{\Omega}_2 \boldsymbol{k}_{\mathrm{pp}}$.

Gathering the above results, it is straightforward to rewrite \Eq{eq:LMfix} as
\bea
    \boldsymbol{G}_{\mathrm{LM}}\left(\begin{array}{c}
        \delta N \\
        k\delta N_1
    \end{array}\right)\approx\boldsymbol{N}\left(\begin{array}{c}
        Z_1 \\
        Z_2
    \end{array}\right),
\eea
where
\bea
    \boldsymbol{N}=\left(\dot{\boldsymbol{G}}-\boldsymbol{G}\boldsymbol{k}_{\mathrm{cg}}\right)\boldsymbol{G}^{-1}\boldsymbol{G}_{\mathrm{p}}-\left(\dot{\boldsymbol{G}}_{\mathrm{p}}+\boldsymbol{G}_{\mathrm{p}}\boldsymbol{\Omega}_2 \boldsymbol{k}_{\mathrm{pp}}\right)+\boldsymbol{G k}_{\mathrm{cg}}.
\eea
We note that the above expression relating the Lagrange multipliers to the physical degrees of freedom holds providing that the gauge is fixed.            

\section{Non-pathological synchronous gauge}
\label{app:synchronous}

In this appendix, we show how to define a non-pathological gauge in which $\delta N$ and $\delta N_1$ are both vanishing. To this end, we start from \Eq{eq:LMphysdof} and identify the conditions that $\boldsymbol{G}$ and $\boldsymbol{G}_{\mathrm{p}}$ should satisfy to make $\boldsymbol{N}=\boldsymbol{0}$, which ensures that the two Lagrange multipliers vanish on-shell. Since the matrices $\boldsymbol{G}$ and $\boldsymbol{G}_{\mathrm{p}}$ are built from the components on the gauge vectors in the planes $\mathcal{G}$ and $\mathcal{P}$ respectively, they entirely define the two gauge vectors, hence the gauge choice. We note that in this identification, it is needed to impose the matrix $\boldsymbol{G}$ to be non-singular for the gauge to be non-pathological.

Sufficient conditions for $\boldsymbol{N}$ to equal the null matrix are
\bea
    &&\dot{\boldsymbol{G}}-\boldsymbol{G}\boldsymbol{k}_{\mathrm{cg}}=\boldsymbol{0}, \label{eq:Gsynch} \\
    &&\dot{\boldsymbol{G}}_{\mathrm{p}}+\boldsymbol{G}_{\mathrm{p}}\boldsymbol{\Omega}_2 \boldsymbol{k}_{\mathrm{pp}}=\boldsymbol{G k}_{\mathrm{cg}}. \label{eq:Gpsynch}
\eea
The first equation is easily solved and general solutions read
\bea
    \boldsymbol{G}_\mathrm{sol}=\boldsymbol{m}^{\mathrm{T}\,-1}_\mathrm{sol} \label{eq:Gsolsynch}\, ,
\eea
where $\boldsymbol{m}^{\mathrm{T}}_\mathrm{sol}$ is non-singular (see \App{ssec:dynPhys}).\footnote{We use the fact that if $\boldsymbol{M}$ is solution of $\dd{\boldsymbol{M}}/\dd\tau+\boldsymbol{A}(\tau)\boldsymbol{M}=\boldsymbol{0}$, then $\boldsymbol{M}^{-1}$ is solution of $\dd\boldsymbol{M}^{-1}/\dd\tau-\boldsymbol{M}^{-1}\boldsymbol{A}(\tau)=\boldsymbol{0}$. It is straightforwardly shown by noting that $\dd\boldsymbol{M}^{-1}/\dd\tau=-\boldsymbol{M}^{-1}(\dd\boldsymbol{M}/\dd\tau)\boldsymbol{M}^{-1}$. Similarly, if $\boldsymbol{M}$ is solution of $\dd{\boldsymbol{M}}/\dd\tau+\boldsymbol{M}\boldsymbol{A}(\tau)=\boldsymbol{0}$, then $\boldsymbol{M}^{-1}$ is solution of $\dd\boldsymbol{M}^{-1}/\dd\tau-\boldsymbol{A}(\tau)\boldsymbol{M}^{-1}=\boldsymbol{0}$.} We stress that $\boldsymbol{G}_\mathrm{sol}$ is non-singular, hence the gauge is guaranteed to be non-pathological. We also note that $\boldsymbol{G}^{-1}_\mathrm{sol}$ can be interpreted as the fundamental matrix generating the homogeneous evolution of the gauge degrees of freedom, \ie \Eqs{eq:P1dot} and \eqref{eq:P2dot}. The second equation is solved by first solving the homogeneous problem which can be cast as
\bea
    \frac{\dd}{\dd\tau}\boldsymbol{G}_{\mathrm{p}}^{-1}-\boldsymbol{\Omega}_2 \boldsymbol{k}_{\mathrm{pp}}\boldsymbol{G}_{\mathrm{p}}^{-1}=0.
\eea
This equation is a Hamilton equation generated by the Hamiltonian $\boldsymbol{k}_{\mathrm{pp}}$, hence it necessarily has a non-singular solution in the group $\mathrm{Sp}(2,\mathbb{R})$. We note this solution $\boldsymbol{G}_{\mathrm{p},\,\mathrm{hom}}$ and it is no more than the fundamental matrix generating the dynamics of the physical degrees of freedom. Then, the general solutions of \Eq{eq:Gpsynch} are 
\bea
    \boldsymbol{G}_{\mathrm{p},\,\mathrm{sol}}=\left[\int^\tau\dd\tau'\boldsymbol{G}_\mathrm{sol}(\tau')\boldsymbol{k}_{\mathrm{cg}}(\tau')\right]\,\boldsymbol{G}_{\mathrm{p},\,\mathrm{hom}}, \label{eq:Gpsolsynch}
\eea
which finalises the proof that one can set $\boldsymbol{N}=\boldsymbol{0}$ with a non-singular $\boldsymbol{G}$ matrix.

As a consequence, the gauge defined by setting $\boldsymbol{G}$ and $\boldsymbol{G}_{\mathrm{p}}$ to be given by \Eqs{eq:Gsolsynch} and \eqref{eq:Gpsolsynch} is non-pathological and leads to $\delta N\approx0\approx\delta N_1$. In that sense, it can be dubbed as synchronous. However, it is not {\it a priori} synchronised with the homogeneous and isotropic FLRW space-time but rather {\it a posteriori}. Indeed, the Lagrange multipliers are now vanishing on-shell, which is unlike the standard (and pathological way) of defining the synchronous gauge by directly imposing $\delta N=0=\delta N_1$ off-shell. The foliation is synchronised with the FLRW background as a result of synchronising the gauge vectors with the dynamics of the gauge degrees of freedom and the physical degrees of freedom through $\boldsymbol{G}_\mathrm{sol}$ and $\boldsymbol{G}_{\mathrm{p},\,\mathrm{sol}}$, which are given by solutions of the dynamical equations of the cosmological perturbations. Hence, a non-pathological implementation of the synchronous gauge is primarily synchronised with perturbative variables of the cosmological space-time, rather than with background variables. 

It is worth noting that the above-defined gauge corresponds in fact to an entire class of non-pathological synchronous gauges since the matrices $\boldsymbol{G}_\mathrm{sol}$ and $\boldsymbol{G}_{\mathrm{p},\,\mathrm{sol}}$ are both defined up to two constants of integration. Moreover, we have only exhibited a set of solutions which makes $\boldsymbol{N}$ vanish but other solutions may exist, hence other non-pathological synchronous gauges.

Finally, though conceptually interesting, it is fair to say that the class of non-pathological synchronous gauge defined above is likely to be of little practical use. Indeed, the full dynamics of the physical degrees of freedom needs to be solved in order to define such gauges. Because of this, it cannot be used to make the dynamics of cosmological perturbations easier to solve.

\section{Pathology of the uniform-expansion gauge}
\label{app:UnifExpGauge}

We show in this appendix four different implementations of the uniform-expansion gauge, which are all pathological.

\subsubsection*{Implementation 1}

Let us start with the common way of fixing the uniform-expansion gauge. It is based on the expression given in \Eq{eq:NintStandard}, which implies that imposing $\delta\gamma_1=0$ and $\delta N_1=0$ as gauge conditions guarantees the perturbation of the integrated expansion rate to be zero. In our language, it corresponds to
\bea
	\vec{G}_1=\vec{e}^{\,\phi}_1, \,\,\,&\vec{G}_2=\vec{0}\,\,\,&\mathrm{and}\,\,\,\boldsymbol{\lambda}=\left(\begin{array}{cc} 0 & 0 \\ 0 & 1 \end{array}\right).
\eea
The second gauge condition is the same as the second condition in the Newtonian gauge. Thus it leads to a vector $\vec{J}_2$ aligned with $\vec{e}_2^{\,\phi}$. For time derivatives of the gauge degrees of freedom to be removed, the projection of $\vec{e}_2^{\,\phi}$ onto $\mathcal{G}$ has to be aligned with the projection of $\vec{e}^{\,\phi}_1$ onto $\mathcal{G}$. However, the study of the spatially-flat gauge in \Sec{sec:examples:gauges} showed that it is not the case. Hence, the conditions $\delta\gamma_1=0$ and $\delta N_1=0$ lead to a pathological gauge.

\subsubsection*{Implementation 2}

Alternatively, one can start from the Hamiltonian expression of the perturbations of the integrated expansion rate given in \Eq{eq:NintHam}. As a consequence, imposing $\delta N=0=\delta\Theta$ as gauge conditions guarantees a vanishing $\delta\mathcal{N}_\mathrm{int}$. Such a gauge is fixed by introducing 
\bea
	\vec{G}_1=-\frac{\theta}{4}\,\vec{e}^{\,\phi}_1+v^{1/3}\,\vec{e}^{\,\pi}_1,\,\,\,&\vec{G}_2=\vec{0}\,\,\,&\mathrm{and}\,\,\,\boldsymbol{\lambda}=\left(\begin{array}{cc}
		0 & 0 \\
		1 &0
	\end{array}\right).
\eea
The rank of $\boldsymbol{\lambda}$ equals 1 and we write the second gauge condition as $\vec{G}'_2\cdot\dz+\vec{J}_2\cdot\dzdot=0$, where
 \bea
 	\vec{G'}_2&=&N \left(\frac{\sqrt{3}}{2}\frac{1}{v^{2/3}}\,\vec{e}^{\,\phi}_1 -\frac{1}{\pi_\phi}\,\vec{e}^{\,\pi}_0 \right), \label{eq:Gp2uniexp} \\
	 \vec{J}_2&=&\left(\frac{v}{\pi_\phi}\right)\,\vec{e}^{\,\phi}_0. \label{eq:J2unif}
\eea
A direct calculation shows that
\bea
	\vec{J}_2\cdot(\boldsymbol{\Omega}_6\vec{D}_k)=0&\,\,\,\mathrm{and}\,\,\,&\vec{J}_2\cdot(\boldsymbol{\Omega}_6\vec{S}_k)=1,
\eea
as well as
\bea
	\vec{G}_1\cdot(\boldsymbol{\Omega}_6\vec{D}_k)=0&\,\,\,\mathrm{and}\,\,\,&\vec{G}_1\cdot(\boldsymbol{\Omega}_6\vec{S}_k)=\frac{v^{2/3}}{\sqrt{3}}\left[\frac{3}{2}\left(\frac{\pi_\phi}{v}\right)^2+\Mp^2\left(\frac{k}{v^{1/3}}\right)^2\right].
\eea
Consequently, the projection of $\vec{G}_1$ and $\vec{J}_2$ onto the plane of gauge degrees of freedom are aligned one with each other, and this implementation of the uniform-expansion gauge is free of time-derivatives of the gauge degrees of freedom. Let us now turn to the matrix $\boldsymbol{G}_{\mathrm{LM}}$ which has to be non-singular. Its first row is given by $\vec{G}_1\cdot(\boldsymbol{\Omega}_6\vec{S}_k)\neq0$ and $\vec{G}_1\cdot(\boldsymbol{\Omega}_6\vec{D}_k)=0$ and its second row by the projections of $\vec{G}'_2-\dot{\vec{J}}_2+\dot{\alpha}\vec{G}_1$ onto the plane of gauge degrees of freedom. Thus, this matrix is non-singular providing that $(\vec{G}'_2-\dot{\vec{J}}_2+\dot{\alpha}\vec{G}_1)\cdot(\boldsymbol{\Omega}_6\vec{D}_k)$ is not equal to zero. However, a direct calculation shows that $\vec{G}'_2-\dot{\vec{J}}_2+\dot{\alpha}\vec{G}_1$ has a zero projection onto $\boldsymbol{\Omega}_6\vec{D}_k$ and the matrix $\boldsymbol{G}_\mathrm{LM}$ is singular. This shows that implementing the uniform-expansion gauge by imposing $\delta N=0=\delta\Theta$ leads to a pathological gauge.

\subsubsection*{Implementation 3}

Another strategy consists in imposing $\Theta\delta N+N\delta\Theta=0$ as a first gauge condition to be complemented with another gauge condition. This condition has to be of the form $\vec{G}_1\cdot\dz=0$ otherwise the gauge is pathological. As a result, the uniform-expansion gauge can be implemented introducing 
\bea
	\vec{G}_2=\frac{N\sqrt{3}}{v^{2/3}\Mp^2}\left(-\frac{\theta}{4}\,\vec{e}^{\,\phi}_1+v^{1/3}\,\vec{e}^{\,\pi}_1\right) &\,\,\,\mathrm{and}\,\,\,&\boldsymbol{\lambda}=\left(\begin{array}{cc} 0 & 0 \\ \Theta & 0 \end{array}\right),
\eea
while the vector $\vec{G}_1$ is left unspecified. First, the time-derivative of the gauge degrees of freedom have to be removed for the gauge to be non-pathological. The second gauge condition leads to a derivative condition on the phase space where
\bea
	\vec{G}'_2 &=&\frac{N\sqrt{3}}{v^{2/3}\Mp^2}\left(\frac{\theta}{2}\,\vec{e}^{\,\phi}_1-\frac{\sqrt{3}}{2}\frac{v^{2/3}\theta}{\pi_\phi}\,\vec{e}^{\,\pi}_0+v^{1/3}\,\vec{e}^{\,\pi}_1\right), \\
	\vec{J}_2 &=& \frac{3\theta}{2\Mp^2} \frac{v}{\pi_\phi} \vec{e}_0^{\,\phi}.
\eea
This gives a first constraint on the vector $\vec{G}_1$ since its projection onto the plane of gauge degrees of freedom has to be aligned with the projection of $\vec{J}_2$ onto that same plane (note that without loss of generality, we can choose these two projections to be equal). Second, the matrix $\boldsymbol{G}_\mathrm{LM}$ has to be non-singular. Since $\vec{G}^{\,\mathcal{G}}_1$ has to be aligned with $\vec{J}^{\,\mathcal{G}}_2$, the first row of $\boldsymbol{G}_\mathrm{LM}$ is proportional to the projections of $\vec{e}^{\,\phi}_0$ onto the plane of gauge degrees of freedom. They read
\bea
	\vec{e}_0^{\, \phi} \cdot (\boldsymbol{\Omega}_6\vec{S}_k) = \frac{\pi_\phi}{v} \quad &\mathrm{and}& \quad 
\vec{e}_0^{\, \phi} \cdot (\boldsymbol{\Omega}_6\vec{D}_k) = 0.
\eea
Thus, the matrix $\boldsymbol{G}_\mathrm{LM}$ is not singular if  $\vec{G}_2' - \dot{\vec{J}}_2 + \dot{\alpha} \vec{G}_1$ has a non-zero projection onto $\boldsymbol{\Omega}_6\vec{D}_k$. However, a direct computation shows that $\vec{G}_2' \cdot(\boldsymbol{\Omega}_6\vec{D}_k)=0$ and similarly for $\dot{\vec{J}}_2$ and $\vec{G}_1$. Hence, this implementation of the uniform-expansion gauge is also pathological. Note that constraining the vector $\vec{G}_1$ to have a non-zero projection onto $\boldsymbol{\Omega}_6\vec{D}_k$ is necessary for $\det(\boldsymbol{G}_\mathrm{LM})\neq0$. However, this means that the projection of $\vec{G}_1$ onto the plane of gauge degrees of freedom is not aligned anymore with the projection of $\vec{J}_2$ onto $\mathcal{G}$. As a consequence, time-derivative of gauge degrees of freedom are not entirely removed and the gauge remains pathological.

\subsubsection*{Implementation 4}

Finally, let us stress that the above implementation assumes gauge conditions such that $\mathrm{Rank}(\boldsymbol{\lambda})=1$. Instead, one could start from a gauge with $\mathrm{Rank}(\boldsymbol{\lambda})=0$ and design the gauge vectors such that a vanishing integrated-expansion appears as a dynamical consequences of the gauge conditions.  For instance, we introduce the two following gauge vectors
\bea
	\vec{G}_1&=&-\frac{\theta}{4}\,\vec{e}^{\,\phi}_1+v^{1/3}\,\vec{e}^{\,\pi}_1, \label{eq:U-Eg1} \\
	\vec{G}_2&=&N^{-1}\boldsymbol{\nabla}_\tau\vec{G}_1+C_k(\tau)\vec{G}_1, \label{eq:U-Eg2}
\eea
where $C_k(\tau)$ can be any arbitrary function of time and scale. A direct calculation leads to
\bea
	N^{-1}\boldsymbol{\nabla}_\tau\vec{G}_1&=&-\frac{\sqrt{3}}{2}v^{2/3}V_{,\phi}\,\vec{e}^{\,\phi}_0-\frac{1}{6}\left[{5}\left(\frac{\pi_\phi}{v}\right)^2+V-{\Mp^2}\left(\frac{k}{v^{1/3}}\right)^2\right]\,\vec{e}^{\,\phi}_1-\frac{\sqrt{2}\Mp^2k^2}{12v^{2/3}}\,\vec{e}^{\,\phi}_2 \nonumber \\
	&&+\frac{\sqrt{3}}{2}\frac{\pi_\phi}{v^{4/3}}\,\vec{e}^{\,\pi}_0+\frac{v^{1/3}\theta}{2\Mp^2}\,\vec{e}^{\,\pi}_1.
\eea
Let us first show that perturbations of the integrated expansion rate equal zero in that gauge. First, it is easily checked that the first gauge condition leads to $\delta\Theta=0$. Second, the perturbed lapse function is obtained from \Eq{eq:LMfix}. Since $\vec{G}_1\cdot(\boldsymbol{\Omega}_6\vec{D}_k)$ is vanishing, it boils down to 
\bea
	\delta N&=&-\left[\vec{G}_1\cdot(\boldsymbol{\Omega}_6\vec{S}_k)\right]^{-1}\,\left(\boldsymbol{\nabla}_\tau\vec{G}_1\,\cdot \dz \right).
\eea
The right-hand side of the above is proportional to $N\left[\vec{G}_2-C_k(\tau)\vec{G}_1\right]\cdot\dz$, which is vanishing once the gauge conditions are imposed. As a consequence, the perturbed lapse function equals zero, which is here a dynamical consequence of the gauge choice rather than an imposed gauge condition. Combined with $\delta\Theta=0$, this proves that $\dd\delta\mathcal{N}_\mathrm{int}/\dd\tau=0$ and so is the integrated-expansion rate. However, a direct calculation shows that $\vec{G}_2\cdot(\boldsymbol{\Omega}_6\vec{D}_k)=0$. Since $\vec{G}_1$ is also orthogonal to $(\boldsymbol{\Omega}_6\vec{D}_k)$, the matrix $\boldsymbol{G}_{\mathrm{LM}}$ is singular and this implementation of the uniform-expansion gauge is pathological too.

\bibliographystyle{JHEP}
\bibliography{SepUniv}

\end{document}